\documentclass[useAMS,usenatbib]{mn2e}
\usepackage{epsf}
\usepackage{epsfig}
\newcommand{\apj}{ApJ}

\newcommand{\apjs}{ApJS}

\newcommand{\planck}{\textit{Planck}}

\newcommand{\calsim}{\textsc{bahamas}}
\newcommand{\athena}{\textsc{athena}}

\usepackage{aas_macros}
\usepackage{hyperref} 
\hypersetup{colorlinks,citecolor=blue,linkcolor=blue,urlcolor=blue}

\title[{CMB--large-scale structure tension}]{The BAHAMAS project: the CMB--large-scale structure tension and the roles of massive neutrinos and galaxy formation}
\author[I.~G.~McCarthy et~al.]{Ian G.~McCarthy$^1$\thanks{E-mail:i.g.mccarthy@ljmu.ac.uk}, Simeon Bird$^2$, Joop Schaye$^3$, Joachim Harnois-Deraps$^4$, \newauthor Andreea S. Font$^1$, Ludovic van Waerbeke$^5$\\
$^{1}$Astrophysics Research Institute, Liverpool John Moores University, 146 Brownlow Hill, Liverpool L3 5RF\\
$^{2}$Department of Physics and Astronomy, Johns Hopkins University, 3400 N.~Charles Street, Baltimore, MD 21218, USA\\
$^{3}$Leiden Observatory, Leiden University, P. O. Box 9513, 2300 RA Leiden, the Netherlands\\
$^4$Scottish Universities Physics Alliance, Institute for Astronomy, University of Edinburgh, Blackford Hill, EH9 3HJ, Scotland\\
$^5$Department of Physics and Astronomy, University of British Columbia, 6224 Agricultural Road, Vancouver, BC, V6T 1Z1, Canada
}
\begin{document}

\date{Accepted ... Received ...}

\pagerange{\pageref{firstpage}--\pageref{lastpage}} \pubyear{2016}

\maketitle

\label{firstpage}

\begin{abstract}
  Recent studies have presented evidence for tension between the constraints on $\Omega_{\rm m}$ and $\sigma_8$ from the cosmic microwave background (CMB) and measurements of large-scale structure (LSS).  This tension can potentially be resolved by appealing to extensions of the standard model of cosmology and/or untreated systematic errors in the modelling of LSS, of which baryonic physics has been frequently suggested.  We revisit this tension using, for the first time, carefully-calibrated cosmological hydrodynamical simulations, which thus capture the back reaction of the baryons on the total matter distribution.  We have extended the \calsim~simulations to include a treatment of massive neutrinos, which currently represents the best motivated extension to the standard model.   We make synthetic thermal Sunyaev-Zel'dovich effect, weak galaxy lensing, and CMB lensing maps and compare to observed auto- and cross-power spectra from a wide range of recent observational surveys.  We conclude that: i) in general there is tension between the primary CMB and LSS when adopting the standard model with minimal neutrino mass; ii) after calibrating feedback processes to match the gas fractions of clusters, the remaining uncertainties in the baryonic physics modelling are insufficient to reconcile this tension; and iii) if one accounts for internal tensions in the \planck~CMB dataset (by allowing the lensing amplitude, $A_{\rm Lens}$, to vary), invoking a non-minimal neutrino mass, typically of 0.2-0.4 eV, can resolve the tension.  This solution is fully consistent with separate constraints from the primary CMB and baryon acoustic oscillations.
\end{abstract}


\section{Introduction}
\label{sec:intro}

It has long been recognized that measurements of the growth of large-scale structure (LSS) can provide powerful tests of our cosmological framework (e.g., \citealt{Peebles1980,Bond1980,Blumenthal1984,Davis1985,Kaiser1987,Peacock1994}).  Importantly, growth of structure tests are independent of, and complementary to, constraints that may be obtained from analysis of the temperature and polarization fluctuations in the cosmic microwave background (CMB) and to so-called geometric probes, such as Type Ia supernovae and baryon acoustic oscillations (BAOs) \citep{Albrecht2006}. 

The consistency between these various probes has been heralded as one of the strongest arguments in favour of the current standard model of cosmology, the $\Lambda$CDM model.  The successes of the model, which contains only six adjustable degrees of freedom, are numerous and impressive.  However, the quality and quantity of observational data used to constrain the model has been undergoing a revolution and a few interesting `tensions' (typically at the few sigma level) have cropped up recently that {\it may} suggest that a modification of the standard model is in order.

One of the tensions surrounds the measured value of Hubble's constant, $H_0$.  Local estimates prefer a relatively high value of $73\pm2$ km/s/Mpc \citep{Riess2016}, whereas analysis of the CMB and BAOs prefer a relatively low value of $67\pm1$ km/s/Mpc \citep{Planck2015_cmb}.  A separate tension arises when one compares various LSS joint constraints\footnote{The joint constraint is often parametrised as $S_8 \equiv \sigma_8 \sqrt{\Omega_{\rm m}/0.3}$.} on the matter density, $\Omega_{\rm m}$, and the linearly-evolved amplitude of the matter power spectrum, $\sigma_8$, with constraints on these quantities from \planck~measurements of the primary CMB.  In particular, a number of LSS data sets (e.g., \citealt{Heymans2013,Planck2015_clusters,Hildebrandt2017}) appear to favour relatively low values of $\Omega_{\rm m}$ and/or $\sigma_8$ compared to that preferred by the CMB data.  (We summarize these constraints in detail in Section \ref{sec:tension}.)  Our focus here is on this latter tension.

There are three (non-mutually exclusive) possible solutions to the aforementioned CMB-LSS tension: i) there are important and unaccounted for systematic errors in the measurements of the primary CMB data; and/or ii) there are remaining systematics in either the LSS measurements or in the physical modelling of the LSS data (e.g., inaccurate treatment of non-linear or baryon effects); and/or iii) the standard model is incorrect.  While exploration of measurement systematics in both the CMB and LSS data is clearly a high priority, significant focus is also being devoted to the question of LSS modelling systematics, as well as to making predictions for possible extensions to the standard model of cosmology.  In the present study, we zero in on these modelling issues.

We first point out that the different LSS tests (e.g., Sunyaev-Zel'dovich power spectrum, cosmic shear, CMB lensing, cluster counts, galaxy clustering, etc.) are just different ways of characterising the `lumpiness' of the matter distribution and how these lumps cluster in space.  On very large scales (i.e., in the linear regime), perturbation theory is sufficiently accurate to calculate the matter distribution.  However, most of the tests mentioned above probe well into the non-linear regime.  The standard approach to modelling the matter distribution is therefore either to calibrate the so-called `halo model' using large dark matter cosmological simulations, or to use such simulations to empirically correct calculations based on linear theory (as in, e.g., the HALOFIT package; \citealt{Smith2003,Takahashi2012}). 

If the matter in the Universe was composed entirely of dark matter, such approaches would likely be highly accurate (assuming the analytic models could be accurately calibrated).  However, baryons contribute a significant fraction of the matter density of the Universe and recent simulation work has shown that feedback processes associated with galaxy and black hole formation can have a significant effect on the spatial distribution of baryons, which then induces a non-negligible back reaction on the dark matter (e.g., \citealt{vanDaalen2011,vanDaalen2014,Velliscig2014,Schneider2015,Mummery2017,Springel2017}).  Until quite recently such effects have typically been ignored when modelling LSS data, which might be expected to lead to significant biases in the inferred cosmological parameters \citep{Semboloni2011}.  Recent cosmic shear studies (e.g., \citealt{Hildebrandt2017}), however, have attempted to account for the effects of baryons in the context of the halo model.

A separate modelling issue, which has so far attracted significantly less attention, is that the different LSS tests typically use quite different modelling approaches.  For example, modelling of the galaxy cluster counts typically involves using parametrisations of the halo mass function from dark matter-only simulations, while modelling of galaxy clustering normally involves using the so-called Halo Occupation Distribution (HOD) approach that takes relatively weak guidance from simulations, and modelling of weak lensing often uses linear theory with non-linear corrections.  These differences likely reflect the fact that different aspects of the matter distribution are being probed by the different tests, but it does raise the important question of how appropriate it is to compare/combine the results of different LSS tests when they do not assume the same underlying matter distribution for a given cosmology.

Cosmological hydrodynamical simulations are the only method capable of self-consistently addressing the modelling limitations discussed above.  Such simulations start from cosmological initial conditions and follow the evolution of matter into the non-linear regime, solving simultaneously for the gas, stellar, black hole, and dark matter evolution in the presence of an evolving cosmological background.  The back reaction of the baryons onto the dark matter is therefore modelled self-consistently.  As all of the important matter components are followed, it is possible to create virtual observations to make like-with-like comparisons with the full range of LSS tests, whether they are based on galaxies, the hot gas, or lensing produced by the total matter distribution.  Hydro simulations therefore offer a means to address the issue of the lack of consistency in the modelling in different LSS fields.  

As the simulations track star formation and black hole accretion, they also offer a means to account for the effects of `cosmic feedback'.  This is a difficult problem though, as the feedback originates on scales that are too small to resolve with the kind of large-volume simulations required to do LSS cosmology.  Therefore, one must employ physically-motivated `subgrid' prescriptions to take these processes into account.  Recent studies have highlighted that many aspects of the simulations are more sensitive to the details of the subgrid modelling than one might hope (e.g., \citealt{Schaye2010,LeBrun2014,Sembolini2016}), calling into question their ab initio predictive power.  On the positive side, however, one can learn about these processes by assessing which models give rise to systems that resemble those in the real Universe.  Remarkable progress has been made in this regard recently, to the point where it is now possible to produce simulations that are difficult to distinguish from the real Universe in many respects.

Note that although current large-volume simulations lack the resolution to directly simulate the initiation of outflows on small scales (typically below scales of 1 kpc), the effects of feedback on larger scales can be directly simulated.  This is relevant for LSS cosmology, where the typical length scales are $>$ 1 Mpc. Thus, if we can calibrate physically-motivated prescriptions for the small-scale physics against observational constraints on some judiciously-chosen properties, we can strongly increase the predictive power of the simulations for other observables.  In other words, with calibration of physical feedback models we can strongly reduce the main theoretical limitation in current LSS cosmology tests.

This calibration approach is now being adopted by several groups in the theoretical galaxy formation field and has yielded significant progress (e.g., \citealt{Vogelsberger2014,Schaye2015,Crain2015,Pillepich2018}).  The emphasis of these projects has been on simulating, at relatively high resolution, the main galaxy population (stellar masses of $\sim10^{8-11}$ M$_\odot$).  The simulations were calibrated on important galaxy properties (stellar masses and sizes in the case of EAGLE; \citealt{Schaye2015}) and it has been shown that they are able to reproduce other properties of the galaxy population quite well.

For LSS cosmology, much larger (and many more) simulations are required than considered previously.  Additionally, while having realistic galaxy properties is clearly desirable, it is not sufficient to judge whether the feedback effects on LSS have been correctly captured in the simulations.  That is because most of the baryons are not in the form of stars/galaxies, but in a diffuse, hot state.  Thus, the simulations should reproduce the hot gas properties well if we are to trust the predictions for LSS.

In \citet{McCarthy2017} (hereafter M17) we introduced the \calsim~simulations, which were designed specifically with LSS cosmology in mind.  The stellar and AGN feedback prescriptions were carefully calibrated to reproduce the observed baryon fractions of massive systems (see Section \ref{sec:sims}), but M17 demonstrated that the simulations also reproduced an extremely wide range of observations, including the various observed mappings between galaxies, hot gas, total mass, and black holes.  For example, the simulations reproduce the observed X-ray and thermal Sunyaev-Zel'dovich effect scaling relations of galaxy groups and clusters (including their intrinsic scatter), the thermodynamical radial profiles of the intracluster medium (density, pressure, etc.), the stellar mass--halo mass relations of galaxies and its split into centrals and satellites, the radial distribution of satellite stellar mass in groups and clusters, and the evolution of the quasar luminosity function.

Here we employ the \calsim~simulations to revisit the claimed tension between LSS and the primary CMB.  We focus here on comparisons to the thermal Sunyaev-Zel'dovich (tSZ) effect, cosmic shear, CMB lensing, and their various cross-correlations.  We also extend \calsim~to include a contribution from massive neutrinos to the dark matter, which has previously been proposed in a number of studies (e.g., \citealt{Battye2014,Beutler2014,Wyman2014}) as a solution to the aforementioned tension.  We constrain the summed mass of neutrinos, $M_\nu$, through the various LSS tests.  In terms of the neutrino simulations, our approach to choosing the other relevant cosmological parameters (e.g., $H_0$, $\Omega_{\rm m}$, etc.) is to take guidance from primary CMB constraints and to assess which range of $M_\nu$, if any, can resolve the CMB-LSS tension (see Section \ref{sec:cosmology}).

The present paper is organized as follows.  In Section \ref{sec:tension}, we summarize the CMB-LSS tension and motivate our cosmological parameter selection strategy.  In Section \ref{sec:sims}, we summarize the technical details of the \calsim~simulations and its calibration strategy.  In Section \ref{sec:degen}, we explore the possible degeneracy between our feedback calibration strategy and cosmological parameter determination.  In Section \ref{sec:results} we present our main results, based on comparing synthetic observations of the simulations to a wide variety of LSS observables.  Finally, in Section \ref{sec:discuss} we summarize and discuss our findings.

\begin{figure*}
\includegraphics[width=0.7\textwidth]{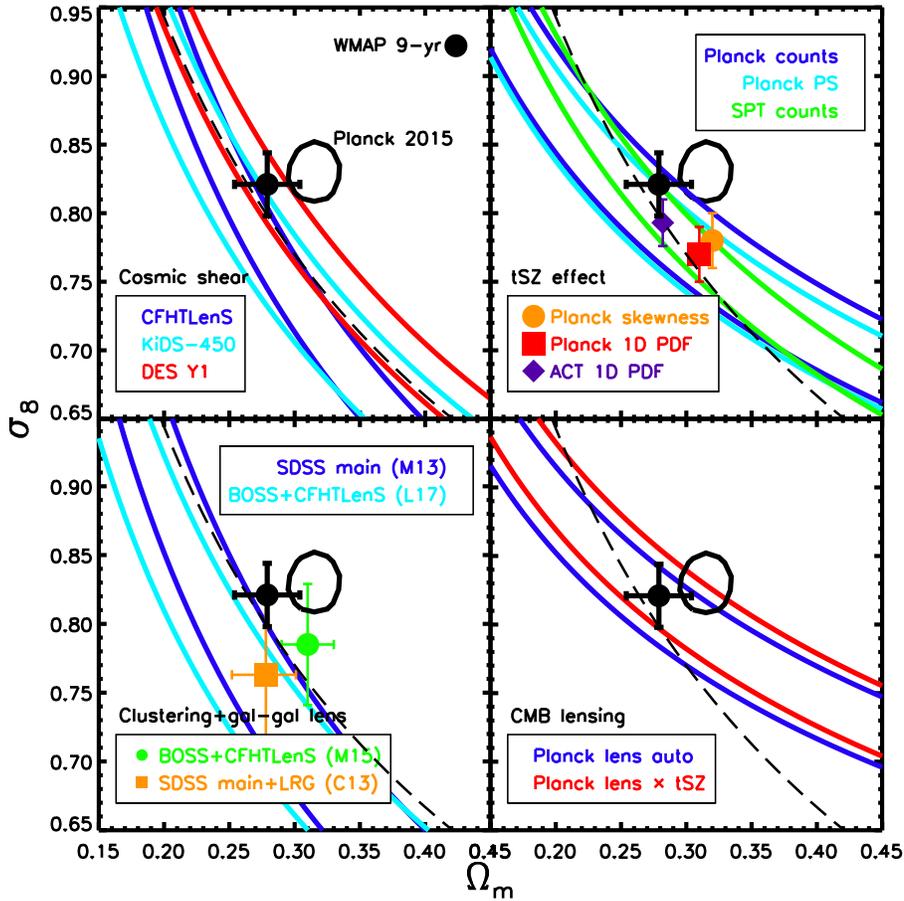}
\caption{\label{fig:sig8_omegam}
  Summary of recent LSS constraints in the $\sigma_8-\Omega_{\rm m}$ plane, compared with \planck~2015 primary CMB constraints (TT+lowTEB, closed contour repeated in each panel) and WMAP 9-yr primary CMB constraints (filled black circle with thick error bars).  {\it Top left:} Cosmic shear results from CFHTLenS, DES, and KiDS.  {\it Top right:} Various tSZ effect tests, including \planck~2015 cluster number counts, angular power spectrum, 1-point PDF, and a combined analysis of the skewness and bi-spectrum of \planck~2015 Compton $y$ map, a 1-point PDF constraints from the Atacama Cosmology Telescope (ACT), and tSZ cluster count constraints from the South Pole Telescope (SPT).  {\it Bottom left:} Combined galaxy clustering plus galaxy-galaxy lensing constraints from SDSS main galaxy catalog (M13), SDSS main galaxy catalog plus Luminous Red Galaxies (C13), SDSS BOSS galaxy clustering plus CFHTLenS lensing (M15), and SDSS BOSS galaxy clustering plus CFHTLenS and CS82 weak lensing data (L17).  {\it Bottom right:} Constraints from the \planck~CMB lensing autocorrelation function and from the cross-correlation function between \planck~CMB lensing and \planck~Sunyaev-Zel'dovich effect maps.  The curves represent the best-fit power laws (derived by the original authors) describing the degeneracy between $\sigma_8$ and $\Omega_{\rm m}$ for the different datasets.  There are two curves for each dataset, representing the $\pm$1-sigma uncertainties in the best-fit amplitude of the power law.  To help compare the different LSS tests, we show in each panel, as the black dashed curve, a power law of the form $S_8 \equiv \sigma_8 (\Omega_{\rm m}/0.3)^{1/2} = 0.77$.  The various LSS constraints consistently (at the $\approx$1-3 sigma level) point to lower values of $\sigma_8$ at fixed $\Omega_{\rm m}$ (or lower values of $\Omega_{\rm m}$ at fixed $\sigma_8$) compared to that derived from the most recent primary CMB data from \planck.    
}
\end{figure*}

\section{CMB-LSS tension and previous constraints on neutrino mass}
\label{sec:tension}

A number of recent studies, which used simple analytic modelling\footnote{Here we collectively refer to halo model-based modelling, Halo Occupation Distribution (HOD) modelling, and linear theory+non-linear corrections, as in the HALOFIT package often used to predict lensing.  Note that none of these methods self-consistently treat the evolution of baryons and dark matter, they are usually guided by the results of dark matter-only simulations.} of LSS, have found that there is presently tension between the constraints in the $\sigma_8-\Omega_{\rm m}$ plane derived from various LSS tests and that derived from the CMB, particularly so for the recent \planck~results.  (Note that $\sigma_8$ is defined as the linearly-evolved present-day amplitude of the matter power spectrum on a scale of $8 h^{-1} {\rm Mpc}$; i.e., it is the root mean square of the mass density in a sphere of radius $8 h^{-1} {\rm Mpc}$ in linear theory.)

We summarize recent LSS constraints in Fig.~\ref{fig:sig8_omegam}.  The four panels correspond to different LSS observables, including cosmic shear, tSZ effect statistics, galaxy clustering plus galaxy-galaxy lensing, and CMB lensing.  In the top left panel we show recent cosmic shear results from the CFHTLenS (\citealt{Kilbinger2013}; see also \citealt{Heymans2013}), DES \citep{Troxel2017}, and KiDS \citep{Hildebrandt2017} surveys.  In the top right panel we show various tSZ effect tests, including cluster number counts \citep{Planck2015_clusters,deHaan2016}, the power spectrum, 1-point PDF, and a combined analysis of the skewness and bi-spectrum of the \planck~Compton $y$ map \citep{Planck2015_sz}.  Also shown are independent 1-point PDF constraints from ACT data \citep{Hill2014b}.  In the bottom left panel we show recent combined galaxy clustering plus galaxy-galaxy lensing constraints using the SDSS main galaxy catalog \citep{Mandelbaum2013}, SDSS main galaxy catalog plus Luminous Red Galaxies \citep{Cacciato2013}, SDSS BOSS galaxy clustering plus CFHTLenS lensing \citep{More2015}, and SDSS BOSS galaxy clustering plus CFHTLenS and CS82 weak lensing data \citep{Leauthaud2017}.  In the bottom right panel we show constraints from modelling the \planck~CMB lensing autocorrelation function \citep{Planck2015_lensing} and the cross-correlation function between \planck~CMB lensing and \planck~tSZ effect maps \citep{Hill2014a}.  The curves represent the best-fit power laws (derived by the original authors) describing the degeneracy between $\sigma_8$ and $\Omega_{\rm m}$ for the datasets.  There are two curves for each dataset, representing the $\pm$1-sigma uncertainties in the best-fit amplitude of the power law.  Note that for some of the tSZ effect tests (data points with errors), $\Omega_{\rm m}$ was held fixed at the (\planck) primary CMB best-fit value and only $\sigma_8$ was constrained by the data.  Note also that, with the exception of the DES Y1 analysis, all of the LSS results presented in Fig.~\ref{fig:sig8_omegam} were derived assuming either massless neutrinos or adopt the minimum mass ($\approx 0.06$ eV) allowed by oscillation experiments.  The DES Y1 analysis allowed the summed neutrino mass to be a free parameter.

The various LSS constraints consistently, at the $\approx$1-3 sigma level, prefer lower values of $\sigma_8$ at fixed $\Omega_{\rm m}$ (or lower values of $\Omega_{\rm m}$ at fixed $\sigma_8$) compared to that derived from the most recent primary CMB data from \planck.  The consistency amongst the different LSS tests is rather remarkable, given the very different nature of the tests involved, which probe different aspects of the matter distribution (i.e., galaxies vs.\ hot gas vs.\ total matter) at different redshifts and on different scales, each with their own differing sets of systematic errors.  And note that the constraints shown in Fig.~\ref{fig:sig8_omegam} do not form an exhaustive list.  For example, other recent LSS tests, such as those based on the cross-correlations between CMB lensing and galaxy overdensity \citep{Giannantonio2016}, CMB lensing and cosmic shear \citep{Liu2015,HarnoisDeraps2017}, and cosmic shear and the tSZ effect \citep{Hojjati2015,Hojjati2017}, also find qualitative evidence for tension (and in the same sense), but we do not plot them in Fig.~\ref{fig:sig8_omegam} since they have not formerly quantified their best-fit cosmological parameter values and their uncertainties.  

The role that remaining systematics in either the analysis of the CMB (e.g., \citealt{Spergel2015,Addison2016,Planck2016}) or that of LSS (such as the neglect of important baryon physics, which we will consider here) plays in this tension has yet to be fully understood.  In spite of this, various extensions of the standard model have already been proposed to try to reconcile the apparent tension.  One of the most interesting and well-motivated proposed solutions is that of a non-negligible contribution from massive neutrinos.  Neutrinos affect the growth of LSS in two ways: i) by altering the expansion history of the Universe, as neutrinos are relativistic at early times (and therefore evolve like radiation) but later become non-relativistic (evolving in the same way as normal matter); and ii) their high streaming motions allow them to free-stream over large distances, resisting gravitational collapse and slowing the growth of density fluctuations on scales smaller than the free-streaming scale.  The latter effect is the more important one for LSS.  Note that the CMB is also somewhat sensitive to the presence of massive neutrinos, via the change in the expansion history (which alters the distance to the surface of last scattering and therefore the angular scale of the acoustic peaks) and also via their free-streaming effects on high-redshift LSS that gives rise to CMB lensing.

Neutrinos are a well-motivated addition to the standard model of cosmology as the results of atmospheric and solar oscillation experiments imply that the three active species of neutrinos have a {\it minimum} summed mass, $M_\nu$, of 0.06 eV (0.1 eV) when adopting a normal (inverted) hierarchy (see \citealt{Lesgourgues2006} for a review).  As we will show later, even adopting the minimum allowed mass has noticeable effects on LSS, which should be within reach of upcoming surveys such as Advanced ACTpol, \textsc{Euclid}, and LSST.

Previous studies combining simple physical modelling of LSS with primary CMB constraints (sometimes also including BAO, $H_0$ and/or SNIa constraints) have indeed found a preference for a non-zero summed neutrino mass, at the level $M_\nu\approx0.3$-$0.4$ eV with a typical statistical error of $\approx0.1$ eV (e.g., \citealt{Battye2014,Beutler2014,Wyman2014}).  Note that the CMB alone (TT+lowP) constrains $M_\nu\la0.70$ eV \citep{Planck2015_cmb}, whereas for LSS alone, $M_\nu$ is usually highly degenerate with $\sigma_8$ and $\Omega_{\rm m}$.  Combining the CMB with LSS allows one to break this degeneracy and obtain much tighter constraints on $M_\nu$ than either of the individual probes can provide.

However, a number of important objections have been raised about massive neutrinos as a solution to the CMB-LSS tension.  For example, \citet{Planck2015_cmb} note that in order to preserve the fit to the CMB, raising the value of the summed neutrino mass (from the minimum of 0.06 eV adopted in their analysis) requires lowering the value of Hubble's constant, $H_0$, in order to preserve the observed acoustic peak scale.  Lowering Hubble's constant would then exacerbate the tension that exists between the CMB(+BAO) constraints on $H_0$ and cosmic distance ladder-based estimates (e.g., \citealt{Riess2016}).  In addition, \citet{MacCrann2015} have argued that when one considers the full $n$-parameter space in the standard model, adding massive neutrinos does not, in any case, significantly resolve the tension between the CMB and LSS in the $\sigma_8-\Omega_{\rm m}$ plane (the individual constraints on $\sigma_8$ and $\Omega_{\rm m}$ do weaken, but the joint constraint runs nearly parallel to, but offset from, the LSS constraints; see their Fig.\ 5).  Finally, \citet{Planck2015_cmb} find that the combination of the CMB with BAO (the latter of which places strong constraints on $H_0$ and $\Omega_{\rm m}$) places strong (95\%) upper limits of $M_\nu \la 0.21$ eV (but see \citealt{Beutler2014} for different conclusions), while \citet{Lyman2015} (see also \citealt{Lyman2017}) find that the combination of \planck~CMB data with measurements of the Lyman-alpha forest power spectrum at $2 \la z \la 4$ constrains $M_\nu < 0.12$ eV (95\% C.~L.).  Both of these constraints are lower than what previous LSS studies claim is required to resolve the aforementioned CMB-LSS tension.

\begin{figure*}
\includegraphics[width=0.99\textwidth]{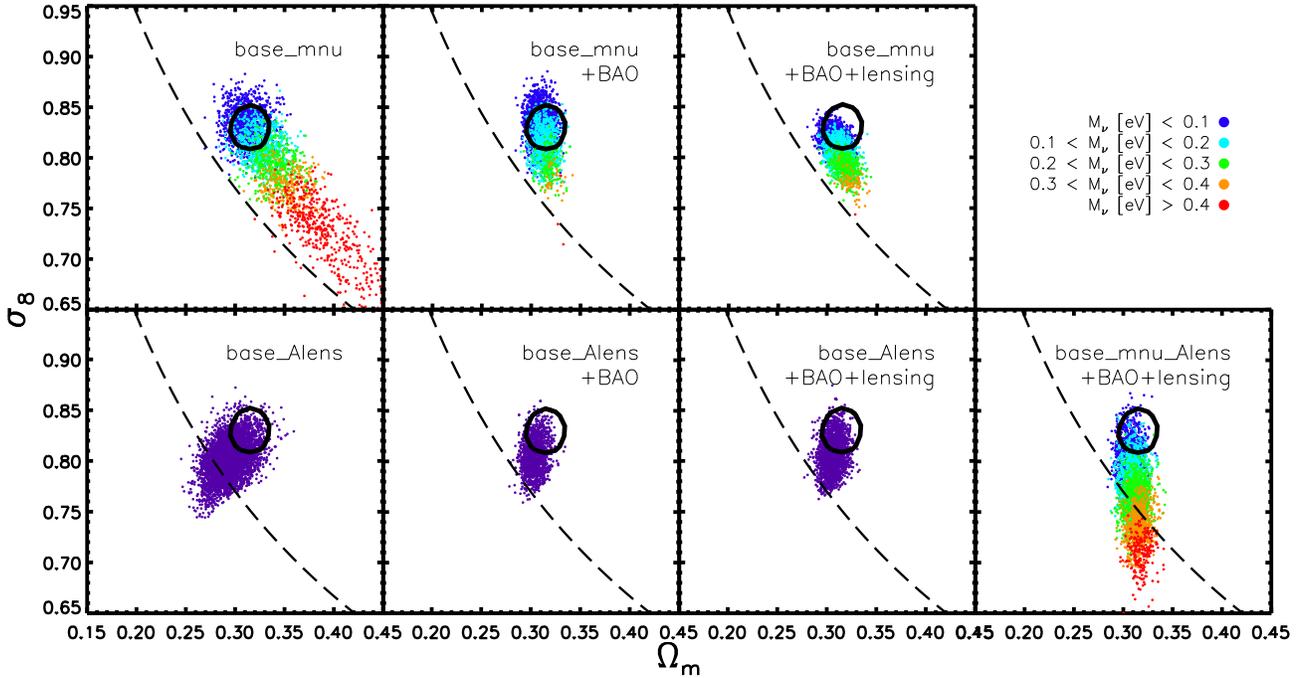}
\caption{\label{fig:cmb_mnu}
  Constraints in the $\sigma_8-\Omega_{\rm m}$ plane extracted from different sets of \citet{Planck2015_cmb} Markov chains.  {\it Top left:} the case of a standard 6 parameter $\Lambda$CDM model (base) + a single parameter characterising the summed mass of neutrinos (mnu) where only primary CMB (Planck TT+lowTEB) is used to constrain the model.  {\it Top middle:} Adopts the same model and uses the same CMB data but also adds external BAO constraints.   {\it Top right:} Adds further constraints from modelling of the \planck~CMB lensing power spectrum.  In all of these cases $A_{\rm Lens}$ is fixed to unity.  In the three left-most panels in the bottom row, $A_{\rm Lens}$ is allowed to vary while the summed mass of neutrinos is fixed to 0.06 eV (i.e., the minimum allowed by oscillation experiments).  These three panels mirror those in the top row in terms of the data sets used to constrain the cosmological parameters.  {\it Bottom right:} Both $A_{\rm Lens}$ and $M_\nu$ are allowed to vary.  In all panels the black circular and black dashed curves have the same meaning as in Fig.\ref{fig:sig8_omegam}.  The dots represent randomly extracted parameter sets from the Markov chains (taking into account their weighting) and are coloured by the summed mass of neutrinos, $M_\nu$, for cases where this parameter is allowed to vary.  The constraints on $\sigma_8-\Omega_{\rm m}$ and on $M_\nu$ depend strongly on whether one includes external data sets (particularly BAO) and on whether the lensing amplitude scale factor, $A_{\rm Lens}$, is fixed or marginalized over.
}
\end{figure*}

\subsection{Implications of remaining CMB systematics}
\label{sec:alens}

It is important to emphasise that the \planck~CMB constraints on the summed mass of neutrinos, whether in combination with other probes such as BAO or not, depend upon whether one takes account of known residual systematics in the primary CMB data.  In particular, it has been shown in a number of previous studies (e.g., \citealt{Planck2013,Addison2016,Planck2016}) that sizeable (1-2 sigma) shifts in the best-fit parameters can occur depending on which range of multipoles one analyses in the primary CMB data and we show below that this has significant implications for the constraints on $M_\nu$.  \citet{Planck2016} argue that these shifts are due to both an apparent deficit of power at low multipoles ($\ell \la 30$) and an enhanced `smoothing' of the peaks and troughs in the TT power spectrum at high multipoles ($\ell \ga 1000$), similar to that induced by gravitational lensing.  The latter appears to be most relevant for shifts in $\sigma_8$ (and therefore for the constraints on $M_\nu$), and hence for the CMB-LSS tension.

\citet{Addison2016} have shown that one can mitigate the effects of the enhanced smoothing by allowing the amplitude of the CMB lensing power spectrum, $A_{\rm Lens}$, to be free when fitting the TT power spectrum (see also \citealt{Calabrese2008}), rather than fixing its natural value of unity\footnote{The lensing amplitude can be directly calculated using linear theory given a set of cosmological parameters.  The amplitude can then be scaled by a fixed value of $A_{\rm Lens}$.  The natural (unscaled) value corresponds to $A_{\rm Lens}=1$.}.  Allowing $A_{\rm Lens}$ to be a free parameter, the \planck~data prefers a higher value of $A_{\rm Lens}\approx1.2\pm0.1$, which is consistent with the apparent extra smoothing (relative to a model with $A_{\rm Lens}=1.0$) visible in the TT power spectrum.  We stress here that this does not imply that the CMB lensing calculation is in error.  It more likely reflects some other subtle unaccounted for systematic issue.  In any case, marginalizing over $A_{\rm Lens}$ appears to be a reasonable and practical way to resolve the issue and results in best-fit cosmological parameters that are much less sensitive to the choice of multipole range over which one fits the data \citep{Addison2016}.

To demonstrate the importance of these issues for cosmological parameter selection, we show in Fig.~\ref{fig:cmb_mnu} how allowing $M_\nu$ and $A_{\rm Lens}$ to vary (separately and together) impacts the CMB constraints in the $\sigma_8-\Omega_{\rm m}$ plane.  We focus first on the top row, for which $M_\nu$ is allowed to vary while $A_{\rm Lens}$ is held fixed to unity.  The left panel shows the case of a standard 6 parameter $\Lambda$CDM model (base) + a single parameter characterising the summed mass of neutrinos (`mnu') where only the primary CMB (\planck~TT+lowTEB) is used to constrain the model.  The middle panel adopts the same model and uses the same CMB data but also adds external BAO constraints.   The right panel in the top row adds further constraints from modelling of the \planck~CMB lensing power spectrum, measured using the four-point function.

%Finally, the bottom right panel represents the same setup as the bottom left panel, except that the $A_{\rm Lens}$ parameter (i.e., the lensing amplitude used in the TT modelling) has been marginalized over, rather than fixing it to unity.

Focusing on the top left panel, we see that a wide range of $M_\nu$ values are allowed by the {\it Planck} primary CMB data.  Furthermore, the constraints on the $\sigma_8-\Omega_{\rm m}$ plane are much weaker in comparison to the case where $M_\nu$ is fixed to the minimum value of 0.06 eV (compare coloured dots to the solid black contour).  However, as noted previously by \citet{MacCrann2015} (see also \citealt{Joudaki2017b}), allowing $M_\nu$ to vary does not bring the CMB constraints on $\sigma_8-\Omega_{\rm m}$ into significantly better agreement with those of LSS, as the degeneracies from the two sets of constraints run approximately parallel to one another other (compare the coloured dots to the dashed curve).  Furthermore, as noted by \citet{Planck2015_cmb}, higher values of $M_\nu$ generally result in lower values of $H_0$ (not shown), in order to preserve the angular scale of the CMB acoustic peaks, thereby increasing the previously mentioned tension with local $H_0$ determinations.

The inclusion of external constraints from BAO observations (top middle panel of Fig.~\ref{fig:cmb_mnu}) greatly reduces the allowed range of $M_\nu$ while also pegging the $\sigma_8-\Omega_{\rm m}$ constraints back close to those derived from the standard model with $M_\nu=0.06$ eV held fixed (compare coloured dots to solid black contour).  It is important to note that the addition of BAO data also strongly constrains $H_0$, to $67\pm1$ km/s.  

The further introduction of external constraints based on the modelling of the observed CMB lensing power spectrum (top right panel) does not allow for significantly higher summed neutrino masses, but it does result in a downward $\approx$1-sigma shift in $\sigma_8$.  That the constraints shift down slightly is not surprising, as we have already noted that the analysis of the CMB lensing power spectrum alone leads to a $\sigma_8-\Omega_{\rm m}$ relation that is lower in amplitude than preferred by the primary CMB (\citealt{Planck2015_lensing}; see also bottom right panel of Fig.~\ref{fig:sig8_omegam}).  It is interesting to note that the primary effect of incorporating the CMB lensing constraints is a downward shift in $\sigma_8$ only, whereas it might have been anticipated that that there would be a shift in both $\sigma_8$ and $\Omega_{\rm m}$, given the degeneracy between these two quantities for CMB lensing (Fig.~\ref{fig:sig8_omegam}).  However, opposing constraints from the external BAO datasets strongly pin down the values of $\Omega_{\rm m}$ and $H_0$ (not shown) while placing no direct constraints on $\sigma_8$.  The combination of BAO and CMB lensing therefore helps to break the $\sigma_8-\Omega_{\rm m}$ degeneracy in the CMB lensing constraints.

In all the cases considered above, the lensing amplitude $A_{\rm Lens}$ was held fixed to unity when modelling the primary CMB TT data.  In the three left-most panels of the bottom row in Fig.~\ref{fig:cmb_mnu}, we consider the case where $A_{\rm Lens}$ is allowed to vary while the summed neutrino mass is held fixed to 0.06 eV, mirroring the data sets used in the three panels in the top row.  Here we see that marginalising over $A_{\rm Lens}$ results in a preference for lower values of $\Omega_{\rm m}$ and $\sigma_8$.  When BAO constraints are included, the main effect of marginalizing over $A_{\rm Lens}$ is a downward shift in $\sigma_8$.  Comparing these constraints to those derived from LSS in Fig.~\ref{fig:sig8_omegam}, it is clear that allowing $A_{\rm Lens}$ to vary already goes a good distance towards resolving the overall tension between the primary CMB and LSS and completely resolves it for some specific cases (e.g. DES Y1, CMB lensing constraints), although it should be borne in mind that many of the constraints in Fig.~\ref{fig:sig8_omegam} do not include potentially important baryonic effects.

While allowing $A_{\rm Lens}$ to vary does reduce the tension, it does not completely remove it for the case where the summed neutrino mass is held fixed at the minimum value allowed by ocsillation experiments.  Furthermore, since there is no strong a priori reason why the summed mass of neutrinos should be the minimum value, this parameter should be allowed to vary and to be constrained by astrophysical experiments.  In the bottom right panel of  Fig.~\ref{fig:cmb_mnu}, we therefore show the constraints on $M_\nu$ and $\sigma_8-\Omega_{\rm m}$ when $A_{\rm Lens}$ is marginalized over (i.e., both $M_\nu$ and $A_{\rm Lens}$ are allowed to vary).  Interestingly, while $\Omega_{\rm m}$ is still well determined (due to the addition of BAO), the constraints on $\sigma_8$ and $M_\nu$ are significantly broader compared to the case where $A_{\rm Lens}$ is fixed to unity.  Thus, if one takes into account the apparent residual systematics remaining in the high-multipole primary CMB data, by marginalizing over $A_{\rm Lens}$, massive neutrinos may potentially provide a full reconciliation of the primary CMB and LSS data sets.  We say `may' as it has yet to be demonstrated that current LSS cosmological constraints (e.g., those described in Fig.~\ref{fig:sig8_omegam}) are robust to the modifications induced by baryonic physics, such as AGN feedback.  This is far from clear at present and is one of the main issues that we seek to address with \calsim.

With regard to the recent constraints on $M_\nu$ using measurements of the Lyman-alpha forest power spectrum by \citet{Lyman2015} and \citet{Lyman2017}, we first point out that the Lyman-alpha forest alone only constrains $M_\nu \la 1$ eV.  The strong upper limits placed on $M_\nu$ in these studies ($M_\nu < 0.12$ eV) come from the combination with the \planck~primary CMB data.  Both of the studies mentioned above use the fiducial \planck~CMB Markov chains which adopt $A_{\rm Lens} = 1$, finding an upper limit on the summed neutrino mass that is only just above the minimum value allowed by neutrino oscillation experiments.  We speculate that if the Lyman-alpha forest measurements were instead combined with the \planck~chains for the case where $A_{\rm Lens}$ is allowed to vary, that the derived constraints on $M_\nu$ may actually be in tension with neutrino oscillation experiments.  (This is just because marginalizing over $A_{\rm Lens}$ tends to lower the best-fit value of $\sigma_8$ from the primary CMB, which would in turn reduce the best-fit value of $M_\nu$.)  Such a tension would suggest that there are still relevant systematic errors in the Lyman-alpha forest data and/or modelling (e.g., \citealt{Rogers2017}).  

Finally, it is worth noting that the Lyman-alpha forest constraints on the spectral index, $n_{\rm s}$, are in tension with constraints from \planck, with the Lyman-alpha forest data preferring a relatively low value of $n_{\rm s} = 0.938 \pm 0.010$ \citep{Lyman2015} while the \planck~CMB data constrains $n_{\rm s} = 0.9655 \pm 0.0062$ \citep{Planck2015_cmb}, representing a $\approx3$-sigma difference.  This indicates that the Lyman-alpha forest data does actually prefer less small-scale power than predicted given the standard model of cosmology with primary CMB constraints.  It is the shape of the Lyman-alpha power spectrum that allows one to individually constrain $M_\nu$ and $n_{\rm s}$ (or, alternatively, the running of spectral index, ${\rm d} n_{\rm s} / {\rm dln} k$).  Even a subtle scale-dependent bias could have significant implications for the individual constraints on $M_\nu$, $\sigma_8$, and $n_{\rm s}$.

\section{Simulations}
\label{sec:sims}

\subsection{BAHAMAS}
\label{sec:bahamas}

We use the \calsim~suite of cosmological hydrodynamical simulations to predict the various LSS diagnostics (e.g., cosmic shear, tSZ power spectrum, etc.) in the context of massive neutrino cosmologies.  Here we provide a brief summary of the simulations, including their feedback calibration strategy, but we refer the reader to M17 for further details.

The \calsim~suite of cosmological hydrodynamical simulations consists of $400 \ {\rm Mpc}/h$ comoving on a side, periodic box simulations containing $2 \times 1024^3$ particles.  We use 11 runs from that suite here, which vary the cosmological parameter values, including the summed mass of neutrinos, as discussed in detail in Section \ref{sec:nu_sims}. The Boltzmann code {\small CAMB}\footnote{\url{http://camb.info/}} (\citealt{Lewis2000}; April 2014 version) was used to compute the transfer functions and a modified version of {\small N-GenIC} to create the initial conditions, at a starting redshift of $z=127$.  {\small N-GenIC} has been modified by S.\ Bird to include second-order Lagrangian Perturbation Theory corrections and support for massive neutrinos\footnote{\url{https://github.com/sbird/S-GenIC}}.  Note that when producing the initial conditions, we use the separate transfer functions computed by {\small CAMB} for each individual component (baryons, neutrinos, and CDM), whereas in most existing cosmological hydro simulations the baryons and CDM adopt the same transfer function, corresponding to the total mass-weighted function.  Note also that we use the same random phases for each of the simulations, implying that comparisons between the different runs are not subject to cosmic variance complications.

\begin{figure*}
\includegraphics[width=0.995\columnwidth]{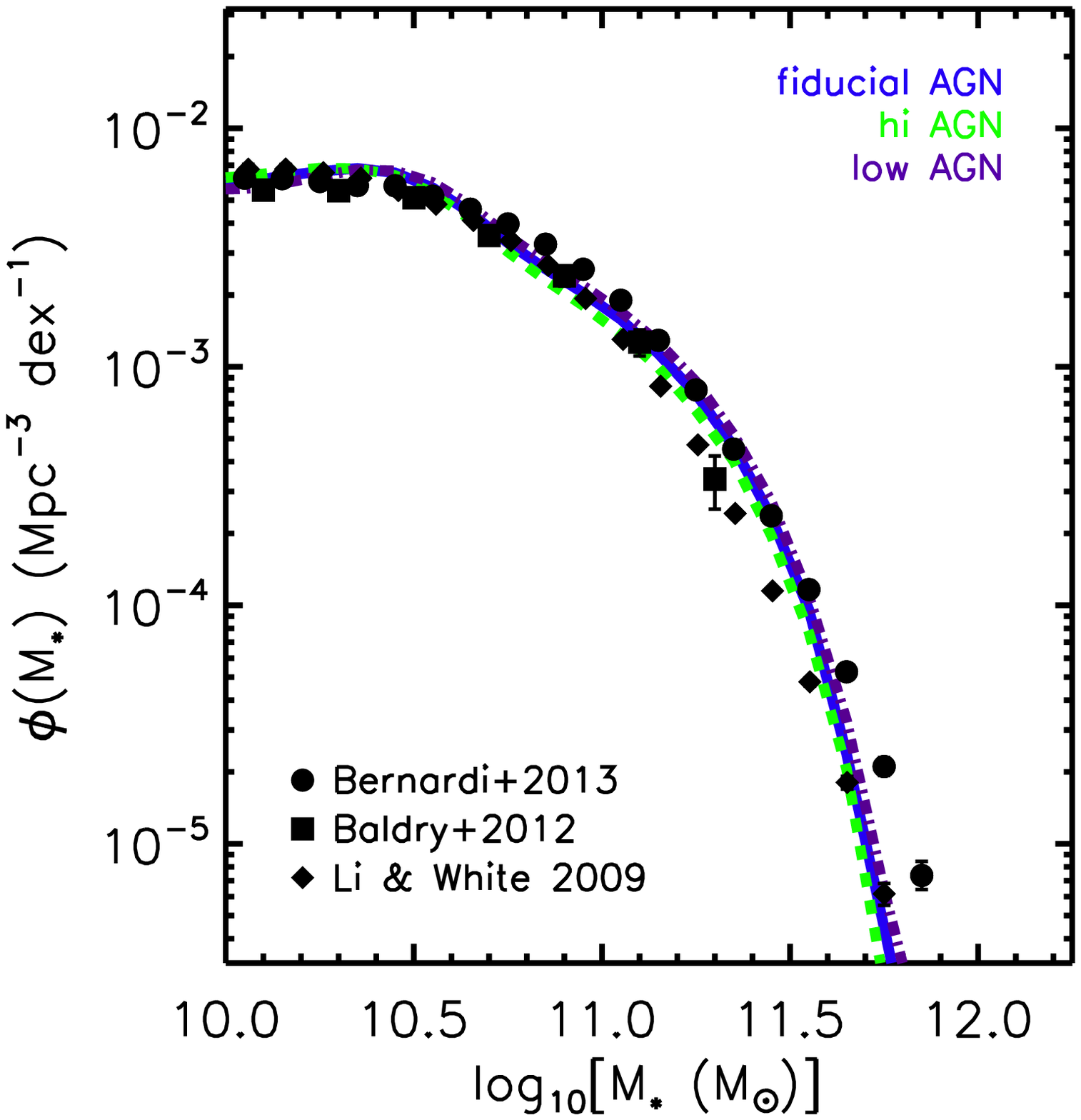}
\includegraphics[width=0.995\columnwidth]{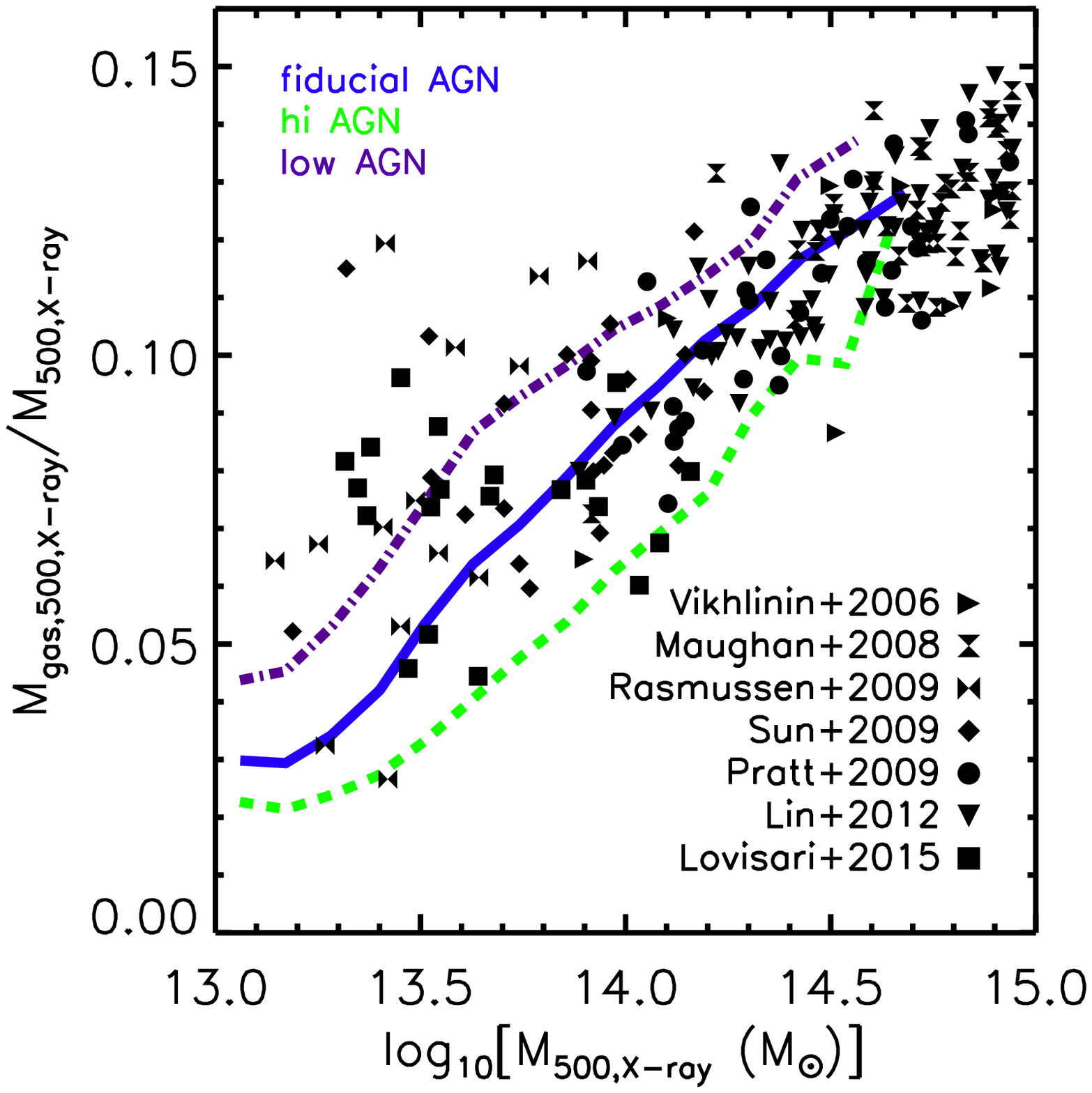}
\caption{\label{fig:bar_vary_agn}
Comparison of the predicted local galaxy stellar mass function (left) and hot gas mass fraction$-$total halo mass trends (right) of the fiducial \calsim~model (solid blue) with that predicted by simulations where the subgrid AGN heating temperature is raised (`hi AGN' - dashed green) or lowered (`low AGN' - dot-dashed purple) by 0.2 dex, all in the context of a WMAP9 cosmology.  Stellar masses in the left panel are computed within a 30 kpc aperture in the simulations, while halo masses and gas fractions in the right panel are derived from a synthetic X-ray analysis of a mass-limited sample (all haloes with $M_{\rm 500,true} > 10^{13} \ m_\odot$).  See M17 for further details.  The curves in the right panel correspond to the median relations (the simulations predict a similar amount of intrinsic scatter as seen in the data, see Fig.~\ref{fig:calibrated}).  As shown by M17, varying the AGN heating temperature has very little effect on the GSMF but does affect the gas mass fractions.  Varying the heating temperature by $\pm0.2$ dex yields predictions that effectively skirt the upper and lower bounds of the observed trend.  We will use these additional simulations to help quantify the level of error in our cosmological constraints due to imperfect feedback calibration.
}
\end{figure*}

The simulations were carried out with a version of the Lagrangian TreePM-SPH code \textsc{gadget3} \citep[last described in][]{Springel2005b}, which was modified to include new subgrid physics as part of the OWLS project \citep{Schaye2010}.  The gravitational softening is fixed to $4~h^{-1}$ kpc (in physical coordinates below $z=3$ and in comoving coordinates at higher redshifts) and the SPH smoothing is done using the nearest 48 neighbours.  

The simulations include subgrid prescriptions for metal-dependent radiative cooling \citep{Wiersma2009a}, star formation \citep{Schaye2008}, and stellar evolution, mass loss and chemical enrichment \citep{Wiersma2009b} from Type II and Ia supernovae and Asymptotic Giant Branch stars.  The simulations also incorporate stellar feedback \citep{DallaVecchia2008} and a prescription for supermassive black hole growth and AGN feedback (\citealt{Booth2009}, which is a modified version of the model originally developed by \citealt{Springel2005a}).  

As described in M17, we have adjusted the parameters that control the efficiencies of the stellar and AGN feedback so that the simulations reproduce the present-day galaxy stellar mass function (GSMF) for $M_* > 10^{10}$ M$_\odot$ and the amplitude of the gas mass fraction$-$halo mass relation of groups and clusters, as inferred from high-resolution X-ray observations.  (Synthetic X-ray observations of the simulations were used to make a like-with-like comparison in the latter case.)   These two observables were chosen to ensure that the collapsed structures in the simulations have the correct baryon content in a global sense.  The associated back reaction of the baryons on the total matter distribution should therefore also be broadly correct.  M17 demonstrated that this simple calibrated model, where the efficiencies are fixed values (i.e., they do not depend on redshift, halo mass, etc.), reproduces an unprecedentedly wide range of properties of massive systems, including the various observed mappings between galaxies, hot gas, total mass, and black holes.  Note that the number of parameters that dictate the overall feedback efficiency is small.  In particular, we adjusted only two parameters for each of the two forms of feedback (stellar and AGN) to reproduce the GSMF and gas fractions over two orders of magnitude in mass for both diagnostics (see Section \ref{sec:degen} for further discussion of the calibration procedure).

We point out that the parameters governing the feedback efficiencies are {\it not} recalibrated when varying the cosmological parameters away from the fiducial WMAP~9-yr cosmology (with massless neutrinos) adopted in M17.  But, as we will demonstrate in Section \ref{sec:degen}, the internal properties of collapsed structures (stellar masses, gas masses, etc.) are, to first order, insensitive to the variations in cosmology that we consider, even though the abundance of collapsed objects (and density fluctuations in general) depends relatively strongly on the adopted cosmology.  

\subsubsection{Remaining feedback calibration uncertainties}
\label{sec:vary_agn}

Although \calsim~arguably yields the best match of presently available simulations to observational constraints on the baryon content of massive systems, this does not imply that the problem of `baryon physics' for LSS cosmology has been fully resolved.  Firstly, the observational data on which the simulation feedback parameters were calibrated is itself prone to non-negligible uncertainties.  In particular, there is a large degree of intrinsic scatter in the gas fractions of observed X-ray-selected galaxy groups, and there is a danger that X-ray selection itself may bias our view of the overall hot gas content of groups (e.g., \citealt{Anderson2015,Pearson2017}).  A second issue is that, in \calsim, we have adopted a particular parametrisation for the feedback modelling, which corresponds to the simplest case where the feedback efficiency parameters are fixed.  However, more complicated dependencies could be adopted and may more closely represent feedback processes in nature.  While our expectation is that the act of calibrating such models against the observed stellar and gas masses of massive systems will yield LSS predictions similar to those from \calsim, we cannot presently quantify the level of expected differences.  Ultimately, we will only be able to assess the remaining feedback calibration uncertainties on LSS predictions by comparing the results of different (calibrated) simulations.  As already noted, \calsim~is a first attempt to calibrate the feedback for LSS cosmology.

While it may be difficult at present to assess how adopting other feedback parametrisations will affect the LSS predictions, we {\it can} provide a simple assessment of the role of observational uncertainties in the calibration.  Specifically, while the local galaxy stellar mass function is pinned down with sufficient accuracy observationally, the same is not true for the gas fractions of groups and clusters.  As the gas dominates the stars by mass, this uncertainty could propagate through to our cosmological parameter inference.  We have therefore run a number of additional smaller test simulations that vary the subgrid AGN heating temperature so that the predicted gas fractions approximately span those seen in the observations, while leaving the predicted GSMF virtually unchanged.  We have found that varying the AGN temperature by $\pm$0.2 dex approximately achieves this aim and we have therefore run two additional large-volume simulations (L400N1024, WMAP9 cosmology) that vary the heating temperature at this level, which we will use to quantify the error in our LSS cosmology results due to uncertainties in the calibration data. 

We show in Fig.~\ref{fig:bar_vary_agn} the predicted local GSMF and hot gas mass fraction$-$halo mass trends of the fiducial \calsim~model (solid blue), and the trends predicted by simulations where the AGN heating temperature is raised (`hi AGN' - dashed green) or lowered (`low AGN' - dot-dashed purple) by 0.2 dex, all in the context of a WMAP9 cosmology.  Varying the heating temperature by $\pm0.2$ dex yields predictions that effectively skirt the upper and lower bounds of the observed trend, as desired.  These simulations should therefore provide us with a (hopefully) conservative estimate of the error in the calibration due to uncertainties/scatter in the observational data against which the simulations were calibrated.

While these simulations enclose the scatter in the amplitude of the observed gas fraction$-$halo mass relation, there is an apparent difference in the predicted and observed slope of the relation at low mass (the galaxy group regime) that is worth commenting on.  This difference is likely explained by selection effects.  Specifically, for the simulations we select all haloes above a given spherical overdensity mass for analysis, whereas the X-ray constraints in Fig.~\ref{fig:bar_vary_agn} are generally derived from follow-up {\it Chandra} or {\it XMM-Newton} observations of group samples derived from X-ray flux-limited samples.  Naively, we expect galaxy groups that are more gas rich to also be more X-ray luminous, which ought to flatten the observed relation.  We note that recent stacking constraints  on the relation between tSZ effect flux and halo mass (e.g., \citealt{Planck2013_stack,Wang2016,Lim2018,Jakobs2018}), including its slope, are reproduced remarkably well by our simulations (e.g., M17; \citealt{Lim2018,Jakobs2018}), although converting the observed tSZ effect measured within the \planck~beam to an estimate of the gas fraction within the halo virial radius is non-trivial \citep{LeBrun2015}.  Future high-resolution tSZ effect observations of optically-selected groups will be invaluable for nailing down the precise form of the baryon mass--halo mass relation at low masses.

Finally, while we have only varied the feedback prescription in the context of a specific cosmology, we point out that in \citet{Mummery2017} we have shown that the effects of feedback on LSS are separable from those of massive neutrinos.  Thus, it is sufficient for our purposes to propagate the uncertainties in the feedback modelling using a single cosmological model.

\subsection{Massive neutrino implementation in BAHAMAS}
\label{sec:nu_sims} 

To include the effects of massive neutrinos, both on the background expansion rate and the growth of density fluctuations, we use the semi-linear algorithm developed by \citet{Bird2013} (see also \citealt{Bond1980,Ma1995,Brandbyge2008,Brandbyge2009,Bird2012}), which we have implemented in the \textsc{gadget3} code.  The semi-linear code computes neutrino perturbations on the fly at every time step using a linear perturbation integrator, which is sourced from the non-linear baryons+CDM potential and added to the total gravitational force.  As the neutrino power is calculated at every time step, the dynamical responses of the neutrinos to the baryons+CDM and of the baryons+CDM to the neutrinos are mutually and self-consistently included.  Note that because the integrator uses perturbation theory, the method does not account for the non-linear response of the neutrino component to itself.  However, this limitation has negligible consequences for our purposes, as only a very small fraction of the neutrinos (with lower velocities than typical) are expected to collapse and the neutrinos as a whole constitute only a small fraction of the total matter density.

In the present simulations, we adopt the so-called `normal' neutrino hierarchy, rather than just assuming degenerate neutrino masses, as done in many previous simulation studies.

\citet{Caldwell2016} and \citet{Mummery2017} have previously used a subset of our neutrino simulations to explore the consequences of free-streaming on collapsed haloes, such as their masses, velocity dispersions, density profiles, concentrations, and clustering.  Here our focus is on comparisons to LSS diagnostics, such as cosmic shear.

In addition to neutrinos, all of the \calsim~runs (i.e., with or without massive neutrinos) also include the effects of radiation when computing the background expansion rate.  We find that this leads to a few percent reduction in the amplitude of the present-day linear matter power spectrum compared to a simulation that only considers the evolution of dark matter and dark energy in the background expansion rate, if one does not rescale the input power by the growth rate so that the present-day power spectrum is correct.

\subsection{Choice of cosmological parameter values}
\label{sec:cosmology}

Large-volume hydrodynamical simulations are still sufficiently expensive that we cannot yet generate large grids of cosmologies with them.  This will inevitably limit our ability to systematically explore the available parameter space associated with the standard model of cosmology, or extensions thereof, and to determine the best-fit parameter values and their uncertainties.  However, there is an emerging consensus that baryon physics plays an important role in shaping the total mass distribution even on very large scales (e.g., \citealt{vanDaalen2011,vanDaalen2014,Velliscig2014,Schneider2015}) and if these effects are ignored, or modelled inaccurately, they are expected to lead to significant biases \citep{Semboloni2011,Eifler2015,HarnoisDeraps2015b}.  It is therefore important that, even with a relatively small range of simulated cosmologies, we make comparisons with the observations to provide an independent check of the results of less expensive (but ultimately less accurate) methods, such as those based on the halo model.  But which cosmologies should we focus on?

To significantly narrow down the available cosmological parameter space, we take guidance from the two most recent all-sky CMB surveys, by the WMAP~and \planck~missions.  In the context of the 6-parameter standard $\Lambda$CDM model of cosmology, comparisons to the primary CMB alone already pin down the best-fit parameter values to a few percent accuracy and the model agrees every well with the CMB data.  However, it must be noted that the best-fit parameters inferred from the WMAP~and \planck~data are not in perfect agreement, differing in some cases at up to the 2-sigma level.  This motivates us to consider two sets of cosmologies, one from each of the CMB missions (see Table \ref{tab:sims}).  Furthermore, as the CMB is not particularly sensitive to possible `late-time' effects (e.g., time-varying dark energy, massive neutrinos, dark matter interactions/decay, etc.), it remains crucially important to make comparisons to the observed evolution of the Universe, including that of LSS, to test our cosmological framework.  

We adopt the following strategy when selecting the values for the various cosmological parameters.  We first choose a number of values for the summed neutrino mass, $M_\nu$, that we wish to simulate.  Here we choose four different values, ranging from 0.06 eV up to 0.48 eV in factors of 2 (i.e., $M_\nu=$ 0.06, 0.12, 0.24, 0.48 eV).  Using the Markov chains of \citet{Planck2015_cmb} corresponding to the bottom right panel of Fig.~\ref{fig:cmb_mnu}; i.e., CMB+BAO+CMB lensing with marginalization over $A_{\rm Lens}$ (see discussion in Section \ref{sec:tension}), we select all of the parameter sets that have summed neutrino masses within $\Delta M_\nu=0.02$ of the target value.  The weighted mean values for each of the other important cosmological parameters is then computed using the supplied weights of each selected parameter set in the chain.  We follow this procedure for each of the summed neutrino mass cases we consider.  We have verified that when selecting the parameter values in this way the predicted CMB TT angular power spectrum (computed by {\small CAMB}) is virtually indistinguishable for the four different massive neutrino cases we consider.  Henceforth, we refer to the simulations whose cosmological parameter values were selected in this way as being `Planck2015/$A_{\rm Lens}$-based'.

Prior to adopting the above strategy for the `Planck2015/$A_{\rm Lens}$-based' simulations, we ran a number of `WMAP9-based' and `Planck2013-based' simulations with massive neutrinos in which all of the cosmological parameters apart from $\Omega_{\rm cdm}$ (i.e., $H_0$, $\Omega_b$, $\Omega_{\rm m}$, $n_{\rm s}$, and $A_s$) were held fixed at their primary CMB maximum-likelihood values (from \citealt{Hinshaw2009} and \citealt{Planck2013_cmb}, respectively) assuming massless neutrinos.  The CDM matter density was reduced to maintain a flat geometry, so that $\Omega_b + \Omega_{\rm m} + \Omega_{\Lambda} + \Omega_\nu = 1$ given the neutrino mass density of the run.  The disadvantage of this strategy is that it will not precisely preserve the predicted CMB angular power spectrum, since the neutrinos are relativistic at recombination but evolve like matter (i.e., are non-relativistic) today.  The deviations in the predicted power spectrum are quite small, though, given that we are only considering cases with $\Omega_\nu \la 0.01$, and would not be easily detectable with either {\it Planck} or WMAP (as noted previously, the {\it Planck} CMB only constraint is $M_\nu \la 0.70$ eV, corresponding to $\Omega_\nu \la 0.017$).  This strategy allows one to see the effects of massive neutrinos in the absence of variations of the other parameters.  For these reasons, we include the `WMAP9-based' and `Planck2013-based' runs in our analysis as well.

A summary of the runs used in the present study is given in Table \ref{tab:sims}.

\begin{table*} 
\caption{\label{tab:sims} Cosmological parameter values for the simulations presented here.  The columns are: (1) The summed mass of the 3 active neutrino species (we adopt a normal hierarchy for the individual masses); (2) Hubble's constant; (3) present-day baryon density; (4) present-day dark matter density; (5) present-day neutrino density, computed as $\Omega_\nu = M_\nu / (93.14 \ {\rm eV} \ h^2)$; (6) spectral index of the initial power spectrum; (7) amplitude of the initial matter power spectrum at a {\small CAMB} pivot $k$ of $2\times10^{-3}$ Mpc$^{-1}$; (8) present-day (linearly-evolved) amplitude of the matter power spectrum on a scale of 8 Mpc/$h$ (note that we use $A_s$ rather than $\sigma_8$ to compute the power spectrum used for the initial conditions, thus the ICs are `CMB normalised').  In addition to the cosmological parameters, we also list the following simulation parameters: (9) dark matter particle mass; (10) initial baryon particle mass.}
\begin{tabular}{*{10}{c}}                                                                 \hline
(1)        & (2)        & (3)               & (4)                & (5)              & (6)      & (7)              &  (8)          &   (9)                       & (10)                       \\
$M_{\nu}$   & $H_0$      & $\Omega_b $  & $\Omega_{\rm cdm} $ &  $\Omega_\nu$     & $n_{\rm s}$    & $A_s$            &  $\sigma_8$   &  $M_{\rm DM}$                & $M_{\rm bar,init}$           \\
(eV)       & (km/s/Mpc) &                   &                    &                  &          & ($10^{-9}$)      &               &  [$10^9 {\rm M_{\odot}}/h$]  & [$10^8 {\rm M_{\odot}}/h$]   \\
\hline
{\bf Planck2015/$A_{\rm Lens}$-based}\\
\hline
0.06     & 67.87 & 0.0482 & 0.2571 & 0.0014 & 0.9701 & 2.309 & 0.8085 & 4.25 & 7.97 \\
0.12     & 67.68 & 0.0488 & 0.2574 & 0.0029 & 0.9693 & 2.326 & 0.7943 & 4.26 & 8.07 \\
0.24     & 67.23 & 0.0496 & 0.2576 & 0.0057 & 0.9733 & 2.315 & 0.7664 & 4.26 & 8.21 \\
0.48     & 66.43 & 0.0513 & 0.2567 & 0.0117 & 0.9811 & 2.253 & 0.7030 & 4.25 & 8.49 \\ 
\hline
{\bf Planck2013-based}\\
\hline
0.0      & 67.11 & 0.0490 & 0.2685 & 0.0    & 0.9624 & 2.405 & 0.8341 & 4.44 & 8.11 \\
0.24     & 67.11 & 0.0490 & 0.2628 & 0.0057 & 0.9624 & 2.405 & 0.7759 & 4.35 & 8.11 \\
\hline
{\bf WMAP9-based}\\
\hline
0.0      & 70.00 & 0.0463 & 0.2330 & 0.0    & 0.9720 & 2.392 & 0.8211 & 3.85 & 7.66 \\
0.06     & 70.00 & 0.0463 & 0.2317 & 0.0013 & 0.9720 & 2.392 & 0.8069 & 3.83 & 7.66 \\
0.12     & 70.00 & 0.0463 & 0.2304 & 0.0026 & 0.9720 & 2.392 & 0.7924 & 3.81 & 7.66 \\
0.24     & 70.00 & 0.0463 & 0.2277 & 0.0053 & 0.9720 & 2.392 & 0.7600 & 3.77 & 7.66 \\
0.48     & 70.00 & 0.0463 & 0.2225 & 0.0105 & 0.9720 & 2.392 & 0.7001 & 3.68 & 7.66 \\
\hline
\end{tabular}
\end{table*}

\subsection{Light cones and map-making}
\label{sec:cones_maps}

\subsubsection{Light cones}
\label{sec:cones}

To make like-with-like comparisons to the various LSS observations, we first construct light cones.  This is done by stacking randomly rotated and translated simulation snapshots along the line of sight (e.g., \citealt{daSilva2000}), back to $z=3$.  Each of our simulations has 15 snapshots between the present-day and $z=3$, output at $z$ =0.0, 0.125, 0.25, 0.375, 0.5, 0.75, 1.0, 1.25, 1.5, 1.75, 2.0, 2.25, 2.5, 2.75, 3.0.  Note that for a {\it WMAP} 9-yr cosmology, the comoving distance to $z=3$ is $\approx4600$ Mpc/h, implying that a minimum of 11 snapshots would need to be stacked along the line of sight, if the snapshots were written out at equal comoving distance intervals (of the box size).  The snapshots, however, are not written out in equal comoving distance intervals, so occasionally we do not use a full snapshot, while for a handful of times we have to use a single snapshot (slightly) more than once\footnote{When constructing cones along the line of sight (i.e., moving out in comoving distance), we use the snapshot that is nearest to the present comoving distance to draw particles/haloes from.  Occasionally, the comoving distance between snapshots is larger than the box size, in which case we first randomly rotate/translate the full box and stack it and then we go back to the same snapshot and randomly rotate/translate again and extract a subvolume of the required size to fill the gap before the next snapshot is used.}.  

For a maximum redshift of $z=3$, which was chosen to achieve convergence in the various LSS diagnostics we consider (such as the tSZ effect power spectrum), the maximum opening angle of the light cone, given the size of the simulation box, is just slightly larger than 5 degrees; i.e., $\theta_{\rm max} = L_{\rm box}/\chi(z=3)$ where $L_{\rm box}$ is the simulation comoving box size (400 Mpc/h) and $\chi(z=3)$ is the radial comoving distance to $z=3$.  We therefore create light cones of $5\times5$ sq.~deg.  (Note that in comoving space, light rays follow straight lines, making the selection of particles and haloes falling within the light cone a trivial task.)  We produce 25 such light cones per simulation, using different (randomly-selected) rotations/translations.  We use the same 25 randomly-selected viewing angles for all the simulations, so that cosmic variance does not play a role when comparing them. 

We have tested our light cone algorithm on smaller box simulations, varying both the number of snapshots that are output and used in the construction of the cones as well as the maximum redshift of the cones.  For all of the tests we consider here, we find that our theoretical predictions (e.g., the predicted $C_\ell$'s for the tSZ effect power spectrum) do not change by more than a few percent when we vary the number of snapshots used in the light cones and maximum redshift of the light cones away from the fiducial values of 15 and $z=3$, respectively.

\subsubsection{tSZ effect maps}
\label{sec:tsz_maps}

To produce tSZ effect Compton $y$ maps, we follow the procedure described in \citet{McCarthy2014}.  The Compton $y$ parameter is defined as:

\begin{equation}
y \equiv \int \sigma_T \frac{k_b T}{m_e c^2} n_e dl \ \ \ ,
\end{equation}

\noindent where $\sigma_T$ is the Thomson cross-section, $k_B$ is Boltzmann's constant, $T$ is the gas temperature, $m_e$ is the electron rest mass, $c$ is the speed of light, and $n_e$ is the electron number density.  Thus, $y$ is proportional to the electron pressure integrated along the observer's line of the sight.

To produce Compton $y$ maps, we first calculate the quantity (see \citealt{Roncarelli2006,Roncarelli2007})

\begin{equation}
\Upsilon \equiv \sigma_T \frac{k_b T}{m_e c^2} \frac{m}{\mu_{e} m_H}
\end{equation}

\noindent for each gas particle selected inside the light cone.   Here $T$ is the temperature of the gas particle, $m$ is the gas particle mass, $\mu_{e}$ is the mean molecular weight per free electron of the gas particle (which depends on its metallicity), and $m_H$ is the atomic mass of hydrogen.  Note that $\Upsilon$ has dimensions of area.

The total contribution to the Compton $y$ parameter in a pixel by a given particle is obtained by dividing $\Upsilon$ by the physical area of the pixel at the angular diameter distance of the particle from the observer; i.e., $y \equiv \Upsilon/L_{\rm pix}^2$.  We adopt an angular pixel size of 10 arcsec, which is generally better than what can be achieved with current tSZ telescopes.

Finally, we map the gas particles to the 2D grid using a simple `nearest grid point' algorithm and integrate (sum) the $y$ parameters of all of the gas particles along the line of sight to produce images.  As in \citet{McCarthy2014}, we have also produced SPH-smoothed $y$ maps (using the angular extent of the particle's 3D smoothing length as the angular smoothing length) for comparison with our default nearest grid method.  We find virtually identical results, in terms of cosmological parameter constraints, for the two approaches for mapping particles to pixels.

\subsubsection{Weak lensing convergence and shear maps}
\label{sec:lensing_maps}

The lensing of images of background sources (e.g., galaxies, CMB temperature fluctuations) by intervening matter (LSS in this case) depends, to first order, on three quantities: the convergence $\kappa$ and two (reduced) shear components, $g_1$ and $g_2$.  

The 3D lensing `convergence' field, $\kappa({\bf x})$, is related to the matter overdensity, $\delta$, via:
\begin{eqnarray}
2\kappa({\bf x}) = \nabla^2 \Phi ({\bf x})= \frac{3}{2}\Omega_{m} H_{0}^2 (1+z) \delta({\bf x})
\label{eq:gravpot}
\end{eqnarray}
where 
\begin{eqnarray}
\delta({\bf x})=\frac{\rho({\bf x})-\bar{\rho}}{\bar{\rho}}
\label{eq:delta}
\end{eqnarray}
Here $\Phi({\bf x})$ is the local peculiar gravitational potential and $\bar{\rho}$ and $\rho({\bf x})$ are the mean and local matter densities, respectively. 

One does not observe the local 3D convergence, however, but instead measures the projected convergence (convolved with the lensing kernel), obtained by integrating over the intervening matter along line of sight back to the source.  The projected convergence, $\kappa(\theta)$, integrated up to a maximum comoving distance $\chi(z_{\rm max})$ (where $z_{\rm max}=3$ here), is given by
\begin{eqnarray}
\kappa(\theta) = \frac{3 \Omega_{\rm m} H_0^2}{2 c^2} \int_0^{\chi(z_{\rm max})} (1+z) s(\chi) \delta(\chi,\theta) d\chi
\label{eq:kappa}
\end{eqnarray}
where the lensing kernel, $s(z)$, is defined as
\begin{eqnarray}
s(\chi) = \chi(z) \int_{z}^{z_{\rm max}} n_s(z')\left[\frac{\chi(z') - \chi(z)}{\chi(z')} \right] dz'
\label{eq:source}
\end{eqnarray}
\noindent and depends on the source redshift distribution, $n_s(z)$.  The amplitude of $n_s(z)$ is specified so that $\int n_s(z) dz = 1$.  Note that in eqns.~\ref{eq:kappa} and \ref{eq:source} we have implicitly assumed a flat Euclidean geometry, as adopted in the simulations.

In the case of a single source plane, where $n_s(z)$ can be represented by a Dirac delta function, $s(z)$ reduces simply to $s(z) = \chi(z) [1 - \chi(z)/\chi(z_s)]$, where $\chi(z)$ and $\chi(z_s)$ are the comoving distances to the lens and source, respectively.  This is an excellent approximation for CMB lensing, where $z_s \approx 1100$ (i.e., last scattering surface), but it is not a good approximation for most galaxy weak lensing surveys, which typically use samples of galaxies that span wide ranges in redshift.  One therefore must use the source redshift distribution function for each individual survey to make comparisons between theory and a particular survey.  When comparing to different surveys in Section \ref{sec:results_shear}, we will specify the particular forms of $n_s(z)$ that we adopt.

To evaluate eqns.~\ref{eq:kappa} and \ref{eq:source} for a given light cone, we first break the light cone up into a number of segments along the line of sight.  By default we adopt a fixed segment width of $\Delta z = 0.05$, which we note is similar to the resolution in $n_s(z)$ adopted in current imaging surveys (e.g., KiDS, DES).  We therefore have $N=60$ such segments between $z=0$ and $z_{\rm max}=3$, for which we calculate the midplane distances/redshifts and widths in comoving distance (i.e., $\chi$, $z$, $\Delta \chi$).  We evaluate the two-dimensional overdensity at the midplane for each segment by collapsing each segment along the line of sight; i.e., integrating the total mass\footnote{As the neutrino component is not represented by particles in \calsim, we add its contribution.  Specifically, under the accurate assumption that the neutrinos do not significantly cluster on scales smaller than their free-streaming length (and we note here that all of the comparisons we make to data probe scales smaller than the free-streaming scale), we can add a uniform mass density term $\rho_{\nu}(z) = \Omega_\nu(z) \rho_{\rm crit}(z)$ to the local density (this is valid over the redshift range we consider, as the neutrinos are non-relativistic at late times).  The neutrino contribution is also included in the mean matter density, required to compute the overdensity, through its contribution to $\Omega_{\rm m}$.} (due to dark matter, gas, stars and neutrinos) to produce a surface mass density map, $\Sigma(\theta)$, from which we can evaluate the overdensity.  The 2D overdensity map, $\delta(\theta)$, is defined as $\delta(\theta)\equiv[\Sigma(\theta)-\bar{\Sigma}]/\bar{\Sigma}$ and we evaluate $\bar{\Sigma}$ analytically\footnote{One could instead evaluate the mean directly from the $\Sigma(\theta)$ map, but we have found that the mean is sometimes poorly determined for small segment widths and/or light cone opening angles.} given $\Omega_{\rm m}$ of the simulation and the width of the segment, $d\chi$.

We can now discretise eqns.~\ref{eq:kappa} and \ref{eq:source} (see, e.g., \citealt{HarnoisDeraps2012}) as
\begin{eqnarray}
\kappa(\theta)= \frac{3 \Omega_{m} H_0^{2}} {2 c^{2} } \sum_{i=1}^{N}  (1+z_{i}) s[\chi(z_i)] \delta_i(\theta) \Delta \chi_i 
\label{eq:kappa_discrete}
\end{eqnarray}
and
\begin{eqnarray}
s[\chi(z_i)] = \chi(z_i) \sum_{j=i}^{N} n_s(z_{j}) [1 - \chi(z_i)/\chi(z_j)] \Delta z
\label{eq:source_discrete}
\end{eqnarray}

\noindent where the sums are done over $i^{\rm th}$ and $j^{\rm th}$ segments (planes), with $i=1$ corresponding to the nearest (to $z=0$) segment and $i=N$ corresponds to the most distant one (i.e., near $z=3$).

Eqns. \ref{eq:kappa_discrete} and \ref{eq:source_discrete} are strictly valid only for the case of small deflection angles, i.e., photons travelling in straight lines in comoving coordinates.  However, this so-called `Born approximation' has been shown previously to be very accurate in the case of weak lensing \citep{Schneider1998,White2004}, which is our focus here.

We compute the $\gamma_1$ and $\gamma_2$ shear maps from the $\kappa$ map using the method of \citet{Clowe2004} (see also \citealt{Bahe2012}).  Specifically, we evaluate the Fourier transform of the complex shear, $\gamma=\gamma_1 + i \gamma_2$, as 
\begin{equation}
\tilde{\gamma} \equiv (\tilde{\gamma_1}, \tilde{\gamma_2}) =  \left(\frac{\hat{k}_1^2-\hat{k}_2^2}{\hat{k}_1^2+\hat{k}_2^2}\tilde{\kappa}, \frac{2\hat{k}_1\hat{k}_2}{\hat{k}_1^2+\hat{k}_2^2}\tilde{\kappa}\right)
\label{eq:fftmethod}
\end{equation}
where $\tilde{\gamma}$ and $\tilde{\kappa}$ are Fourier transforms of $\gamma$ and $\kappa$, and $\hat{k}$ are the appropriate wave vectors.  We then zero pad (to avoid edge effects) and inverse Fourier transform the $\tilde{\gamma}$ maps to obtain the $\gamma_1$ and $\gamma_2$ maps.  Dividing these by the map of $1-\kappa$ yields the reduced shear, $g_1$ and $g_2$, maps.  Note that for the case of perfectly circular background sources, the reduced shear, $g$, is just the observed galaxy ellipticity ($\epsilon$).  

\begin{figure*}
\includegraphics[width=0.995\columnwidth]{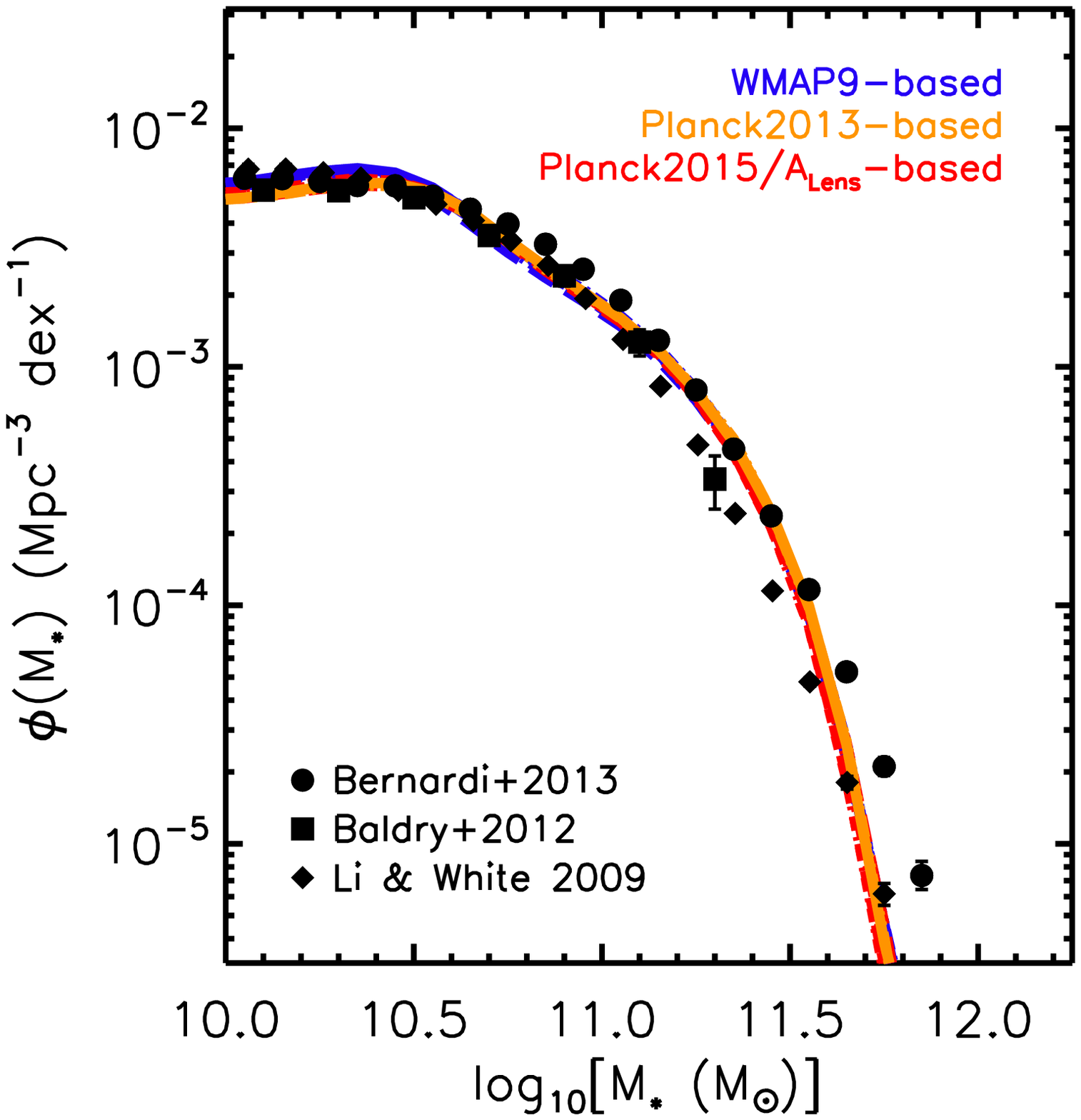}
\includegraphics[width=0.995\columnwidth]{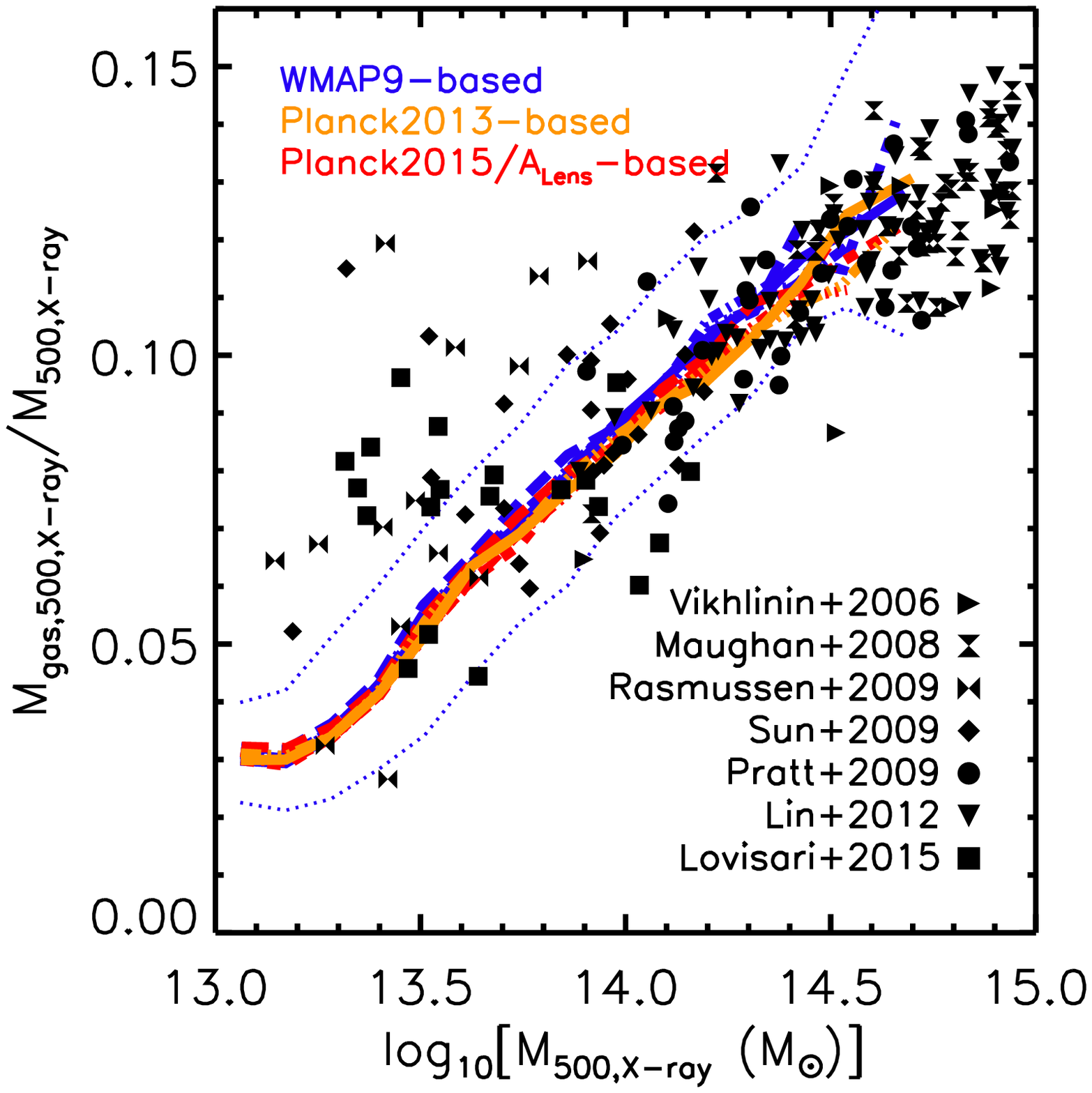}
\caption{\label{fig:calibrated}
The calibrated local GSMF and hot gas mass fraction$-$total halo mass trends, extracted from the 11 different cosmologies considered here (see Table \ref{tab:sims}) using the same (fiducial) feedback model.  Stellar masses in the left panel are computed within a 30 kpc aperture in the simulations, while halo masses and gas fractions in the right panel are derived from a synthetic X-ray analysis of a mass-limited sample (all haloes with $M_{\rm 500,true} > 10^{13} \ m_\odot$).  See M17 for further details.  The solid curves in the right panel represent the median relation, while the dotted red curves enclose the central 68\% of the simulated population for the WMAP9 cosmology with massless neutrinos.  The feedback model, which was calibrated in M17 using simulations run only with the WMAP9 cosmology (with massless neutrinos), produces nearly identical GSMFs and gas fractions for the other cosmologies we include here, implying that there is a negligible degree of degeneracy between cosmology and feedback, at least for the variations in cosmology that we consider here.
}
\end{figure*}

Using the above methodology, what we calculate is the true convergence and shear fields for the simulations.  However, observations cannot necessarily perfectly recover these true quantities.  Leaving aside important observational challenges such as measuring unbiased galaxy shapes and estimating accurate redshifts from photometric data, there is also the potential physical issue of intrinsic alignments (IAs).  That is, to recover the shear field in data one must average together many galaxies to beat down the noise, with the implicit assumption that, in the absence of lensing, there should be no preferential alignment between the galaxy orientations.  If there is a preferential alignment (as might naively be expected from tidal torque theory, if galaxies inhabit the same large-scale environment), this will lead to a bias in the recovered lensing signal.  In principle, we could address this issue by self-consistently lensing the simulated galaxies in our cosmological volumes.  However, this is generally not possible with current large-volume simulations like \calsim, since the resolution is too low to accurately predict and measure simulated galaxy shapes.  One can instead assume a simple physical model of IAs (e.g., \citealt{Bridle2007}) and marginalize over its free parameters when analysing the data, as is typically done in current studies.  For the present study, we neglect the effects of IAs in our modelling.  We note that current observational constraints suggest that its effects are minor; e.g., by neglecting it, the observational constraints on $S_8$ change by less than 1-sigma and do not reconcile the aforementioned CMB-LSS tension (e.g., \citealt{Hildebrandt2017}).  In addition, using the high-resolution EAGLE cosmological hydrodynamical simulations, \citet{Velliscig2015} have shown that the intrinsic alignments of galaxies is far weaker than that of dark matter haloes (which has, to date, been the basis of simple physical models of IAs), particularly if one selects the stars in an observationally-motivated manner.      

\section{How degenerate is cosmology with baryon physics?}
\label{sec:degen}

\calsim~is a first attempt to explicitly calibrate the feedback in large-volume cosmological hydrodynamical simulations in order to minimize the impact of uncertain baryon physics on cosmological studies using LSS.  However, an important question is: {\it to what extent is the calibration of the feedback model parameters dependent upon cosmology?}  If the calibration scheme depends significantly upon cosmology, the implication is that the feedback model parameters would need to be readjusted for each cosmological model that we simulate.  This would obviously complicate the cosmological analysis but may ultimately be necessary.

It is clear that if the feedback model were to be calibrated on the same observational diagnostics that are being used to infer cosmological parameter values (e.g., tSZ effect, cosmic shear, etc.), one should naturally expect there to be degeneracies between the cosmology and feedback parameters.  Recognizing this, with \calsim~we elected instead to calibrate the feedback on {\it internal halo properties} (specifically their stellar and baryon fractions), rather than on the abundance of haloes or the power spectrum of density fluctuations.  The internal properties of haloes ought to be much less sensitive to cosmology, as processes such as violent relaxation, phase mixing, and shock heating will effectively randomize the energies of the dark matter, stars, and gas, thus mostly, though not completely\footnote{The internal structure of dark matter haloes, as characterised by their concentration, is known to depend on cosmology (e.g., \citealt{Bullock2001,Eke2001,Correa2015}), as does the location of halo outer boundary (e.g., \citealt{Diemer2017}).  However, these relations contain significant scatter and in general exhibit a much weaker dependence on cosmology than that of the matter power spectrum.  The introduction of baryons and associated processes will further weaken the link between halo properties and cosmology.}, removing their memory of the background cosmology.  Another important advantage of using the baryon fractions of collapsed haloes is that it provides a direct measure of the effects of expulsive feedback: there are no known forces/processes within the standard model of cosmology other than feedback that can remove a significant fraction of the baryons from collapsed systems\footnote{Some proposed modified theories of gravity and `interacting' dark energy models invoke non-universal couplings, such that the fifth force couples differently with dark matter than it does with baryons (e.g., \citealt{Hammami2015}).  In this case it is possible to affect the baryon fractions of collapsed systems without invoking feedback, but it is far from clear that such models would be able to naturally account for the observed trend of gas fraction with halo mass, which approximately converges to the universal baryon fraction, $\Omega_b/\Omega_{\rm m}$, for the most massive clusters (see Fig.\ref{fig:calibrated})}.

We refer the reader to M17 for the details of the calibration procedure but, briefly, it proceeds as follows.  We first adjusted the stellar feedback wind velocity to reproduce the observed abundance of $M^*$ ($\sim10^{10}$ M$_\odot$) galaxies, which is the minimum mass we can resolve at the fiducial resolution.  A wind velocity of $\approx300$ km/s achieves this for the fiducial resolution. (The stellar feedback mass-loading parameter also affects the abundance of low-mass galaxies, although less so than the wind velocity.  We held the mass-loading parameter fixed at a value of 2.)  Without AGN feedback, the simulations produce far too many massive galaxies; i.e., the well-known overcooling problem.  Adopting the AGN feedback model of \citet{Booth2009}, however, results in a strong quenching of star formation in the most massive galaxies and the resulting GSMF, which agrees well with the observations, and is relatively insensitive to the details of the AGN feedback modelling due to its self-regulating behaviour.  However, the hot gas fractions and thermodynamic profiles of groups and clusters are strongly sensitive to the AGN subgrid heating temperature.  We therefore adjusted the subgrid heating temperature to reproduce the amplitude of the observed local gas mass--total halo mass relation.  This adjustment had no adverse effects on the GSMF.   Note that we calibrated the model on the GSMF and the group/cluster gas fractions only and did not even examine (let alone calibrate on) other observables.  We then subsequently demonstrated that the simulations reproduce a wide range of independent observations, including the profiles and redshift evolution of the gas and stellar content of massive systems (see also \citealt{Barnes2017}).  This was done in M17 in the context of a WMAP9 cosmology with massless neutrinos.

Returning to the present study and the possible cosmology dependence of the calibration scheme, we explicitly verified using small test runs ($100$ Mpc/$h$ on a side boxes) that the stellar and gaseous properties of haloes in the simulations are insensitive to the variations in cosmology we are considering to the required accuracy.  We therefore directly proceeded to run the large-volume ($400$ Mpc/$h$ on a side) boxes necessary for the LSS tests {\it without changing any aspect of the subgrid physics} (feedback or otherwise).  In Fig.~\ref{fig:calibrated} we show the resulting GSMFs and gas fraction$-$halo mass relations for the 11 different cosmologies that we consider here.  As in M17, the stellar masses of simulated galaxies are computed within a 30 kpc (physical) aperture, which approximately mimics what is derived observationally for standard pipeline analysis in SDSS and GAMA (see the appendix of M17 for details).  The halo masses and gas fractions of the simulated groups and clusters in the right panel are derived by performing a synthetic X-ray analysis, as described in M17 (see also \citealt{LeBrun2014}).

\begin{figure*}
\includegraphics[width=0.995\columnwidth]{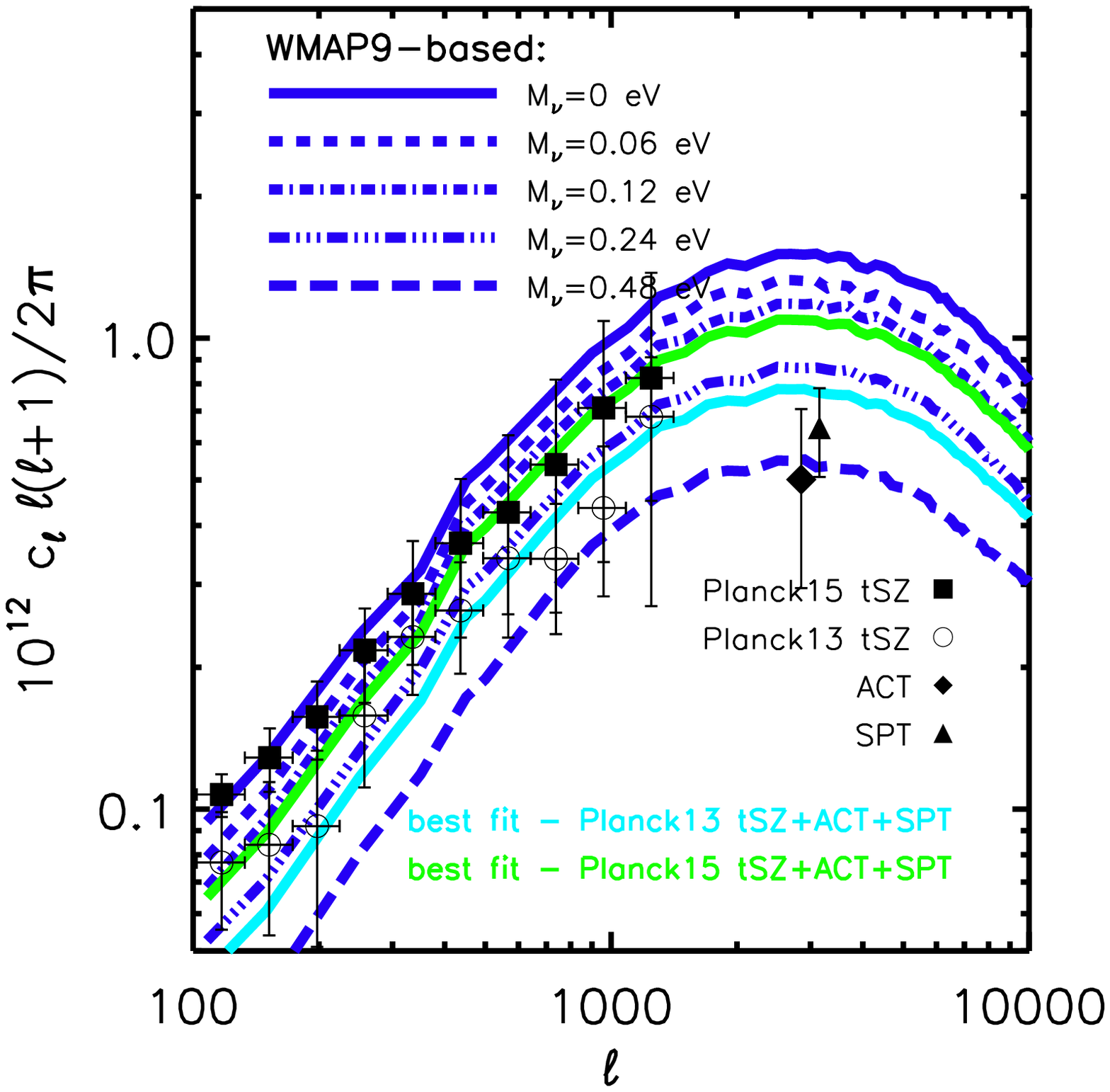}
\includegraphics[width=0.995\columnwidth]{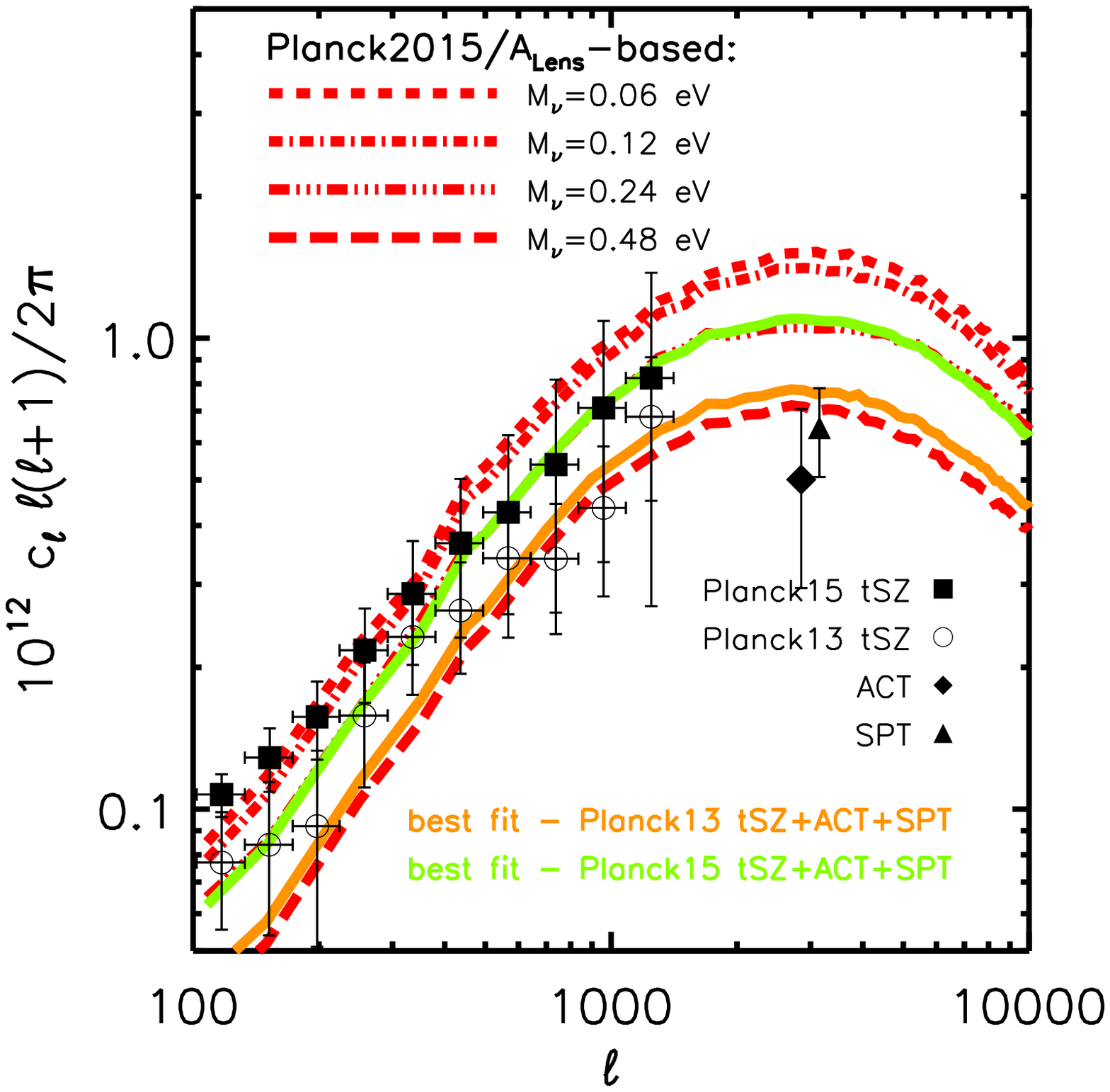}
\caption{\label{fig:cl_y_y}
  Comparison of the observed (data points with 1-sigma errors) and predicted (curves) tSZ effect (Compton $y$) angular power spectra.  {\it Left:} Comparison to the WMAP9-based simulations.  {\it Right:} Comparison to the Planck2015/$A_{\rm Lens}$-based simulations.  The constraints of $M_\nu$ depend strongly on the adopted observational data set.  Both the ACT and SPT measurements at $\ell\approx3000$ are of significantly lower amplitude than expected for a model with minimal neutrino mass, as are the (larger-scale) \planck~2013 tSZ measurements.  All three are consistent with a summed neutrino mass of $\approx 0.3(0.4)$ in the context of the WMAP9-based (Planck2015/$A_{\rm Lens}$-based) simulations.  However, the more recent \planck~2015 tSZ measurements are consistent with the minimal neutrino mass.  See Table \ref{tab:mnu_tsz}.  The origin of this difference is unclear, but is probably related to residual foreground contamination (e.g., the CIB) in the tSZ effect maps (see text for discussion).
}
\end{figure*}

The resulting GSMFs and gas fraction$-$halo mass relations are remarkably similar.  In detail, very small differences are present at the low-stellar mass end of the GSMF, which we attribute to slight differences in the resolution of the simulations (compare the particle masses in Table \ref{tab:sims}), rather than to changes in cosmology.  Very small differences (typically a few percent) are also present in the predicted gas fractions at the high-halo mass end, in the sense that the WMAP9-based simulations predict a slightly higher gas fraction compared to the Planck2013- and Planck2015/$A_{\rm Lens}$-based simulations.  We attribute this difference to the slightly higher universal baryon fraction, $\Omega_b/\Omega_{\rm m}$, in the WMAP9-based cosmologies with respect to the Planck-based cosmologies.  However, this difference is clearly very small compared to the scatter in the observed gas fractions of groups and clusters.  Furthermore, we will demonstrate later, using the two additional runs which vary the AGN feedback (see Section \ref{sec:vary_agn}) and alter the gas fractions by a much larger amount, that our cosmological inferences are negligibly affected by the small differences in the gas fractions of the different simulations.

On the basis of the above, we therefore conclude that our feedback calibration method is sufficiently insensitive to cosmology; i.e., there is no significant degeneracy between uncertainties in the feedback model parameters and the cosmological parameters for the variations in cosmology we consider here (see also \citealt{Mummery2017}).  We emphasise, however, that this insensitivity to cosmology may not hold for much larger variations in the cosmological parameters or for more significant extensions to the standard model of cosmology (e.g., modified gravity).  This should be tested on a case by case basis.

\section{Results}
\label{sec:results}

In this section we present our constraints on the summed mass of neutrinos, $M_\nu$, derived from various statistical measures of the tSZ effect, cosmic shear, and CMB lensing.

\subsection{tSZ effect}
\label{sec:results_tsz}

\subsubsection{Angular power spectrum}
\label{sec:tsz_power}

In Fig.~\ref{fig:cl_y_y} we compare the predicted and observed tSZ effect angular power spectra.  We focus on multipoles of $\ell > 100$, which are accessible with the simulated light cones.

For the observations, we use recent measurements from the South Pole Telescope \citep{George2015} and the Atacama Cosmology Telescope \citep{Sievers2013}, as well as from the \planck~2013 and 2015 data releases \citep{Planck2013_sz,Planck2015_sz}.  The SPT and ACT place independent constraints on the power spectrum at $\ell\sim3000$ and are consistent with each other.  However, there is a clear difference between the reported 2013 and 2015 \planck~power spectra at $\ell \la 1000$, in that the amplitude of the 2015 power spectrum is systematically higher than that of the 2013 power spectrum.  The published uncertainties, which are dominated by systematic foreground subtraction uncertainties (due to point sources and the clustered infrared background, CIB), are also larger for the 2015 measurements.  The larger error bars for the 2015 dataset reflect a more conservative analysis of the foreground uncertainties (B.~Comis, priv.~communication), but the origin of the shift between the 2015 and 2013 power spectra at $\ell\ga100$, or even its presence, was not acknowledged or discussed by \citet{Planck2015_sz}.  For this reason, we examine the constraints using both data sets (independently).

\begin{table} 
\caption{\label{tab:mnu_tsz} Constraints on the summed mass of neutrinos derived from the tSZ effect power spectrum.  The columns are: (1) Observational data set used; (2) Best fit value of $M_\nu$ (eV) with 1-sigma uncertainty; and (3) the reduced chi-squared of the best fit.  We have separated the constraints into two sections, based on whether the WMAP9-based or Planck2015/$A_{\rm Lens}$-based simulations were used for the theoretical modelling.}
\begin{tabular}{lcc}                                                                 
\hline
(1)        & (2)            & (3)            \\
Data set   & $M_{\nu}$ (eV)  & $\chi^2$/DOF   \\
\hline
{\bf Planck2015/$A_{\rm Lens}$-based}\\
\hline
Planck2013+SPT+ACT   & $0.43\pm0.04$    & 0.80 \\
Planck2015+SPT+ACT   & $0.24\pm0.03$    & 3.64 \\
Planck2013 tSZ only      & $0.37\pm0.05$    & 0.57 \\
Planck2015 tSZ only      & $0.07\pm0.03$    & 0.60 \\

\hline
{\bf WMAP9-based}\\
\hline
Planck2013+SPT+ACT   & $0.31\pm0.04$    & 0.76 \\
Planck2015+SPT+ACT   & $0.15\pm0.03$    & 3.30 \\
Planck2013 tSZ only      & $0.27\pm0.04$    & 0.54 \\
Planck2015 tSZ only      & $0.02\pm0.02$    & 0.45 \\
\hline
\end{tabular}
\end{table} 

To place constraints on the summed mass of neutrinos, we first compute the mean tSZ effect power spectrum for each of the simulations, by averaging over the power spectra computed for the 25 light cones for each simulation.  We have produced a simple function that will interpolate (or extrapolate if necessary) from these pre-computed power spectra the value of $C_\ell$ at a specified multipole given a choice of summed neutrino mass.  The interpolator fits a powerlaw of the form $C_\ell = A (1-f_\nu)^B$, where $f_\nu \equiv \Omega_\nu/\Omega_{\rm m}$ and $A$ and $B$ are constants determined by fitting the trend between $C_\ell$ and $f_\nu$ at fixed $\ell$ from the pre-computed spectra.  This is done either in the context of the WMAP9-based simulations or the Planck2015/$A_{\rm Lens}$-based simulations.  To determine the best-fit neutrino mass and its uncertainty, we use the MPFIT package\footnote{\url{https://www.physics.wisc.edu/~craigm/idl/fitting.html}}, which uses the Levenberg-Marquardt technique to quickly solve the least-squares problem.  Note that, as no covariance matrices were published in the tSZ observational studies, we neglect any correlation between the different multipole bins.  \citet{Planck2013_sz} and \citet{Planck2015_sz} state that they have adopted a multipole binning scheme designed to minimize the covariance between the bins.  If there is residual covariance remaining then our analysis will underestimate the statistical uncertainties in the derived value of $M_\nu$.  

\begin{figure}
\includegraphics[width=0.995\columnwidth]{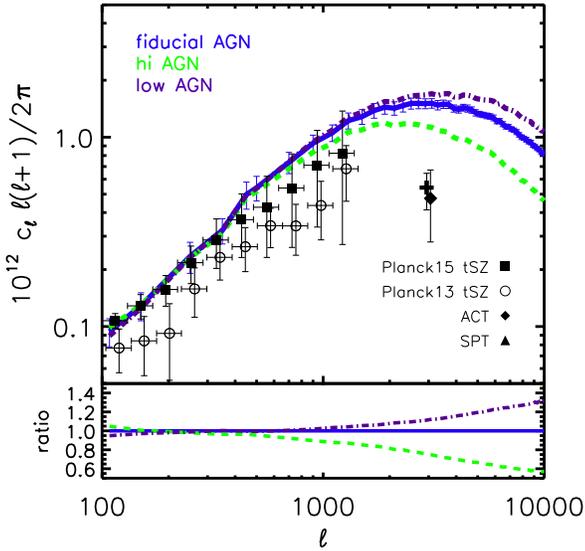}
\caption{\label{fig:cl_y_y_agn}
The sensitivity of the predicted tSZ effect power spectrum to uncertainties in feedback modelling, in the context of the WMAP9-based cosmology with massless neutrinos.  We compare the fiducial \calsim~model with two runs which vary the AGN feedback (see Fig.~\ref{fig:bar_vary_agn}).  The angular power spectrum is insensitive to baryon physics at $\ell\la500$ but can vary at the tens of percent level on scales of a few arcminutes and smaller ($\ell\ga3000$).  Note that the uncertainties are {\it not} sufficiently large to reconcile the differing constraints on $M_\nu$ derived from the \planck~2015 data and ACT and SPT data.
}
\end{figure}

Note that when fitting the models to the data, we neglect the uncertainty in the theoretical predictions.  This is because we find that, for a given simulation, the error on the calculated mean power spectrum (estimated by dividing the scatter about the mean from different sight lines by the square root of the number of sight lines) is generally considerably smaller than the observational measurement errors\footnote{The only statistic that we investigate for which the cone-to-cone scatter is larger than the measurement uncertainties is the Compton $y$ one-point PDF, which we discuss in Section \ref{sec:ypdf}.}.  For comparisons to future surveys, however, a more careful treatment of simulation statistical errors will be required.

It is clear from Fig.~\ref{fig:cl_y_y} that the constraints on $M_\nu$ will be sensitive to both the choice of simulations (WMAP9-based vs. Planck2015/$A_{\rm Lens}$-based) and the observational data sets that are employed.  The former is of course expected, given that the other relevant cosmological parameters (e.g., $\Omega_{\rm m}$, $H_0$, $A_s$) differ for the two sets of simulations (owing to differences in the best-fit parameters derived from the \planck~and WMAP primary CMB data).  The latter (i.e., the choice of tSZ dataset) is, however, more worrying and it is clear that the inferred value of $M_\nu$ will be strongly dependent upon this choice.

In Table~\ref{tab:mnu_tsz} we present the constraints on $M_\nu$ from the tSZ power spectrum analysis.  Using \planck~2013 tSZ data, with or without additional ACT and SPT constraints, leads to a strong preference for a non-minimal neutrino mass, with a best fit of $M_\nu\approx0.3 (0.4)$ eV when using the WMAP9-based (Planck2015/$A_{\rm Lens}$-based) simulations.  The quality of the fits are very good in the context of either \planck~2013 tSZ data alone or in combination with ACT and SPT data, indicating approximate consistency between the \planck~2013 and ACT/SPT data.  These results are consistent with the previous findings of \citet{McCarthy2014}, who showed using the cosmo-OWLS simulations that the predicted tSZ effect power spectrum for a \planck~2013 cosmology was of significantly higher amplitude than the \planck~2013 and ACT/SPT power spectrum measurements.

When fitting to \planck~2015 tSZ data only (not shown), however, the picture changes significantly, with a preference for a minimal contribution from massive neutrinos (see Table \ref{tab:mnu_tsz}).  The quality of the fit in this case is also good (i.e., the standard model is a good fit).  This is consistent with the recent findings of \citet{Dolag2016}, who compared the results of their Magneticum simulations (which were scaled to a \planck~2015 cosmology with minimal neutrino mass) to the \planck~2015 tSZ data and also found relatively good agreement.  However, we note that the \planck~2015 tSZ data is in apparent conflict with the SPT and ACT constraints, as a simultaneous fit to all three data sets leads to a poor reduced $\chi^2$.  This statement assumes that the simulated tSZ power spectra have approximately the correct shape.

\begin{figure*}
\includegraphics[width=0.995\columnwidth]{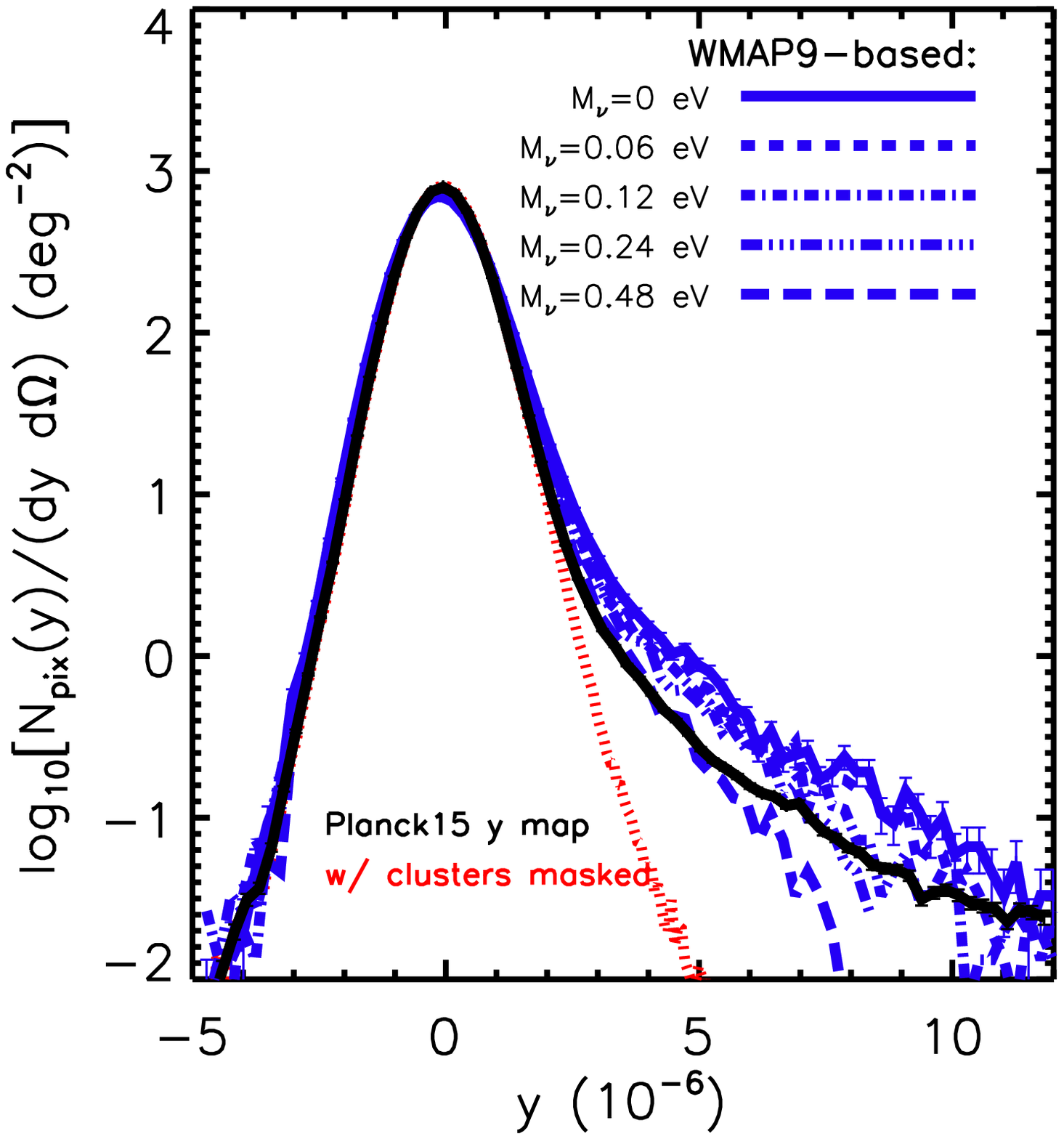}
\includegraphics[width=0.995\columnwidth]{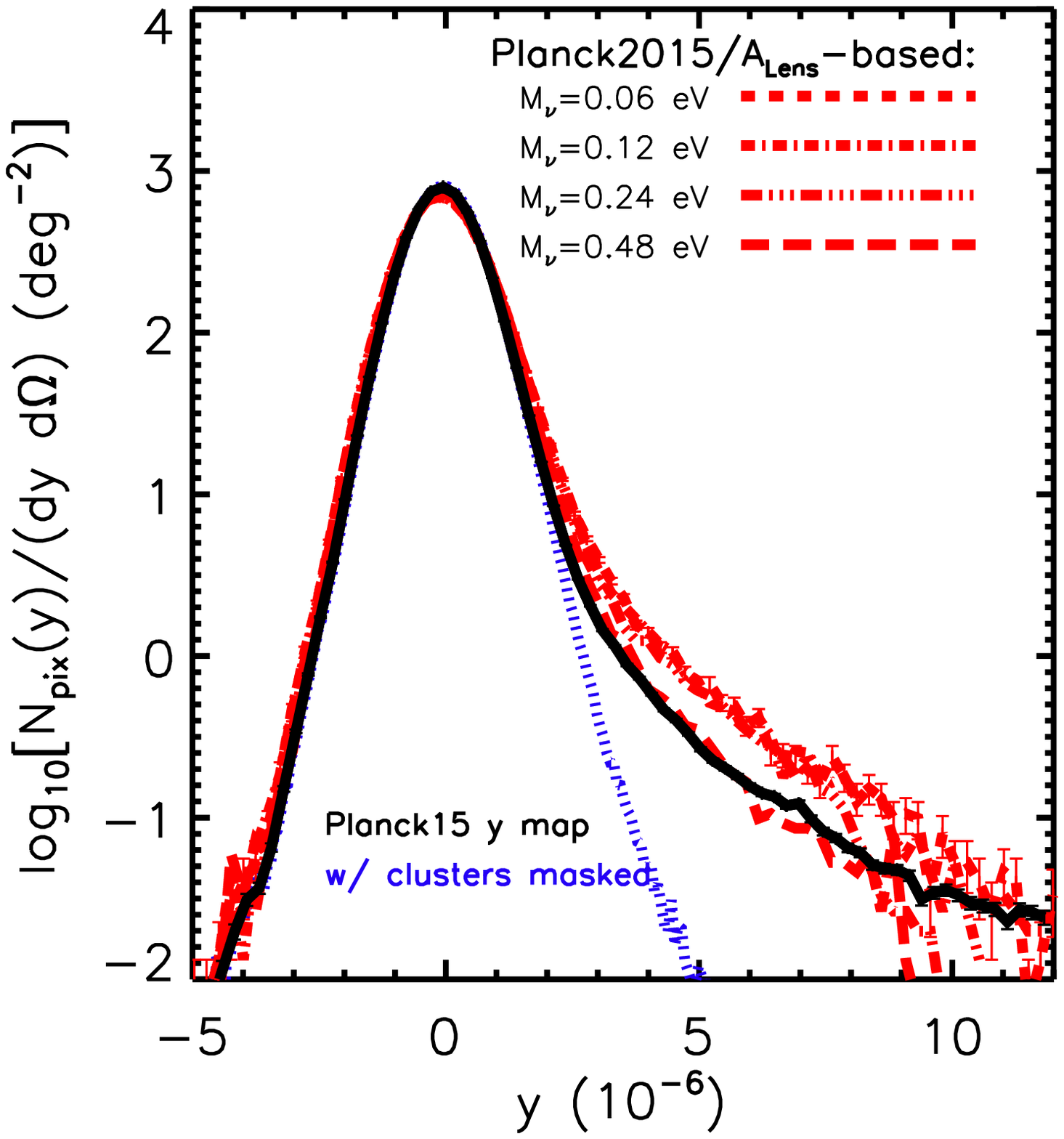}
\caption{\label{fig:y_pdf}
  Comparison of the observed (black curve) and predicted tSZ effect (Compton $y$) normalised one-point probability distribution function.  {\it Left:} Comparison to the WMAP9-based simulations.  {\it Right:} Comparison to the Planck2015/$A_{\rm Lens}$-based simulations. The black data points in both panels represent the one-point PDF derived from the \planck~2015 Compton $y$ (NILC) map.  To make a like-with-like comparison, the simulated tSZ effect maps were rebinned, convolved with the \planck~beam, filtered in Fourier space (as in \citealt{Planck2015_sz}), and had realistic noise/contamination added (see text).  The high $y$ tail of the one-point PDF is sensitive to the abundance of massive clusters and therefore to the neutrino mass.
}
\end{figure*}

Given that the tSZ effect is probing baryons, it is interesting to ask how sensitive the constraints are to uncertainties in the feedback modelling.  A related question is, can these uncertainties accommodate the apparent conflict (in amplitude) between the \planck~2015 tSZ measurements and the ACT and SPT measurements?  To address these questions, we show in Fig.~\ref{fig:cl_y_y_agn} the effects of varying the importance of the AGN feedback, as described in Section \ref{sec:vary_agn}.  (This is done in the context of a WMAP9 cosmology with massless neutrinos.)  Note that the AGN variations bracket the observed gas fractions of groups and clusters (see Fig.~\ref{fig:bar_vary_agn}).  From this comparison we conclude that the tSZ power spectrum is virtually insensitive to feedback modelling uncertainties on large scales, corresponding to multipoles of $\ell \la 500$ or so (see also \citealt{Komatsu1999,Battaglia2010,McCarthy2014}).  At smaller scales, the modelling uncertainties become more significant, but we find that they are insufficiently large to reconcile the \planck~2015 tSZ measurements with the ACT and SPT measurements.  We therefore conclude that uncertainties in the feedback modelling do not make the case for massive neutrinos any more or less compelling.  

It is clear that, at present, systematic errors in the measurements of the power spectrum associated with foreground contamination, particularly at large scales, are the main impediment to arriving at a robust constraint on $M_\nu$ from the tSZ effect power spectrum\footnote{\label{new_tsz} As we were preparing this article for submission, a re-analysis of the tSZ effect power spectrum derived from \planck~2015 data by \citet{Bolliet2017} was posted.  Using a more sophisticated modelling approach for the power spectrum covariance matrix (by including the trispectrum), they derive a tSZ effect power spectrum that is of significantly lower amplitude (at $\ell \ga 300$) than reported previously in \citet{Planck2015_sz}.  The new measurements are very similar to those previously reported in \citet{Planck2013_sz}, which may be somewhat fortuitous.}.

%However, we note that, if the true systematic errors could be roughly estimated by comparing the 2013 and 2015 \planck~tSZ data points, then the derived constraints on $M_\nu$ would be similar to what we derive below from a wide range of independent LSS tests. 

\subsubsection{One-point probability distribution function}
\label{sec:ypdf}
  
The tSZ effect signal on the sky is not a Gaussian random field and is therefore not fully described by the two-point angular power spectrum.  Recognizing this, previous studies have examined what cosmological constraints can be obtained by looking at other moments of the tSZ signal, including the one-point probability distribution function (PDF) and the tSZ bi-spectrum (e.g., \citealt{Wilson2012,Hill2014b,Planck2013_sz}).  Here we compare \planck~measurments of the one-point PDF \citep{Planck2015_sz} (we use the \planck~NILC map) to that derived from the \calsim~simulations.  We plan to examine the bi-spectrum and other higher-order statistics in future work.

In Fig.~\ref{fig:y_pdf} we compare the predicted and observed one-point PDFs, defined as $P(y) = N_{\rm pix}(y)/(dy d\Omega)$, where $d\Omega$ is the solid angle in deg$^2$.  The one-point PDF just describes the frequency of pixels (per solid angle) as a function of the Compton $y$; i.e., it is derived by making a histogram of the $y$ values.  To derive the simulated one-point PDFs, we do the following.  We first rebin the simulated Compton $y$ maps so that the pixels are of the same size as the \planck~map.  The \planck~map is in Healpix format with $N_{\rm side}=2048$ resolution, corresponding to a pixel length of $\theta_{\rm pix} \approx 1.72$ arcmin.  Therefore, the simulated maps must be degraded by approximately a factor of 10.  We then convolve the simulated maps with a Gaussian kernel with a FWHM of 10 arcmin, as was done in the construction of the \planck~$y$ map.  (Note that this was not necessary for the angular power spectrum analysis in Section \ref{sec:tsz_power}, as the beam was deconvolved when computing the observed power spectrum, but it has not been deconvolved for the observed one-point PDF analysis.)  To minimize the impact of noise, \citet{Planck2015_sz} further filtered their $y$ map in harmonic space in order to emphasise the multipole range where the tSZ signal is large compared to the instrumental noise.  We apply the same filtering scheme to our simulated $y$ maps using the \planck~filter (kindly provided by B.~Comis), which we do in Fourier space rather than harmonic space, adopting the flat-sky approximation $\ell \approx 2 \pi u$, where $u$ is the angular Fourier wavenumber.  We then add realistic noise to our maps, by randomly sampling from the observed PDF (post-filtering) derived from the \planck~map after all of the detected tSZ sources have been masked (see the dotted curves in Fig.~\ref{fig:y_pdf}).  (The PDF of this masked \planck~$y$ map is consistent with being entirely due to noise and imperfect contamination removal.)  Finally, we compute the simulated PDF by averaging over the 25 light cones for each simulation.  Note that the 1-sigma error bars in Fig.~\ref{fig:y_pdf} on the observed and simulated one-point PDFs reflect Poisson uncertainties only.  However, the main uncertainty for the one-point PDF analysis is the cone-to-cone scatter in the simulations, as the high-$y$ tail is dominated by very massive, nearby clusters \citep{Dolag2016}.

Over the range $y \la 3\times10^{-6}$ the one-point PDF is dominated by noise/contamination errors.  We therefore follow the approach of \citet{Planck2015_sz} and restrict our cosmological analysis to pixels with $y \ge 4.5\times10^{-6}$ (but note that the results are not strongly sensitive to the precise threshold).   It is immediately apparent from Fig.~\ref{fig:y_pdf} that the limited volume of the simulations results in relatively noisy predictions of the mean one-point PDF for the brightest pixels.  The cone-to-cone scatter about the mean (not shown) is also significant.  However, in spite of this, it is still evident that relatively high summed neutrino masses are required to reduce the overall amplitude to a level that is comparable to what is observed by \planck.  

Given the relatively noisy predictions, we opt here to simply integrate the one-point PDFs above the noise limit.  This yields a single value, which is the number of pixels with $y \ge 4.5\times10^{-6}$ per degree$^2$, for each simulation.  (This statistic is analogous to tSZ cluster number counts but instead the counts involve the number of bright pixels rather than the number of massive clusters.)  The \planck~map yields a total of $0.877 \pm 0.007$ deg$^{-2}$, where the uncertainty reflects Poisson errors only.  The WMAP9-based simulations yield $1.822\pm0.653$, $1.482\pm0.573$, $1.224\pm0.488$, $0.939\pm0.372$, and $0.402\pm0.209$ deg$^{-2}$ for the the simulations with $M_\nu=0.0$, $0.06$, $0.12$, $0.24$, and $0.48$ eV, respectively.  The Planck2015/$A_{\rm Lens}$-based simulations yield $1.637\pm0.614$, $1.514\pm0.548$, $1.242\pm0.463$, and $0.555\pm0.292$ deg$^{-2}$ for the the simulations with $M_\nu=0.06$, $0.12$, $0.24$, and $0.48$ eV, respectively.  The quoted errors for the simulations reflect the error on the mean from the 25 light cones, computed as the RMS divided by the square root of 25.  The Poisson uncertainties for the simulations are approximately an order of magnitude smaller than the error on the mean.

Using a simple interpolation scheme in analogy to that for the power spectrum analysis, we find a best-fit neutrino mass of $M_\nu=0.29_{-0.19}^{+0.09} (0.36_{-0.19}^{+0.11})$ eV for the WMAP9-based (Planck2015/$A_{\rm Lens}$-based) simulations.  These constraints are insensitive to uncertainties in the feedback modelling, as we find that the AGN variation runs only modify the predicted `tSZ pixel counts' by $\pm$5\%, while the error on the mean of a given simulation is typically 40\%.  

The derived constraints on $M_\nu$ are consistent with the power spectrum analysis when adopting the \planck~2013 power spectrum results and are also similar to what one would infer using ACT or SPT power spectrum data alone (see Fig.~\ref{fig:cl_y_y}).  On the other hand, the best-fit $M_\nu$ from the integrated one-point PDF is significantly higher than what we derive from power spectrum analysis using the \planck~2015 power spectrum data.  This is interesting as the present analysis uses the same Compton $y$ map as the \planck~2015 power spectrum analysis.  The origin of this difference is unclear.  It may reflect differences in the effects of remaining foreground contamination for the two tests (naively we expect the power spectrum to be more susceptible to these uncertainties).  

\begin{figure*}
\includegraphics[width=0.85\textwidth]{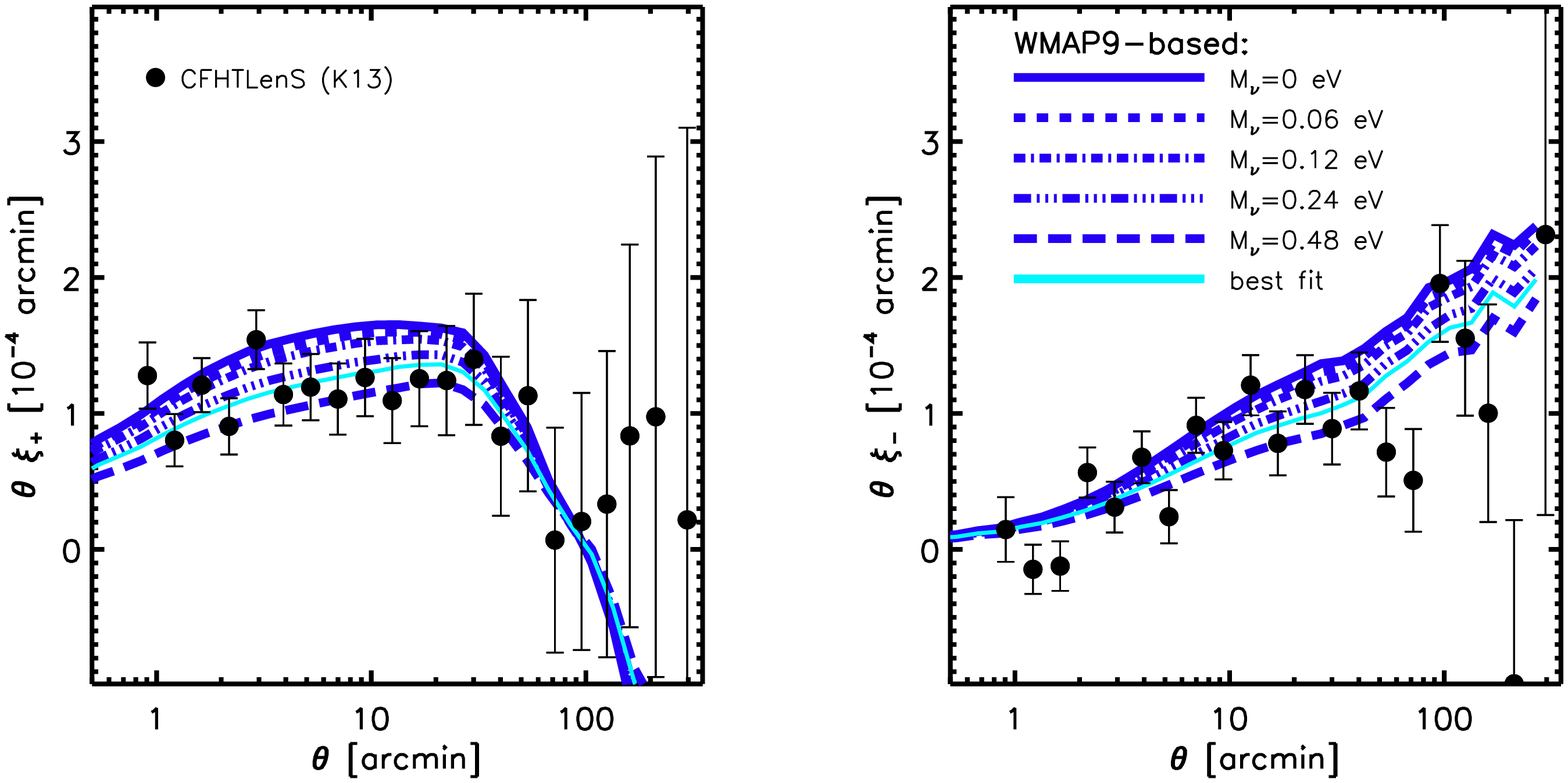} \\
\includegraphics[width=0.85\textwidth]{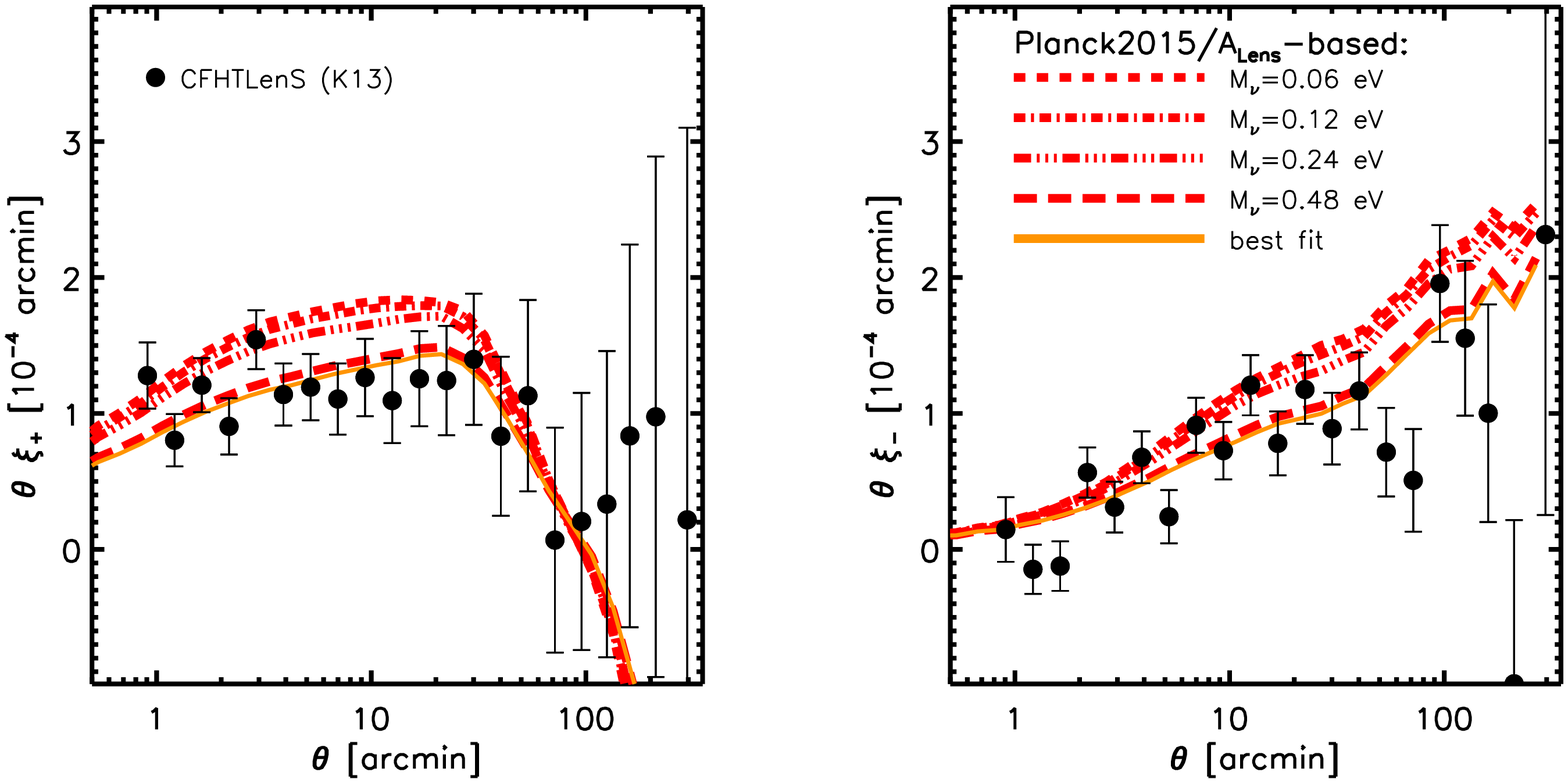}
\caption{\label{fig:cfht_2d}
Comparison of the predicted cosmic shear shape correlation functions to the 2D CFHTLenS measurements of \citet{Kilbinger2013} (data points with 1-sigma error bars).  {\it Left:} The $\xi_+$ correlation function, defined as $\xi_+(\theta) = \langle \gamma_{\rm t} \gamma_{\rm t} \rangle + \langle \gamma_{\rm r} \gamma_{\rm r} \rangle$, where $\gamma_{\rm r}$ and $\gamma_{\rm t}$ are the radial and tangential shear components.  {\it Right:} The $\xi_-$ correlation function, defined as $\xi_+(\theta) = \langle \gamma_{\rm t} \gamma_{\rm t} \rangle - \langle \gamma_{\rm r} \gamma_{\rm r} \rangle$.  {\it Top:} Comparison using the WMAP9-based simulations.  {\it Bottom:} Comparison using the Planck2015/$A_{\rm Lens}$-based simulations.  The amplitudes of the observed correlation functions are significantly lower than expected for either a WMAP9 or \planck~2015 cosmology with minimal neutrino mass (see Table \ref{tab:mnu_cosmic_shear}).
}
\end{figure*}

\subsection{Cosmic shear}
\label{sec:results_shear}

Below we present our constraints on $M_\nu$ from analysis of the lensing shape correlation functions (i.e., cosmic shear).  We compare to two of the most recent cosmic shear surveys: CFHTLenS and the KiDS-450 results.  Since the surveys have different galaxy selection criteria with different source redshift distributions, implying that they are probing LSS at somewhat different redshifts, we analyse these data sets independently.

We note that we have also made comparisons to the DES Science Verification cosmic shear results \citep{DES2016}, but the relatively small survey area, which leads to relatively large errors on the correlation function measurements, prevents a useful constraint on the neutrino mass.  In addition, the Science Verification results have now been superseded by the DES Year 1 results \citep{Troxel2017} based on a survey area that is approximately 10 times larger.  However, the DES collaboration have yet to make the Year 1 source redshift distributions and correlation function measurements publicly available.  A comparison with \calsim~will therefore have to be deferred to a later study.  We note, however, that the derived constraints in the $\sigma_8-\Omega_{\rm m}$ plane from DES Y1 cosmic shear data are consistent with the KiDS-450 constraints of \citet{Hildebrandt2017} (see Fig.~\ref{fig:sig8_omegam}).

\subsubsection{Comparison to CFHTLenS}
\label{sec:cfht}

CFHTLenS is a five-band optical imaging survey conducted with the MegaCam CCD imager on the Canada-France-Hawaii Telescope \citep{Heymans2012}.  The completed survey spans approximately 154 deg$^2$.  There have been three separate cosmic shear analyses of the CFHTLenS survey to date, which include the 2D (i.e., a single tomographic bin) analysis of \citet{Kilbinger2013} (hereafter K13), the 3D tomographic analysis of \citet{Heymans2013} (hereafter H13), and a recent update of the 3D analysis by \citet{Joudaki2017a} (hereafter J17a).  In terms of the shear measurements, the main difference between H13 and J17a is that the latter have extended the measurements to somewhat larger scales using a new covariance matrix calibrated with an updated set of N-body simulations.  Also, J17a use 7 tomographic bins spanning the redshift range $0.15 < z < 1.3$, whereas H13 use 6 spanning $0.2 < z < 1.3$. (There are also important differences in the theoretical modelling between the two studies, but this is not relevant here.)

\begin{table} 
\caption{\label{tab:mnu_cosmic_shear} Constraints on the summed mass of neutrinos derived from cosmic shear (i.e., shape correlation function analysis).  The columns are: (1) Observational data set used; (2) Best-fit value of $M_\nu$ (eV) with 1-sigma uncertainty; and (3) the reduced chi-squared of the best fit.  We have separated the constraints into two sections, based on whether the WMAP9-based or Planck2015/$A_{\rm Lens}$-based simulations were used for the theoretical modelling.}
\begin{tabular}{lcc}                                                                 
\hline
(1)        & (2)            & (3)            \\
Data set   & $M_{\nu}$ (eV)  & $\chi^2$/DOF   \\
\hline
{\bf Planck2015/$A_{\rm Lens}$-based}\\
\hline
CFHTLenS 2D (K13)    & $0.53\pm0.08$    & 1.41 \\
CFHTLenS (H13)       & $0.33\pm0.09$    & 1.35 \\
CFHTLenS rev. (J17)  & $0.43\pm0.10$    & 1.74 \\
KiDS-450 (H17)       & $0.52\pm0.09$    & 1.20 \\
\hline
{\bf WMAP9-based}\\
\hline
CFHTLenS 2D (K13)    & $0.32\pm0.09$    & 1.40 \\
CFHTLenS (H13)       & $0.10\pm0.08$    & 1.35 \\
CFHTLenS rev. (J17)  & $0.23\pm0.11$    & 1.74 \\
KiDS-450 (H17)       & $0.30\pm0.10$    & 1.19 \\
\hline
\end{tabular}
\end{table} 

\begin{figure*}
\includegraphics[width=0.995\columnwidth]{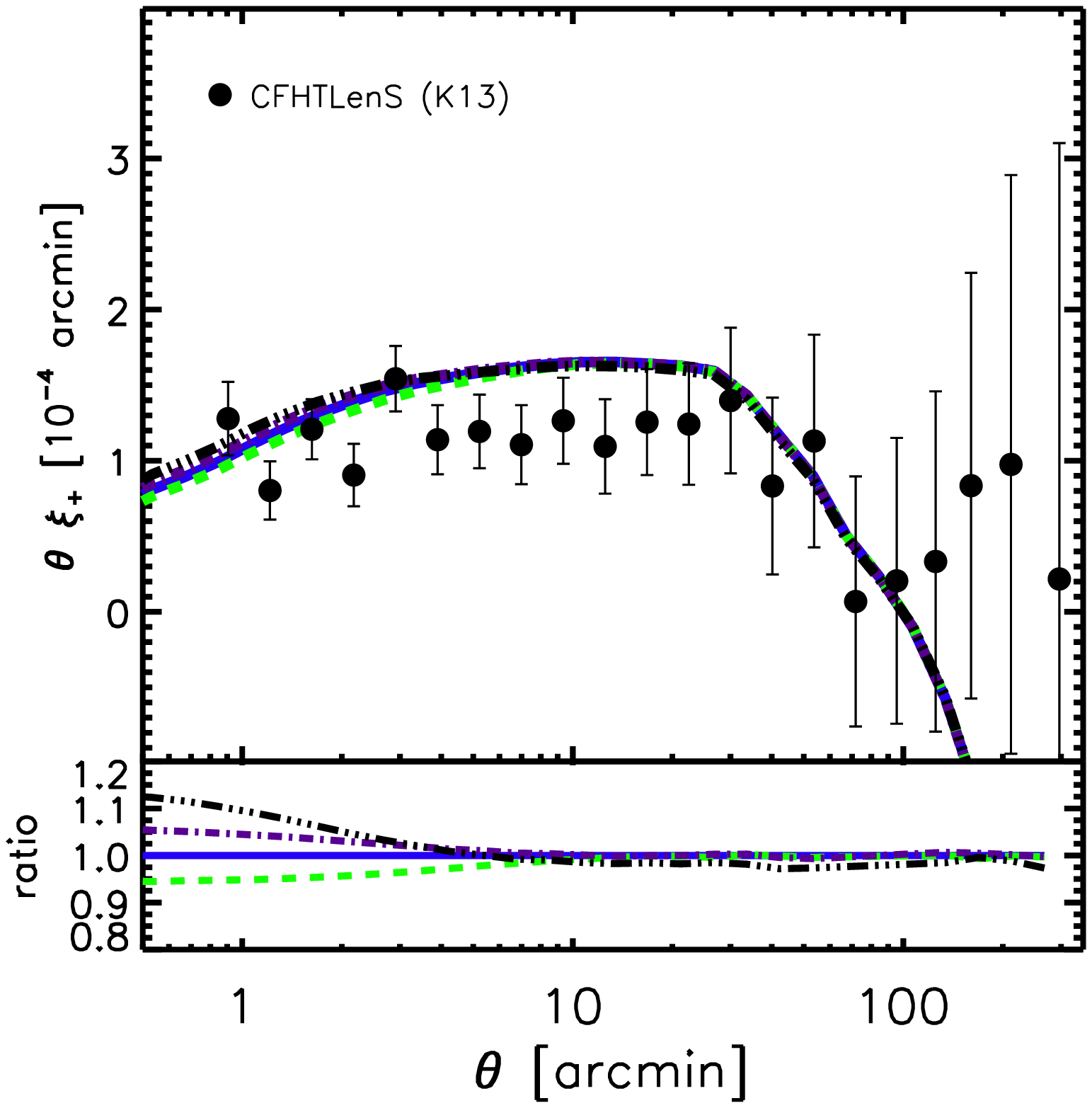}
\includegraphics[width=0.995\columnwidth]{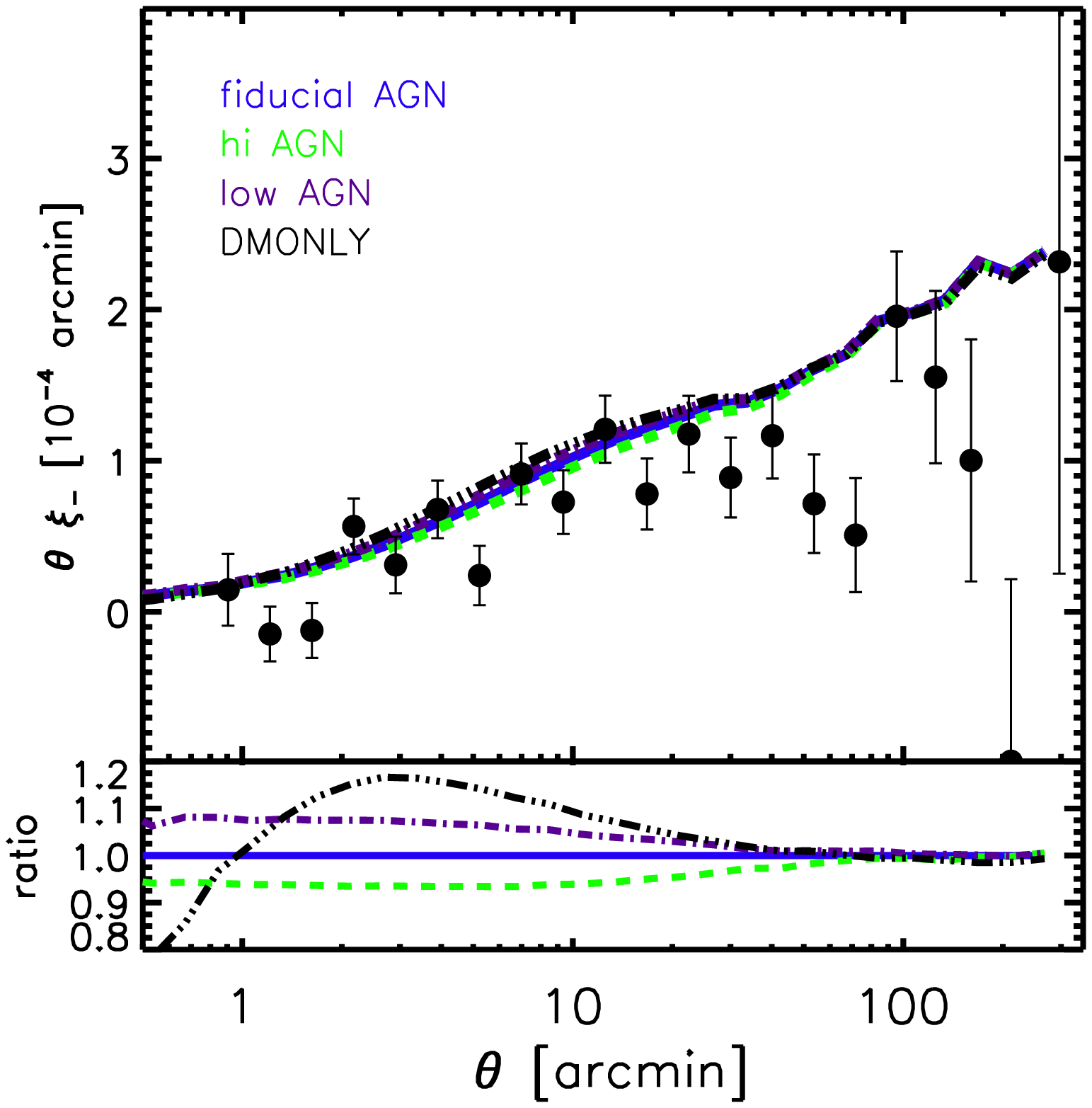}
\caption{\label{fig:cfht_agn}
The sensitivity of the predicted cosmic shear shape correlation functions to uncertainties in feedback modelling, in the context of the WMAP9-based cosmology with massless neutrinos.  The $\xi_+$ correlation function (left panel) is uncertain at the $\approx$5\% level at $\theta\sim1$ arcmin and is essentially unaffected at scales of $\theta \ga 10$ arcmin.  The $\xi_-$ correlation function (right panel) is more sensitive to the inner regions of haloes and sizeable uncertainties persist out to 30-40 arcminutes.  At scales of $\theta \la 1$ arcmin, hydrodynamical simulations predict {\it more} power than a dark matter-only simulation for $\xi_-$, likely owing to stars (e.g., brightest cluster galaxies).  The uncertainty in the baryonic effects is too small to reconcile the standard model with minimal neutrino mass with the observed correlation functions.
}
\end{figure*}

To compute simulated shear maps suitable for comparison to these three datasets, we have retrieved the appropriate background source redshift distributions for the K13, H13, and J17a datasets from the CFHTLenS website\footnote{\url{http://www.cfhtlens.org/}}.  The source redshift distributions are necessary to evaluate the lensing kernel required for computing the simulated shear maps (see eqns.~5-8).  In the present analysis, which is largely a proof of concept that hydro simulations can be used directly for cosmological analyses, we ignore the uncertainty in the photometric redshifts and do not marginalize over a potential offset factor between the estimated and true redshifts.  Going forward, the aim is to directly integrate the \calsim~simulations (via emulators, see Section \ref{sec:discuss}) into the cosmological pipelines being developed for the next generation of LSS surveys.

With shear maps in hand, we compute shear auto- (in the case of 2D) and cross-correlation (tomographic) functions using the publicly available \athena~tree code\footnote{\url{http://www.cosmostat.org/software/athena}} \citep{Kilbinger2014}.  Given a catalog(s) containing the angular coordinates (e.g., RA, DEC) and the complex ellipticities, \athena~returns estimates of the two shape correlation functions $\xi_{\pm}(\theta) = \langle \gamma_{\rm t} \gamma_{\rm t} \rangle \pm \langle \gamma_\times \gamma_\times \rangle$.  We pass \athena~the simulated complex reduced shear maps, $g_1$ and $g_2$, which is equivalent to the complex ellipticity in the absence of shape noise.  Adding realistic shape noise to the simulated shear maps would be straightforward but there is nothing to be gained by doing so, since it would only increase the error bars on the derived $M_\nu$ but without shifting the estimate (i.e., shape noise does not bias the result).  

In analogy to the tSZ effect analysis above, we average the correlation functions over the 25 light cones for each simulation and we produce a function that interpolates (or extrapolates if necessary) the correlation functions for a given choice of angular scale, $\theta$, and summed neutrino mass, $M_\nu$.  We use the MPFIT package, which calls the interpolator, to determine the best fit and 1-sigma errors.  This analysis uses the full publicly-available covariance matrices\footnote{The MPFIT routine `mpfitcovar' uses a singular value decomposition of the covariance matrix to construct a list of uncorrelated deviates, keeping only the largest singular values from the decomposition.  In general, we have found that using the full covariance matrix, rather than just the diagonal elements, leads to only modest shifts (typically less than a few percent) in the best-fit value of $M_\nu$.  However, the 1-sigma uncertainty in $M_\nu$ can increase significantly (by up to 50\%) and the fits are generally of somewhat poor quality (increasing the reduced $\chi^2$ by up to 20-30\%) when allowing for covariance between the different bins.} of each of the CFHTLenS data sets, which we downloaded from the CFHTLenS website.  Note that, when fitting the data, we fit both the $\xi_{+}$ and $\xi_{-}$ functions simultaneously, and for tomographic analyses we fit all redshifts bins simultaneously.  As discussed in Section \ref{sec:lensing_maps}, our analysis neglects intrinsic alignments of background sources.

In Fig.~\ref{fig:cfht_2d} we compare the predicted shape correlation functions to the 2D measurements of K13.  When fitting to the K13 dataset, we use the angular scale range employed in that study, spanning 0.9 to 300 arcmins, with 21 angular bins for both the $\xi_+$ and $\xi_-$ functions and their covariance matrices.  The amplitude of the observed correlation functions is clearly lower than expected for either a \planck~2015 or a WMAP 9 cosmology with minimal neutrino mass.  A summed neutrino mass of $M_\nu \approx 0.3 (0.5)$ eV (see Table \ref{tab:mnu_cosmic_shear}), however, yields relatively good agreement with the data for the WMAP9-based (Planck2015/$A_{\rm Lens}$-based) simulations.  With a reduced-$\chi^2 \approx 1.4$ for the best-fit cases, the `goodness of fit' to the data is reasonable, though clearly not perfect, and is similar to the quality of the fits reported in previous cosmic shear studies that use the halo model or HALOFIT (e.g., K13, H13, J17a).

It is worth briefly commenting on the apparent negative $\xi_+$ correlation predicted by the simulations at large angular scales ($\theta \ga 100$ arcmin).  This negative correlation is a consequence of the finite box size of the simulations; i.e., it is due to a lack of large-scale $k$ modes that are important at large angular scales (see, e.g., \citealt{HarnoisDeraps2015a} for a detailed discussion of this effect).  For the comparisons in this study, this limitation is unimportant because these scales are generally not yet probed by tomographic (3D) analyses (see below) and are only measured with a very low signal-to-noise ratio in 2D tests, such as in the present case.  We have explicitly verified that none of our cosmological results change significantly by excluding these large angular scales.

There has been much interest recently in the possible bias in cosmological constraints from cosmic shear analyses that neglect baryonic feedback (e.g., \citealt{Semboloni2011,Eifler2015}).  This is motivated by previous simulation work, which has found that the matter power spectrum can be modified by up to $\sim$20-30\% by baryonic processes, relative to that of a dark matter-only simulation and that the difference only becomes negligible (i.e., $<$1\%) on relatively large scales of $k\la 0.3 h$/Mpc (e.g., \citealt{vanDaalen2011,Mummery2017}).  Therefore, an important question is, how sensitive are the neutrino mass constraints to uncertainties in the feedback modelling?  To address this question, we show in Fig.~\ref{fig:cfht_agn} the effect of varying the strength of the AGN feedback on the predicted cosmic shear correlation functions.  This is done in the context of the WMAP9-based simulation with massless neutrinos and using the CFHTLenS 2D source redshift distribution.  For comparison, we also show the correlation functions predicted by a dark matter-only simulation with the same cosmology.  

\begin{figure*}
\includegraphics[width=0.75\textwidth]{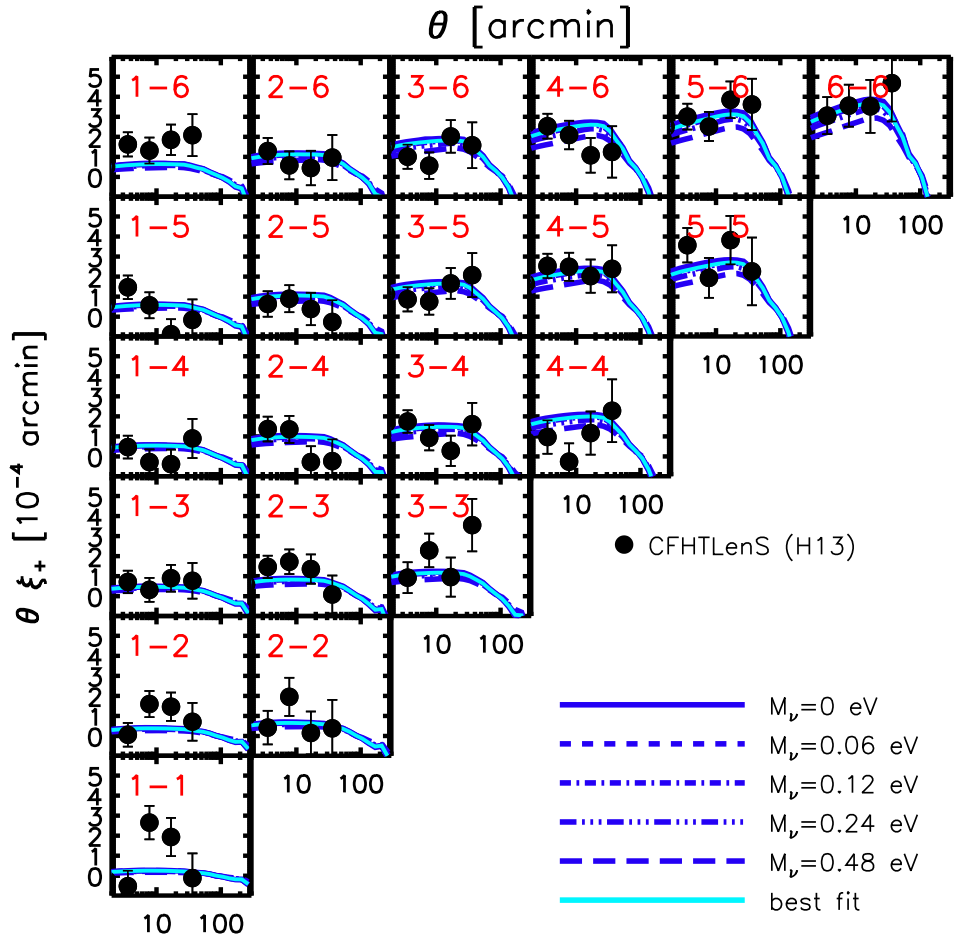}
\caption{\label{fig:cfht_xip_tom}
Comparison of the WMAP9-based predictions (curves) to the $\xi_+$ tomographic CFHTLenS shear measurements of \citet{Heymans2013} (data points with 1-sigma error bars).  (The $\xi_-$ correlation functions and the comparison to the Planck2015/$A_{\rm Lens}$-based simulations can be found in Appendix \ref{sec:append_shear}.)  The red numbers in each panel indicate which tomographic bins are being cross-correlated (e.g., $1$-$6$ indicates that the first and sixth redshift bins are being cross-correlated, where the first bin correspond to the lowest-redshift bin).  The tomographic correlation functions also prefer a non-minimal neutrino mass, though with a somewhat smaller value than preferred by 2D (i.e., a single redshift bin) analysis - see Table \ref{tab:mnu_cosmic_shear}.  A comparison to a re-analysis of the tomographic CFHTLenS data by \citet{Joudaki2017a} (see text and Appendix \ref{sec:append_shear}) yields constraints on $M_\nu$ that lie in between those derived from comparisons to the measurements of \citet{Kilbinger2013} and \citet{Heymans2013}.
}
\end{figure*}

The effects of baryon physics (and variations thereof) becomes noticeable at $\theta<10$ arcmin for the $\xi_+$ correlation function and for $\theta<50$ arcmin for the $\xi_-$ correlation function.  For example, relative to the fiducial \calsim~model, the `high AGN' model predicts a lower value of $\xi_+$ by $\approx$5\% at $\theta\approx1$ arcmin, but is virtually identical to that of the fiducial simulation at $\theta \ga 10$ arcmin.  The `low AGN' model, on the other hand, has a higher value of $\xi_+$, also by $\approx$5\%, at $\theta\approx1$ arcmin, but is virtually the same as the fiducial model beyond 10 arcmin.  {\it Note, however, that the fiducial \calsim~model predicts a value of $\xi_+$ that is $\approx$10\% lower than that of a dark matter-only simulation at $\theta\approx1$ arcmin.}  For the $\xi_-$ correlation function, the effect is even larger and remains large out to $\theta \approx 10$ arcmin.  Interestingly, for the $\xi_-$ correlation function, the hydrodynamical simulations predicted a {\it stronger} correlation (more power) on scales of $\theta\la1$ arcmin compared to a dark matter-only simulation.  This is likely due to the presence of stars (e.g., central galaxies) which begin to dominate the potential well on small physical scales.  This behaviour was previously noted by \citet{Semboloni2011}.

For the present analysis, the uncertainties in the feedback modelling translate into uncertainties in the predicted correlation functions that are still relatively small compared to the observational measurements errors.  Going forward, however, future cosmic shear surveys, such as those to be undertaken with \textsc{Euclid} and LSST, will achieve much more precise estimates of the correlation functions and, therefore, much more care will need to be taken when modelling the effects of baryons, particularly in the $\xi_-$ correlation function.  On the positive side, the differing dependencies of $\xi_+$ and $\xi_-$ to cosmology and baryonic effects suggests that the cosmic shear data itself may be a useful probe of both (e.g., \citealt{Semboloni2013,HarnoisDeraps2015b,Foreman2016}).  Cross-correlation of lensing surveys with surveys of the baryons (e.g., \citealt{vanWaerbeke2014,Hill2014a,Hojjati2017}) offer another interesting way to constrain both cosmology and galaxy formation physics simultaneously. 

In Fig.~\ref{fig:cfht_2d} we explored what neutrino mass constraints can be obtained by using a single redshift bin.  However, this does not exploit the full power of the cosmic shear surveys.  One can subject the model to a more stringent test by breaking the background galaxies up into redshift (`tomographic') bins and performing cross-correlations between the bins.  Such analyses provide additional information about the growth of LSS over cosmic time.

In Fig.~\ref{fig:cfht_xip_tom} we compare the predicted $\xi_+$ correlation functions from the WMAP9-based cosmology simulations with the CFHTLenS tomographic correlation functions of H13.  The $\xi_-$ correlations functions for the WMAP9-based simulations and $\xi_\pm$ for the Planck2015/$A_{\rm Lens}$-based simulations can be found in Appendix \ref{sec:append_shear}.   We adopt the same angular range cuts as employed by H13, which consist of 5 angular bins spanning the range from 1.5 to 30 arcmins for each of the 6 tomographic bins.  Including all of the unique cross-correlations, the data vector contains 210 elements (summing together the $\xi_+$ and $\xi_-$ measurements).  The tomographic analysis yields best-fit values of $M_\nu \approx 0.1 (0.3)$ eV for the WMAP9-based (Planck2015/$A_{\rm Lens}$-based) simulations, respectively (see Table \ref{tab:mnu_cosmic_shear}).  This is somewhat lower (by about 2-sigma) than what we obtained for the non-tomographic comparison to K13 above.  The goodness of fit to the tomographic data, however, is of a similar quality to what we found for the 2D analysis above.  Furthermore, we have examined the impact of uncertainties in the feedback modelling on the tomographic analysis, finding it to be sub-dominant to the current measurement errors.

We note that it is the highest-redshift bins (top right region of Fig.~\ref{fig:cfht_xip_tom}) in the tomographic analysis that show the strongest sensitivity to the summed neutrino mass.  This just reflects the fact that photons from more distant galaxies are more strongly lensed due to intervening matter, as there is a longer path length between source and observer.  It also means that any differences between the simulations are amplified, as the differences accumulate over the longer path length.

\begin{figure*}
\includegraphics[width=0.75\textwidth]{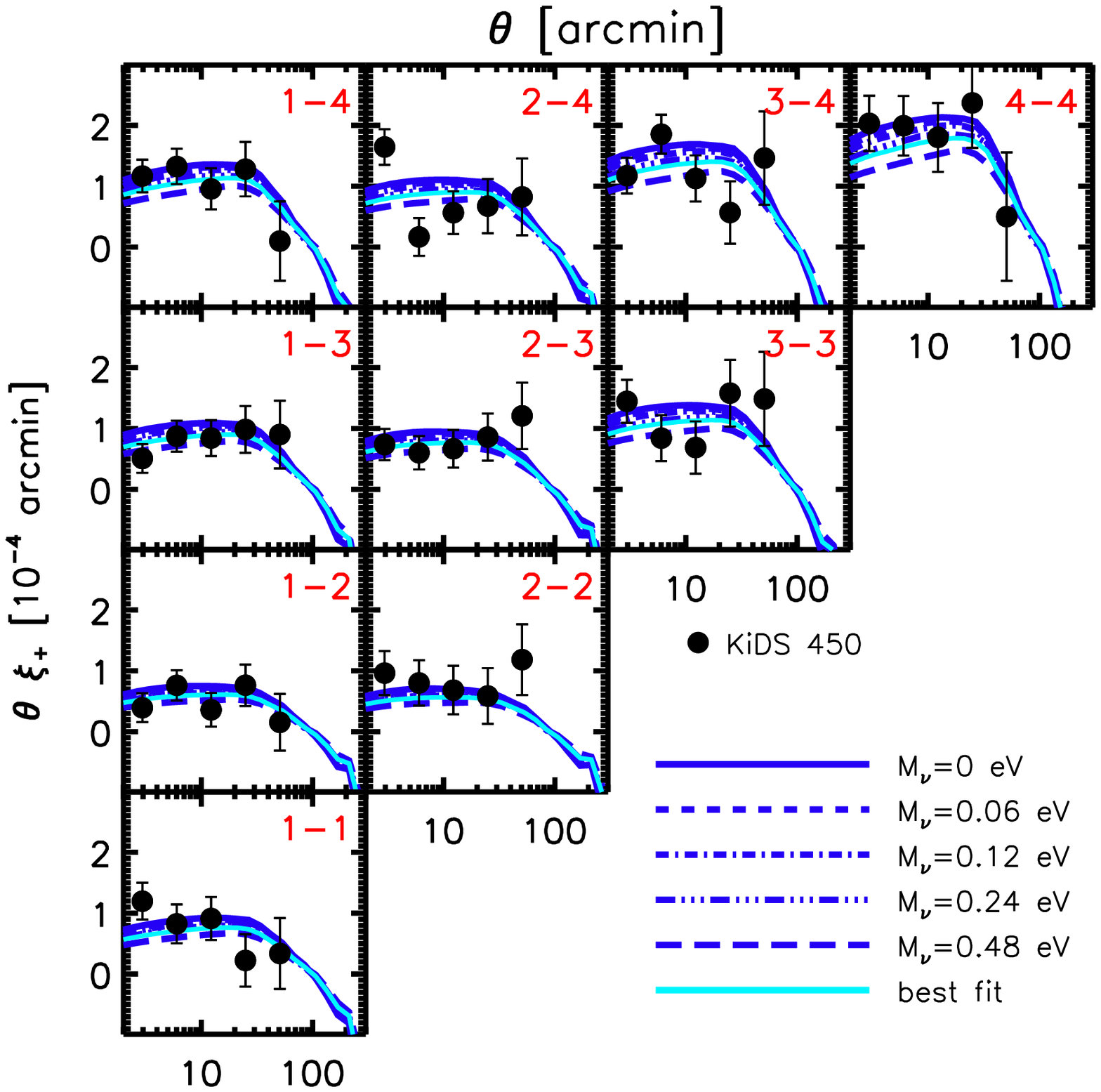}
\caption{\label{fig:kids_xip_tom}
Comparison of the WMAP9-based predictions (curves)to the $\xi_+$ tomographic KiDS-450 shear measurements of \citet{Hildebrandt2017} (data points with 1-sigma errors).  (The $\xi_-$ correlation functions and the comparison to the Planck2015/$A_{\rm Lens}$-based simulations can be found in Appendix \ref{sec:append_shear}.)  The KiDS tomographic correlation functions prefer a non-minimal neutrino mass of $\approx 0.3 (0.5)$ eV in the context of the WMAP9-based (Planck2015/$A_{\rm Lens}$-based) simulations (see Table \ref{tab:mnu_cosmic_shear}).
}
\end{figure*}

We have also made a comparison to the recent re-analysis of CFHTLenS data\footnote{\url{https://github.com/sjoudaki/cfhtlens_revisited}} by J17a (correlation functions can be found in Appendix \ref{sec:append_shear}).  Again, we adopt the same angular range cuts as employed in the observational analysis, resulting in a data vector consisting of 280 elements.  Our analysis yields constraints that lie between those obtained by comparison to K13 and H13: we find $M_\nu \approx 0.2 (0.4)$ eV for the WMAP9-based (Planck2015/$A_{\rm Lens}$-based) simulations, respectively (see Table \ref{tab:mnu_cosmic_shear}).  Here the quality of the fit to the data is worse than for the previous cases.  This was also found by J17a when comparing their models to the data, which they ascribed to a more accurately determined covariance matrix in J17a compared to that used in H13.

Overall, therefore, the CFHTLenS cosmic shear data do tend to prefer a non-mininal neutrino mass, but the constraint on $M_\nu$ can vary by up to 2-sigma depending on which correlation functions are modelled. 

\subsubsection{Comparison to KiDS-450}
\label{sec:kids}

The Kilo Degree Surveys (KiDS) is an ongoing four-band imaging survey being conducted with the OmegaCAM CCD mosaic camera on the VLT Survey Telescope (VST), with the aim of completing $1500$ deg$^2$ split into two approximately equal area regions.  Here we compare to the cosmic shear measurements of \cite{Hildebrandt2017} (hereafter H17), which were derived from $\approx450$ deg$^2$ of the completed imaging data.  H17 split their galaxy sample into 4 tomographic bins spanning the range $0.1 < z < 0.9$.  We obtained the `direct calibration method' (DIR) source redshift distributions for these bins from the KiDS website\footnote{\url{http://kids.strw.leidenuniv.nl/}}, along with the correlation function measurements and covariance matrices.

\begin{figure*}
\includegraphics[width=0.995\columnwidth]{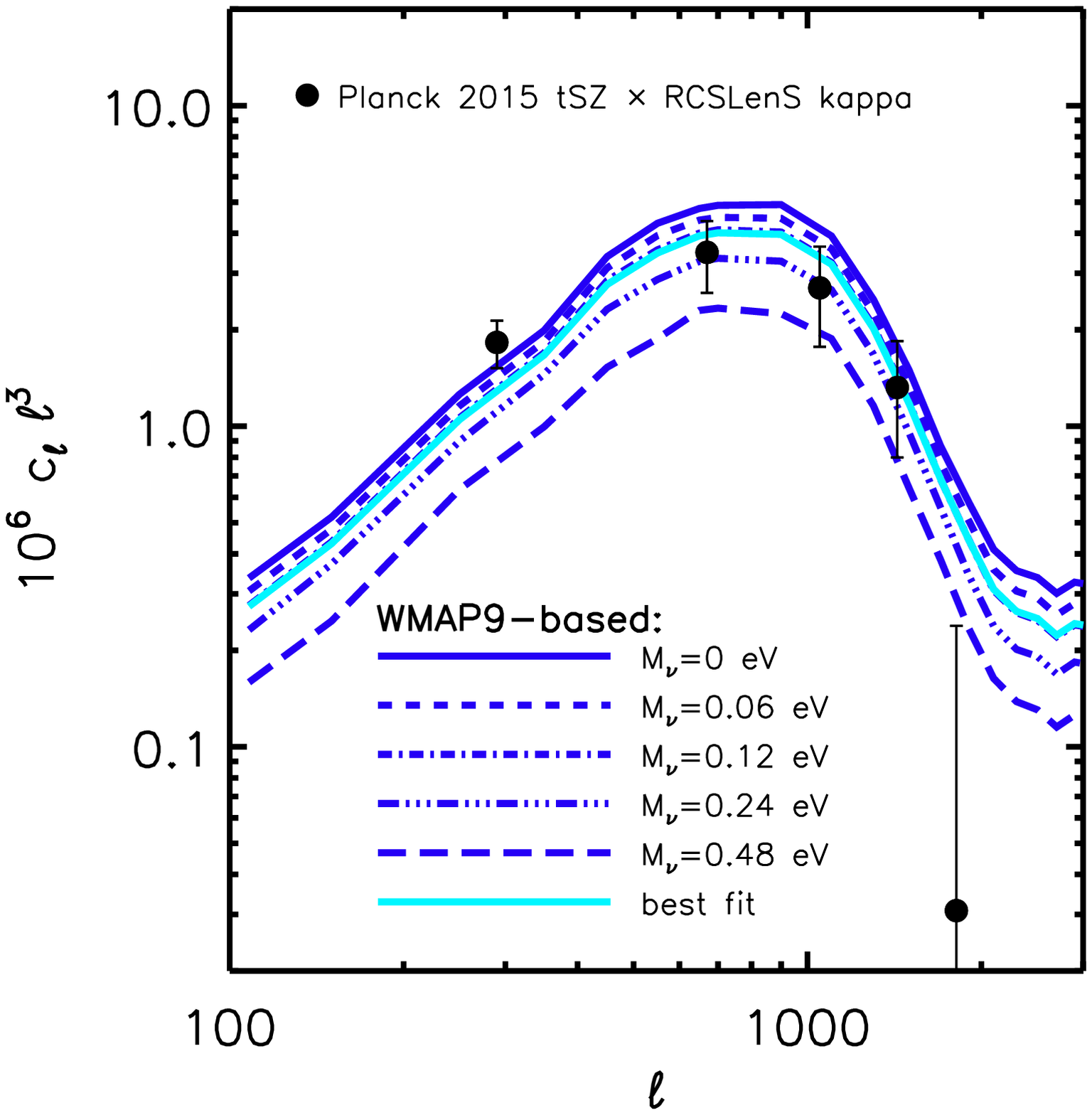}
\includegraphics[width=0.995\columnwidth]{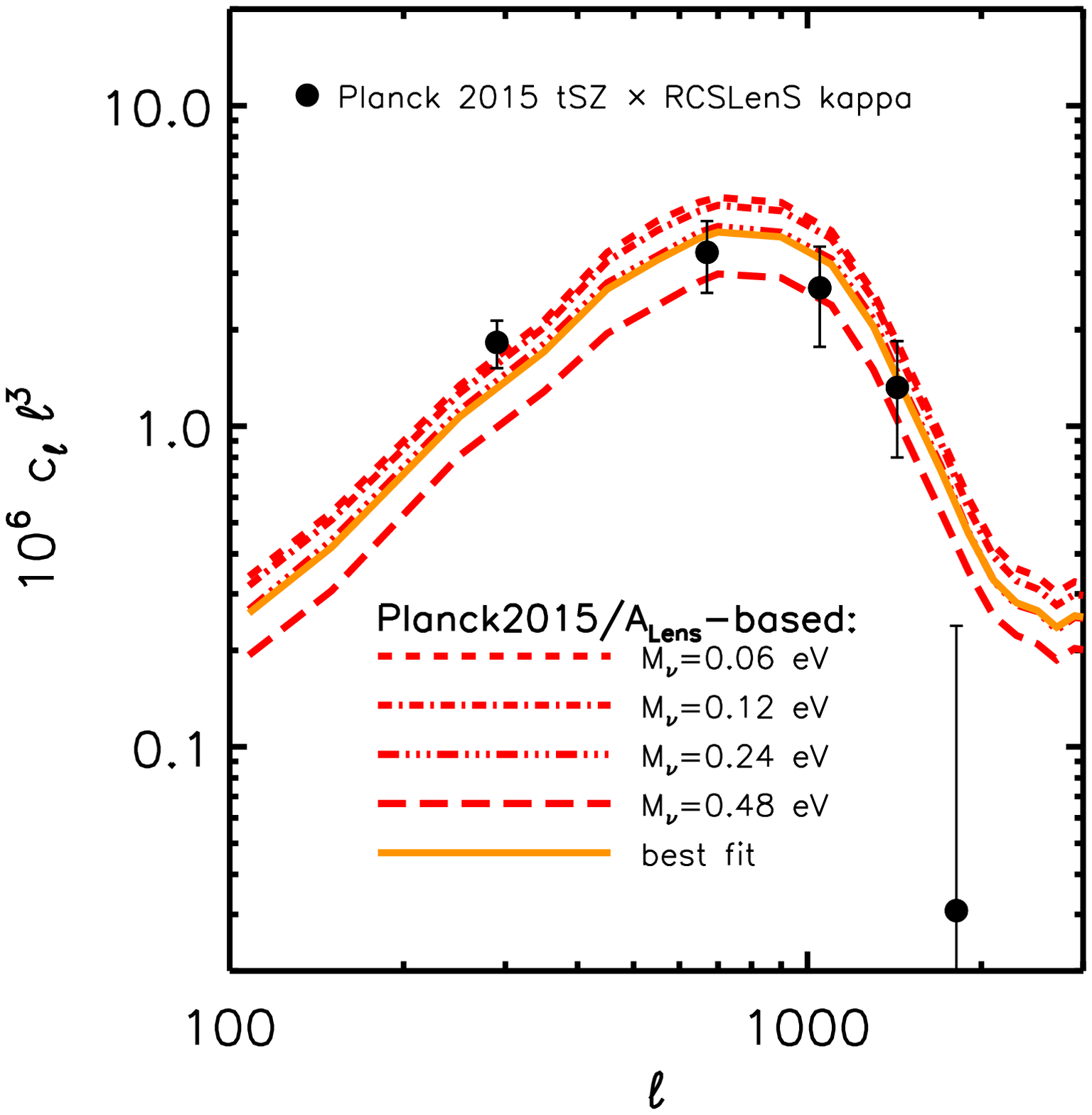}
\caption{\label{fig:cl_y_kappa}
Comparison of the predicted tSZ effect-galaxy weak lensing cross-spectrum (curves) to the measurements of \citet{Hojjati2017} (data points with 1-sigma error bars).  {\it Left:} Comparison using the WMAP9-based simulations.  {\it Right:} Comparison using the Planck2015/$A_{\rm Lens}$-based simulations.  The amplitude of the observed cross-spectrum is lower than expected for either a WMAP-9yr or \planck~2015 cosmology with minimal neutrino mass.  A summed neutrino mass of $M_\nu \approx 0.13 (0.26)$ eV yields a good fit to the data for the WMAP9-based (Planck2015/$A_{\rm Lens}$-based) simulations (see Table \ref{tab:mnu_cross_tests}).
}
\end{figure*}

In Fig.~\ref{fig:kids_xip_tom} we compare the measured $\xi_+$ correlation functions with the WMAP9-based simulations.  The corresponding $\xi_-$ functions, along with the $\xi_\pm$ functions for the Planck2015/$A_{\rm Lens}$-based simulations, can be found in Appendix \ref{sec:append_shear}.  Our constraints on the summed mass of neutrinos  from comparison to the KiDS measurements can be found in Table \ref{tab:mnu_cosmic_shear}.  The comparison with the WMAP9-based (Planck2015/$A_{\rm Lens}$-based) simulations prefers a best-fit summed neutrino mass of $M_\nu \approx 0.3 (0.5)$ eV.  This is broadly consistent with the results obtained from comparison to the 2D CFHTLenS measurements of K13 and the revisited tomographic measurements of J17a.  The quality of the fit to the KiDS dataset, as judged by the reduced $\chi^2$, is very good for both the WMAP9- and Planck2015/$A_{\rm Lens}$-based simulations.

We note that the KiDS team have also attempted to constrain the summed mass of neutrinos using the KiDS correlation functions, in \citet{Joudaki2017b}.  For the theoretical modelling, they employed the halo model-based code of \citet{Mead2016}, which has prescriptions for including the effects of baryon physics calibrated on the previous OWLS simulation results of \citet{vanDaalen2011}, as well as massive neutrinos and other extensions of the standard model of cosmology.  \citet{Joudaki2017b} conclude that the KiDS data alone is fully compatible with a wide range of neutrino masses, as expected.  When jointly fitting the KiDS cosmic shear data and the \planck~CMB data, however, they conclude that their constraints are not competitive with the \planck+BAO joint constraints, quoting that the latter constrain $M_\nu < 0.21$ eV \citep{Planck2015_cmb}.  Here we again point out that the quoted constraints are for the fiducial case with $A_{\rm Lens}$ fixed to unity and that the problem of apparent oversmoothing of the TT power spectrum at high multipoles has not been addressed.  Allowing $A_{\rm Lens}$ to vary, the \planck+BAO data are not only compatible with higher values of $M_\nu$, they actually prefer it (see Fig.~\ref{fig:cmb_mnu}).

\subsection{Cross-correlations}
\label{sec:results_cross}

So far we have examined what constraints on $M_\nu$ may be obtained from separate analyses of the tSZ effect and cosmic shear.  However, one can combine these data sets to perform an additional {\it independent} test of the models, which is the cross-correlation of the tSZ effect with gravitational lensing.  Note that this test is independent of the autocorrelation analyses we have already performed, since the autocorrelations only constrain the (projected) amplitudes of the hot gas and total mass, respectively, but say nothing about their spatial overlap (i.e., their relative phases).  Cross-correlations are also appealing on observational grounds, since they tend to be less sensitive to residual foreground/background contaminants in the individual maps.  

Cross-correlation analyses between the tSZ effect and gravitational lensing are not restricted to galaxy lensing.  Recently, \planck~(e.g., \citealt{Planck2015_lensing}), ACT (e.g., \citealt{Sherwin2017}), and SPT (e.g., \citealt{Omori2017}) have produced the first gravitational lensing maps of fluctuations in the primary CMB.  The first cross-correlation measurements between CMB lensing and the tSZ effect and galaxy weak lensing have also been made and we compare \calsim\footnote{As the simulated light cones extend back only as far $z=3$, they cannot be used to accurately predict the CMB lensing power spectrum (autocorrelation).  To predict the power spectrum, one, at least in principle, needs to account for the lensing due to matter fluctuations all the way back to the last scattering surface.  Cross-correlations between CMB lensing and other lower-redshift signals (e.g., galaxy weak lensing), on the other hand, will only be sensitive to LSS that lies in the overlap region.}~to these measurements to see what constraints may be obtained on $M_\nu$.    

\subsubsection{tSZ effect-galaxy weak lensing}
\label{sec:tsz_shear_cross}

\citet{vanWaerbeke2014} were the first to detect and measure the cross-correlation between the tSZ effect and galaxy weak lensing.  They cross-correlated a custom Compton $y$ map derived from the \planck~2013 data release with a lensing convergence map derived from the CFHTLenS survey, in configuration space.  \citet{Hojjati2015} and \citet{Battaglia2015} subsequently compared these measurements with the predictions of cosmological hydrodynamical simulations (with massless neutrinos), with both studies independently concluding that the observed signal was of lower amplitude than predicted when adopting the best-fit \planck~2013 cosmology.  More recently, \citet{Hojjati2017} measured the configuration-space and Fourier-space cross-correlations between the \planck~2015 Compton $y$ map and galaxy weak lensing measurements from the RCSLenS survey \citep{Hildebrandt2016}.  Hojjati et al.\ note that, although RCSLenS is somewhat shallower than CFHTLenS, it is approximately 4 times larger in area than the latter, which leads to a more precise measurement of the cross-correlation and allowing the measurement to be extended to significantly larger scales.  Here we present a comparison to the Fourier-based cross-correlation measurements of \citet{Hojjati2017}, noting that similar conclusions are obtained from a configuration-space analysis.

\begin{figure}
\includegraphics[width=0.995\columnwidth]{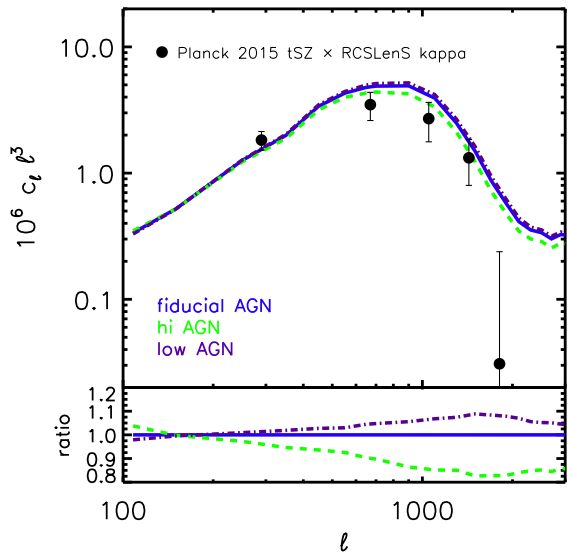}
\caption{\label{fig:cl_y_kappa_agn}
The sensitivity of the predicted tSZ effect-galaxy weak lensing cross-spectrum to uncertainties in the feedback modelling.  The predicted cross-spectrum varies by only $\la5$\% on scales of $\ell \la 500$, but can vary by up to 20\% at $\ell > 1000$.  These uncertainties are sub-dominant compared to the present measurement errors and cannot reconcile the standard model with a \planck~2015 cosmology with minimal neutrino mass with the observations.
}
\end{figure}

In Fig.~\ref{fig:cl_y_kappa} we compare the predicted tSZ-galaxy weak lensing cross-correlations for the WMAP9-based (left panel) and Planck2015/$A_{\rm Lens}$-based (right) simulations with the measurements of \citet{Hojjati2017}.  Note that, to make a like-with-like comparison to the observed cross-correlation, we use the RCSLenS source redshift distribution reported in \citet{Hojjati2017} to compute appropriate galaxy weak lensing maps.  Furthermore, we smooth the simulated galaxy weak lensing convergence and tSZ effect maps with 10 arcmin Gaussian beams, as done for the observational data.  This beam is {\it not} deconvolved from the reported cross-correlation, which is why the power decreases beyond $\ell\sim1000$.  Following our previous analyses, we use MPFIT in conjunction with an interpolation function (that interpolates $C_\ell$ at a given $\ell$ and choice of $M_\nu$) to derive the best-fit value of $M_\nu$ and the associated 1-sigma uncertainties.  This analysis accounts for the small covariance between multipole bins in the measurements, using the covariance matrix of \citet{Hojjati2017}.  Consistent with \citet{Hojjati2017}, we find that the \planck~cosmology with minimal neutrino mass predicts a higher-than-observed amplitude for the cross-correlation.  Allowing the neutrino mass to vary, we find that the data prefer a summed neutrino mass of $M_\nu \approx 0.26 (0.13)$ eV when adopting the Planck2015/$A_{\rm Lens}$-based (WMAP9-based) simulations (see Table \ref{tab:mnu_cross_tests}).  The goodness of fit to the data in both cases is excellent, with a reduced-$\chi^2 \approx 1$.

\begin{figure*}
\includegraphics[width=0.995\columnwidth]{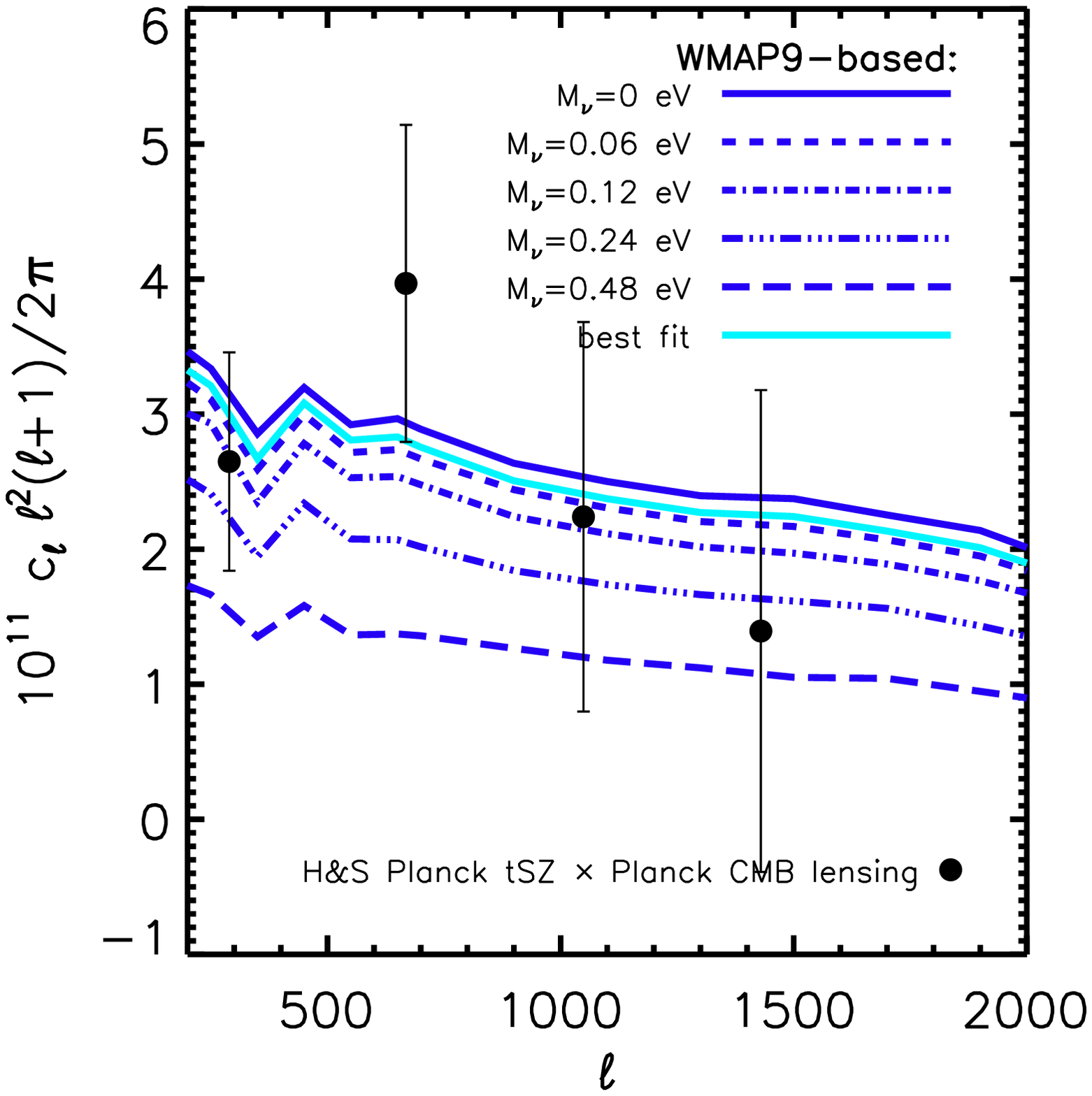}
\includegraphics[width=0.995\columnwidth]{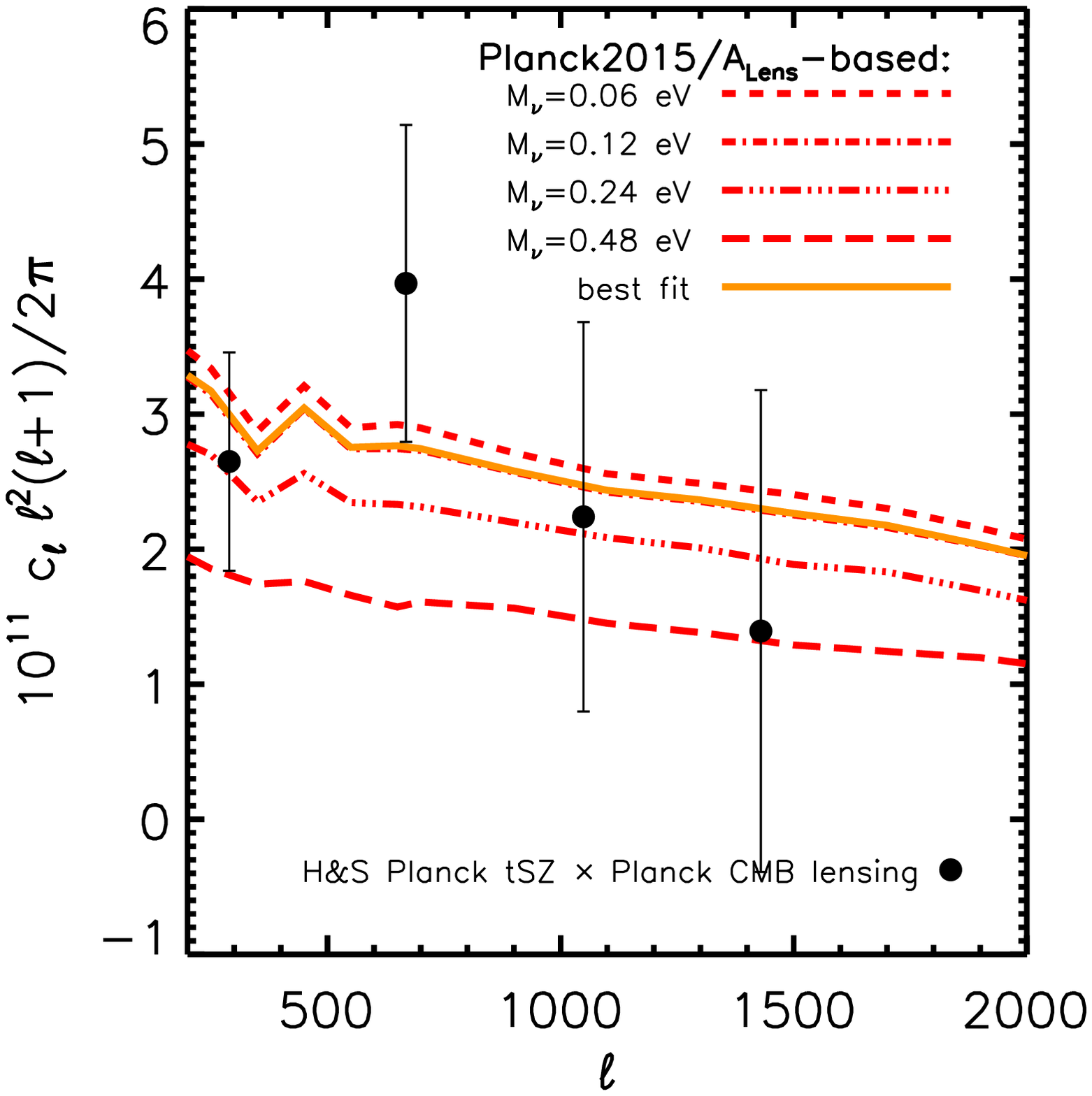}
\caption{\label{fig:cl_y_phi}
Comparison of the predicted tSZ effect-CMB lensing cross-spectrum (curves) to the measurements of \citet{Hill2014a} (data points with 1-sigma error bars).  {\it Left:} Comparison using the WMAP9-based simulations.  {\it Right:} Comparison using the Planck2015/$A_{\rm Lens}$-based simulations.  The amplitude of the observed cross-spectrum is consistent with the minimal neutrino mass case but is also compatible with neutrino masses of up to $\approx 0.2 (0.3)$ at 1-sigma for the WMAP9-based (Planck2015/$A_{\rm Lens}$-based simulations) - see Table \ref{tab:mnu_cross_tests}.  Larger neutrino masses may also be compatible with the data if the observed cross-spectrum has some residual contamination from the CMB lensing-CIB cross-correlation, as argued by \citet{Hurier2015}. 
}
\end{figure*}

An interesting question is, how sensitive is tSZ-galaxy weak lensing cross-correlation to uncertain baryon physics?  \citet{Hojjati2017} explored this question using the cosmo-OWLS suite of simulations \citep{LeBrun2014}, a predecessor to \calsim, finding that the cross-correlation can vary by up to a factor of $2$ in amplitude at $\ell\sim1000$.  However, the models in cosmo-OWLS were not calibrated to match observational data and some of the models in the suite are inconsistent with the observed baryon fractions of groups and clusters (some lie well above the observed relation while others lie below it).  Using the full range of models will therefore likely overestimate the impact of feedback uncertainties.  We revisit this question in Fig.~\ref{fig:cl_y_kappa_agn} using \calsim~which vary the AGN feedback efficiency so that the simulated clusters skirt the upper and lower bounds of the observed cluster gas fractions (Fig.~\ref{fig:bar_vary_agn}).  We find that on large scales, $\ell \la 500$, the cross-correlation is insensitive to baryon physics; i.e., the effects are at the $\la 5$\% level.  At smaller angular scales ($\ell \sim 1000$), we find that the `hi AGN' model predicts an amplitude approximately 15\% lower than to our fiducial model, whereas the `low AGN' model predicts a $\approx$10\% higher amplitude relative to the fiducial model.  

\begin{table} 
\caption{\label{tab:mnu_cross_tests} Constraints on the summed mass of neutrinos derived from cross-correlations between the tSZ effect, CMB lensing, and cosmic shear.  The columns are: (1) Observational data set used; (2) Best fit value of $M_\nu$ (eV) with 1-sigma uncertainty; and (3) the reduced chi-squared of the best fit.  We have separated the constraints into two sections, based on whether the WMAP9-based or Planck2015/$A_{\rm Lens}$-based simulations were used for the theoretical modelling.}
\begin{tabular}{lcc}                                                                 
\hline
(1)        & (2)            & (3)            \\
Data set   & $M_{\nu}$ (eV)  & $\chi^2$/DOF   \\
\hline
{\bf Planck2015/$A_{\rm Lens}$-based}\\
\hline
RCSLenS $\times$ Planck tSZ     &  $0.26\pm0.10$          & 0.91 \\
Planck lensing $\times$ Planck tSZ &  $0.11_{-0.11}^{+0.16}$   & 0.51 \\
KiDS $\times$ Planck lensing &  $0.12\pm0.35$         & 1.00 \\
KiDS (2D) $\times$ Planck lensing &  $<0.49$         & 0.69 \\

\hline
{\bf WMAP9-based}\\
\hline
RCSLenS $\times$ Planck tSZ     &  $0.13\pm0.09$         & 1.07 \\
Planck lensing $\times$ Planck tSZ &  $0.04_{-0.04}^{+0.14}$  & 0.47 \\
KiDS $\times$ Planck lensing &  $<0.34$               & 1.01 \\
KiDS (2D) $\times$ Planck lensing &  $<0.32$               & 0.73 \\
\hline
\end{tabular}
\end{table} 

We emphasize that much of the difference between the different feedback variation models at small scales has been removed as a result of the convolution with the 10 arcminute beam, suitable for a comparison with \planck~data.  As shown by \citet{Hojjati2017}, the differences between the models would be much more significant at higher resolution.  Therefore, if the goal is to probe baryon physics, cross-correlation of higher-resolution tSZ effect maps (such as those obtained with SPT and ACT and their imminent successors, such as SPT-3G and Advanced ACTpol) with cosmic shear surveys offers a very promising avenue to explore.

\subsubsection{tSZ effect-CMB lensing}
\label{sec:tsz_lensing_cross}

\citet{Hill2014a} is the only study we are aware of to date to examine the cross-correlation between the tSZ effect and CMB lensing, which they did in harmonic space.  They cross-correlated a custom Compton $y$ map derived from the \planck~2013 data release with the \planck~2013 CMB lensing map.  CMB lensing measurements are currently of relatively low significance compared to those of galaxy lensing.  Nevertheless, \citet{Hill2014a} derived a competitive constraint on $S_8$ from the tSZ-CMB lensing cross-spectrum, reporting a value of $S_8$ that lies between the best-fit WMAP9 and \planck~2015 values (i.e., they found no significant evidence for a tension with the primary CMB).

In Fig.~\ref{fig:cl_y_phi} we compare the predicted tSZ-CMB lensing cross-correlations for the WMAP9-based (left panel) and Planck2015/$A_{\rm Lens}$-based (right) simulations with the rebinned measurements of \citet{Hill2014a} (see their figure 15).  For the simulated CMB lensing maps, we adopted a single source plane at $z=1100$ (see Section \ref{sec:lensing_maps}) when computing the lensing convergence maps.  Note that, although \citet{Hill2014a} smoothed their maps with a 10 arcmin Gaussian prior to analysis, they deconvolved the beam when computing the cross-correlation function.  We therefore use our raw (unsmoothed) simulated maps to compute the predicted cross-correlation.  We also note that \citet{Hill2014a} actually cross-correlated a map of the lensing potential, $\phi$, rather than the convergence, $\kappa$, with the tSZ effect $y$.  In multipole space, the lensing potential and convergence are related via $\phi = 2 \kappa / [(\ell+1) \ell]$, so we multiply our $\kappa$-$y$ cross-spectrum by this factor to convert to a $\phi$-$y$ cross-spectrum.

The tSZ-CMB lensing cross-correlation data tend to prefer a low value of $M_\nu$ that is consistent with the minimum neutrino mass (see Table \ref{tab:mnu_cross_tests}).  However, the measurements are still relatively noisy and can accommodate neutrino masses of up to $0.18 (0.27)$ eV (at 1-sigma) when adopting WMAP9 (Planck2015) `priors'.  The goodness of fit to the data in both the WMAP9- and Planck2015/$A_{\rm Lens}$-based cases is excellent.

As discussed in \citet{Hill2014a}, the CIB is a major source of contamination for the tSZ effect-CMB lensing cross-correlation, as the CIB itself is strongly correlated with CMB lensing \citep{Holder2013,vanEngelen2015}.  While \citet{Hill2014a} have taken steps to clean their maps of CIB contamination, \citet{Hurier2015} argue that their adopted cleaning method will not completely remove it and he estimated that the amplitude of the tSZ-CMB lensing cross-correlation of \citet{Hill2014a} may be biased high by $\approx$20 per cent at $\ell \sim 1000$.  Applying a $-20$ percent shift to the observed cross-correlation, the best-fit value of $M_\nu$ increases to $0.16\pm0.13$ ($0.24\pm0.15$) eV when using the WMAP9-based (Planck2015/$A_{\rm Lens}$-based) simulations, bringing it into very good agreement with the tSZ-galaxy lensing cross-correlation constraints in Section \ref{sec:tsz_shear_cross}, but somewhat lower than preferred by the tSZ effect-only and cosmic shear-only constraints in Sections \ref{sec:results_tsz} and \ref{sec:results_shear}, respectively.

It is worth briefly commenting on why we have not applied a similar shift to the previous auto- and cross-correlations including the tSZ effect that we examined in Sections \ref{sec:tsz_power} and \ref{sec:tsz_shear_cross}.  The power spectra used in those analyses were derived from the \planck~team's tSZ maps which were constructed using a detailed component separation algorithm that (at least in principle) accounts for CIB contamination.  \citet{Hill2014a}, however, derived their own custom tSZ effect map from the \planck~temperature maps (prior to the \planck~2015 data release) and used their own custom CIB cleaning methodology, which \citet{Hurier2015} re-examined and estimated that a $-20$ percent correction was required.  The fact that the tSZ effect (auto-)power spectrum has changed significantly between different releases by members of the \planck~team (including the new study of \citealt{Bolliet2017}) suggests that the issue of CIB contamination has not been fully resolved for the tSZ effect auto-correlation, but the level of remaining bias is difficult to assess.   For the tSZ effect-galaxy weak lensing cross-spectrum presented by \citet{Hojjati2017}, our expectation is that CIB contamination should be minimal, since the tSZ-galaxy lensing signal is strongly weighted to low redshifts, particularly for RCSLenS data (see also \citealt{Battaglia2015}), whereas the CIB signal is dominated by objects at higher redshifts of $1 \la z \la 5$ (e.g., \citealt{Hurier2015}).

\begin{figure}
\includegraphics[width=0.995\columnwidth]{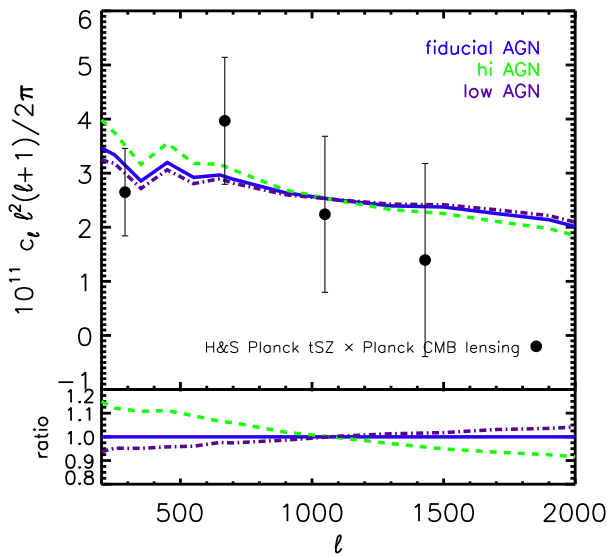}
\caption{\label{fig:cl_y_phi_agn}
The sensitivity of the predicted tSZ effect-CMB lensing cross-spectrum to uncertainties in the feedback modelling, in the context of the WMAP9-based cosmology with massless neutrinos.  The predictions vary by up to 10\%, depending on angular scale.  Increased feedback boosts the signal on large scales, likely as a result of gas ejection \citep{McCarthy2011}.  This effect is also present in the tSZ effect-galaxy lensing cross-spectrum and the tSZ effect power spectrum, though the effect there is smaller in magnitude and confined to the larger scales.  The uncertainties in the feedback modelling are small compared to current measurement uncertainties but will become more relevant for future measurements from, e.g., Advanced ACTpol.
}
\end{figure}

Lastly, in Fig.~\ref{fig:cl_y_phi_agn} we explore the effects of varying the AGN feedback level on the predicted tSZ-CMB lensing cross-correlation function.  While there is a noticeable effect, the uncertainty in the predicted cross-correlation due to uncertainties in the feedback modelling are clearly small compared to current measurement uncertainties.  This situation will likely change in the near future, as much more precise and higher-resolution measurements of both the tSZ effect and CMB lensing will become available from experiments such as Advanced ACTpol.  It is interesting that feedback tends to {\it amplify} the signal on large scales (low multipoles).  We speculate that this is because one is typically probing the outskirts of groups and clusters at large angular scales (see, e.g., the deconstruction of the tSZ effect angular power spectrum by radial ranges in \citealt{Battaglia2012} and \citealt{McCarthy2014}) and AGN feedback tends to boost the pressure beyond the virial radius \citep{LeBrun2015}, which is a consequence of (high-redshift) gas ejection \citep{McCarthy2011}.  We note that this effect is also present in the tSZ-galaxy lensing cross-correlation (see Fig.~\ref{fig:cl_y_kappa_agn}) but is smaller in magnitude.  A detailed comparison of the deconstruction of these two cross-correlation functions into their halo mass, redshift, and radial contributions would be interesting, but we leave this for future work.

\begin{figure*}
\includegraphics[width=0.995\columnwidth]{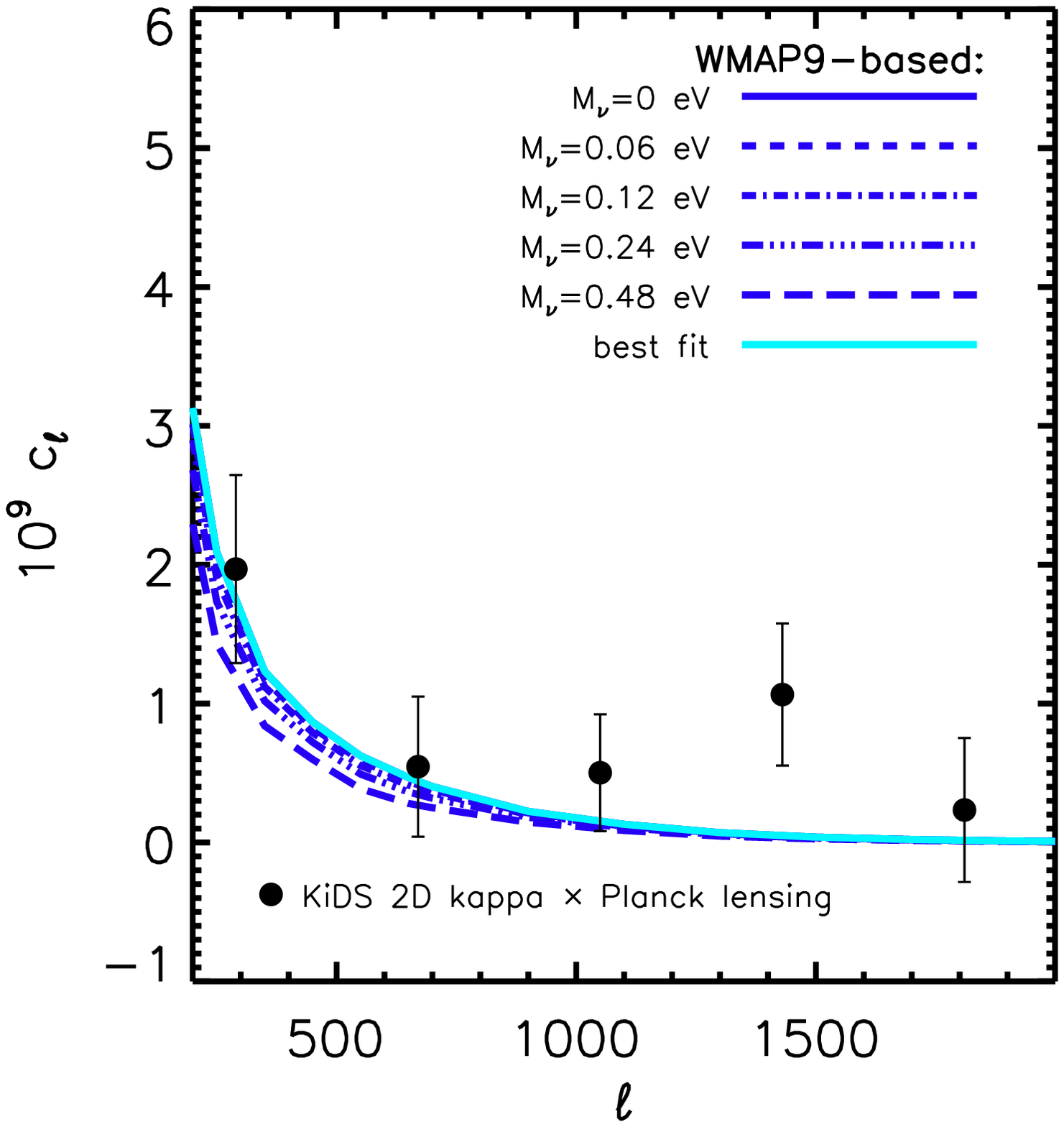}
\includegraphics[width=0.995\columnwidth]{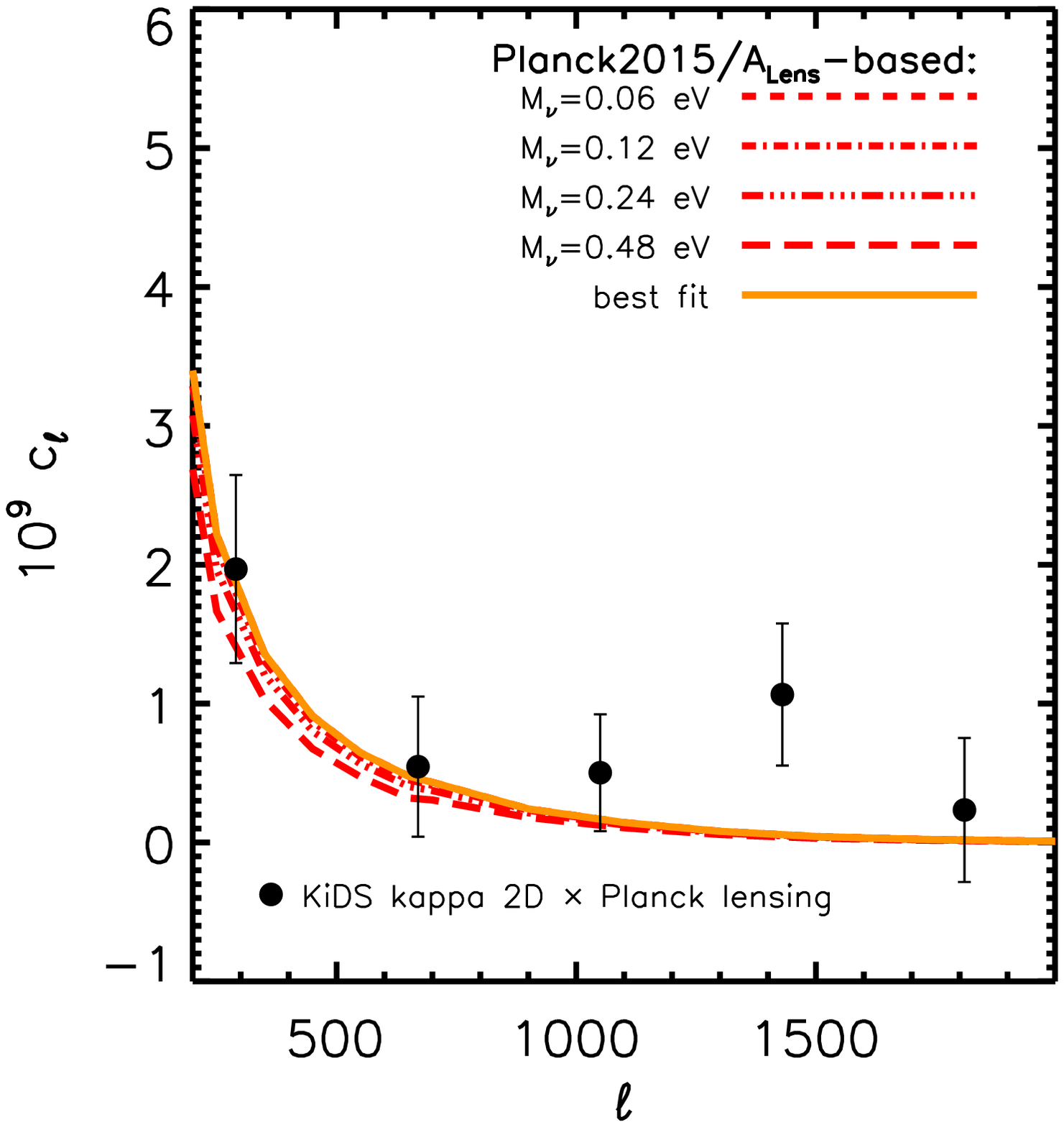}
\caption{\label{fig:cl_kappa_cmb_gal_notom}
Comparison of the predicted galaxy lensing-CMB lensing cross-spectrum (curves) to the measurements of \citet{HarnoisDeraps2017} (data points with 1-sigma error bars).  {\it Left:} Comparison using the WMAP9-based simulations.  {\it Right:} Comparison using the Planck2015/$A_{\rm Lens}$-based simulations.  The theoretical predictions agree well with the measurements, but current measurement errors are too large to distinguish most of the interesting neutrino mass range: the WMAP9-based comparison is compatible with $M_\nu \la 0.3$ eV, while the Planck2015/$A_{\rm Lens}$-based comparison is compatible with $M_\nu \la 0.5$ eV (see Table \ref{tab:mnu_cross_tests}).
}
\end{figure*}

\subsubsection{Galaxy lensing-CMB lensing}
\label{sec:shear_lensing_cross}

As a final test, to close the cross-correlation loop, we examine the cross-correlation between galaxy lensing and CMB lensing.  Measurements of such lensing-lensing cross-correlations have only recently become possible, with the first detection reported by \citet{Hand2015} who cross-correlated the ACT CMB lensing map with the CS82 lensing survey.  More recently, \citet{Liu2015} cross-correlated the CFHTLenS convergence map with the \planck~2013 CMB lensing convergence map, \citet{HarnoisDeraps2016} cross-correlated the CFHTLenS and RCSLenS data with the \planck~2015 CMB lensing map, \citet{Kirk2016} cross-correlated the DES Science Verification data with the SPT CMB lensing map, \citet{Singh2017} cross-correlated SDSS lensing data with the \planck~2015 CMB lensing map,
and \citet{HarnoisDeraps2017} cross-correlated the KiDS-450 data with the \planck~2015 CMB lensing map.  The majority of these studies reported a 1-2 sigma difference in the amplitudes of the observed cross-correlation with respect to that expected for a \planck~2015 CMB cosmology, in the sense that the observed cross-correlations were somewhat lower in amplitude than expected (i.e., consistent with the other LSS constraints we discussed in Section \ref{sec:tension}).  The most sensitive measurements to date are those of \citet{HarnoisDeraps2017} and we compare to their measurements of the galaxy lensing convergence--CMB lensing convergence cross-spectra.

In Fig.~\ref{fig:cl_kappa_cmb_gal_notom} we compare the predicted galaxy lensing-CMB lensing cross-correlations for the WMAP9-based (left panel) and Planck2015/$A_{\rm Lens}$-based (right) simulations with the measurements of \citet{HarnoisDeraps2017}.  (We do not use the full covariance matrices of \citealt{HarnoisDeraps2017} for this analysis, only the diagonal elements, as the Fourier-based cross-correlations show little bin-to-bin covariance which can safely be ignored.)   For the simulated CMB lensing maps, we adopt a single source plane at $z=1100$ when computing the lensing convergence maps.  For the simulations, we use the KiDS source redshift distribution spanning the full redshift range $0.1 < z < 0.9$ (i.e., a single tomographic bin) to compute the predicted convergence maps.  For consistency with the observational analysis of \citet{HarnoisDeraps2017}, we have convolved the predicted CMB lensing maps with a 10 arcmin Gaussian but have not smoothed the simulated convergence maps.

\begin{figure}
\includegraphics[width=0.995\columnwidth]{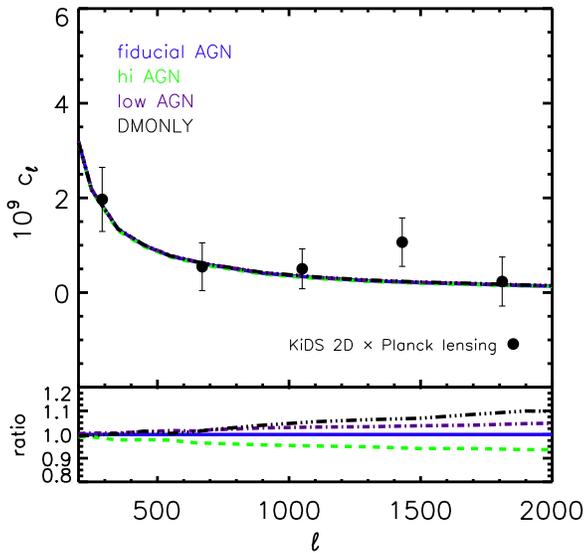}
\caption{\label{fig:cl_kappa_cmb_gal_notom_agn}
The sensitivity of the predicted galaxy lensing-CMB lensing cross-spectrum to uncertainties in the feedback modelling, in the context of the WMAP9-based cosmology with massless neutrinos.  The variations are confined to a few percent on large scales of $\ell \la 500$ but reach $5$\% at $\ell\sim1000$.  These differences are small compared to current measurement uncertainties.
}
\end{figure}

The predicted cross-correlations agree well with the observed ones, with a best-fit reduced-$\chi^2 \approx 0.7$ for both the WMAP9-based and Planck2015/$A_{\rm Lens}$-based simulations.  However, we find that the current measurement errors are too large to distinguish most of the interesting neutrino mass range: the WMAP9-based comparison is compatible with $M_\nu \la 0.3$ eV, while the Planck2015/$A_{\rm Lens}$-based comparison is compatible with $M_\nu \la 0.5$ eV (see Table \ref{tab:mnu_cross_tests}).  

These constraints were derived using a single tomographic bin from KiDS.  Following \citet{HarnoisDeraps2017}, we have also examined what constraints can be obtained by splitting the cosmic shear data into different tomographic bins.  Using the source redshift distributions of the four bins used in the observational analysis, we have computed the corresponding cosmic shear maps and cross-correlated each with the CMB lensing and then performed a joint fit to the four bins (not shown, for brevity).  However, we find that this tomographic analysis does not improve the constraints on the summed neutrino mass compared to the `2D' analysis above.

\begin{figure*}
\includegraphics[width=0.995\columnwidth]{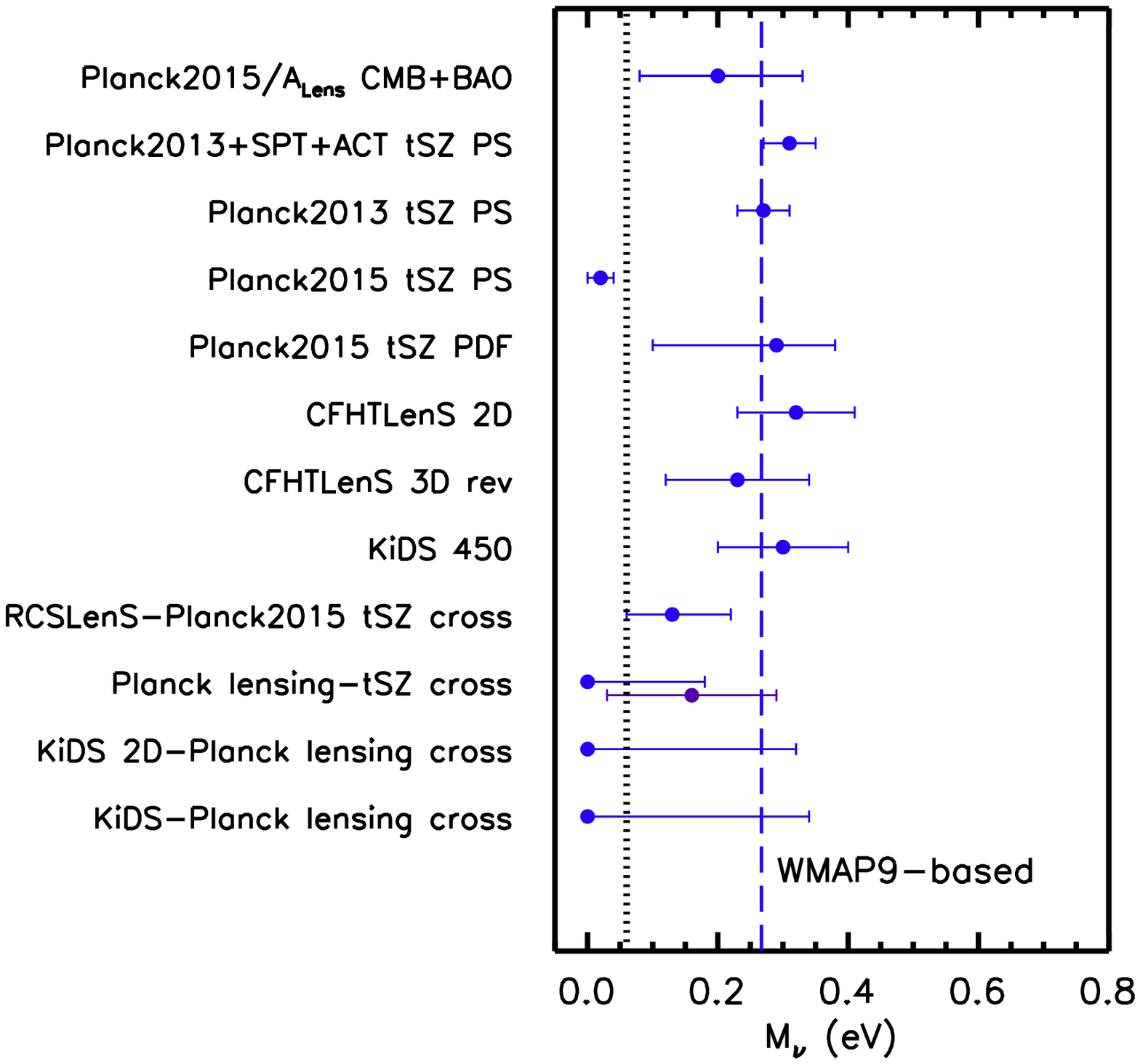}
\includegraphics[width=0.995\columnwidth]{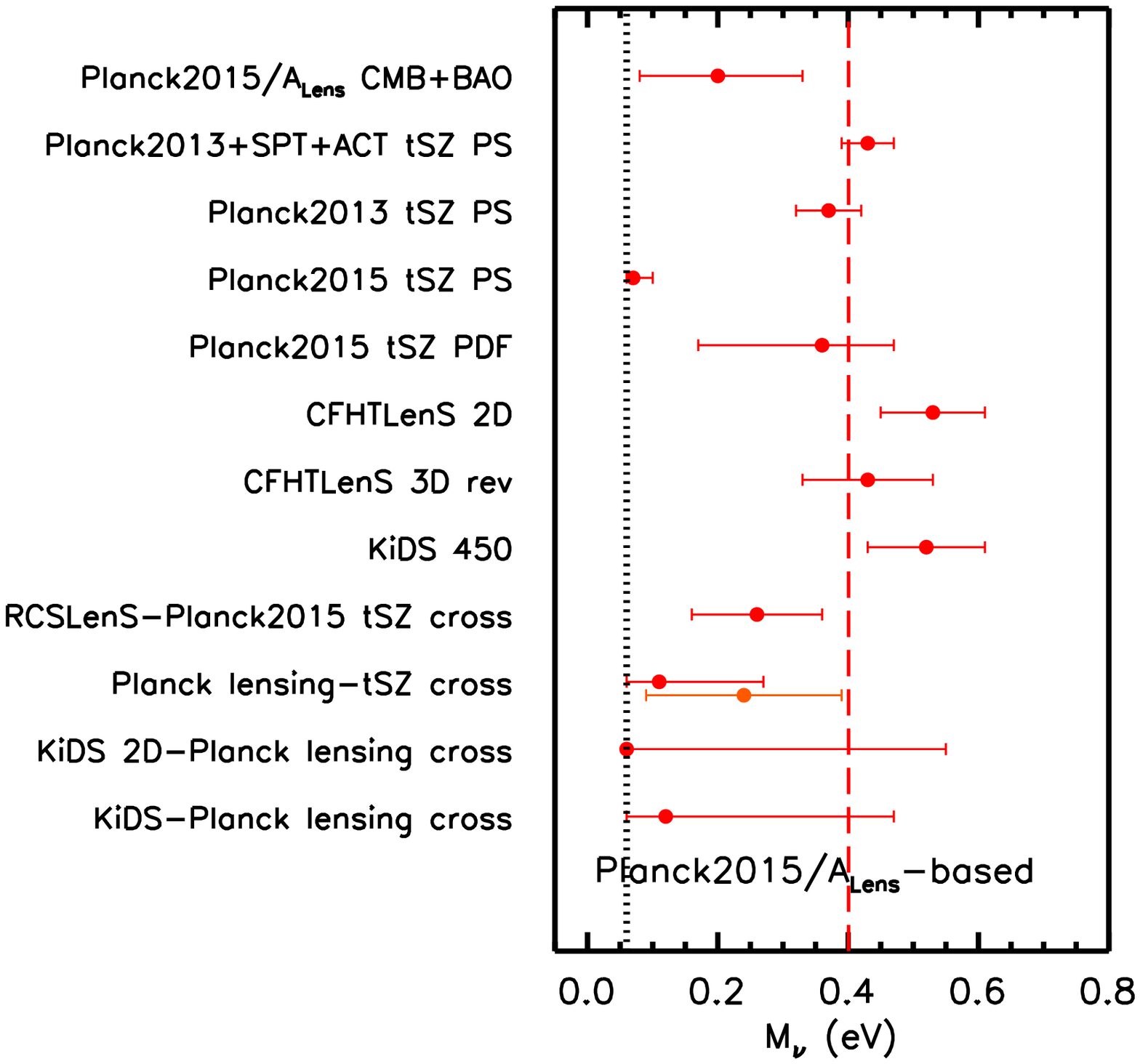}
\caption{\label{fig:mnu_constraints}
A comparison of our 1-sigma constraints on the summed mass of neutrinos, $M_\nu$, via the comparisons to tSZ effect, cosmic shear, and CMB lensing data.  {\it Left:} Constraints obtained when using the WMAP9-based simulations. {\it Right:} Constraints obtained when using the Planck2015/$A_{\rm Lens}$-based simulations.  The vertical dotted line corresponds to the minimum value of $M_\nu$ from neutrino oscillation experiments, assuming a normal hierarchy.  The vertical dashed lines correspond to the best-fit summed neutrino mass when fitting a constant to the individual constraints (excluding the discrepant \planck~2015 tSZ PS constraints).  
For the `Planck lensing-tSZ cross' test, two sets of constraints are shown, corresponding to comparisons with the measurements of \citet{Hill2014a} with and without taking into account possible residual CIB contamination, as suggested by \citet{Hurier2015} (see Section \ref{sec:shear_lensing_cross}).  The `Planck2015 CMB+BAO' constraint in both panels corresponds to the 1-sigma constraint on $M_\nu$ that we derive from the \planck~chains with marginalization over $A_{\rm Lens}$ (see Section \ref{sec:alens}).  If the CMB constraints on the other parameters originate from WMAP-9yr data, the majority of the LSS tests prefer $M_\nu \la 0.3$ eV.  When adopting the \planck~2015 constraints (with marginalization over $A_{\rm Lens}$) on the other parameters, the LSS tests are compatible with higher values, although there is considerable scatter in the best-fit value of $M_\nu$ between the different tests.  We have shown that this scatter is likely {\it not} due to theoretical uncertainties (e.g., baryon effects).  Overall, our results indicate that a non-minimal neutrino mass (i.e., $M_\nu > 0.06$ eV) is preferred, particularly if one combines the recent \planck~CMB constraints with LSS.  
}
\end{figure*}

Finally, in Fig.~\ref{fig:cl_kappa_cmb_gal_notom_agn} we explore the sensitivity of the theoretical predictions to uncertainties in the astrophysical modelling.  At $\ell\sim1000$, the uncertainty in the cross-correlation is $\approx 5$ percent (comparing the three AGN models).  As expected, the uncertainties become somewhat larger at smaller angular scales (high multipoles), but are still only at the level of $\sim5-10\%$, which is smaller than current measurement errors for the CMB lensing--galaxy lensing cross-correlation.  Note, however, that the differences between the fiducial \calsim~model and a dark matter-only simulation are quite a bit larger than this.  Future high-sensitivity and high-resolution measurements of CMB lensing, combined with future cosmic shear measurements (e.g., with \textsc{Euclid} and LSST) may be able to distinguish effects at these levels.  It is interesting to note that feedback affects the lensing-lensing cross-correlation differently than it does for CMB lensing-tSZ and cosmic shear-tSZ cross-correlations, in terms of the angular dependence (compare the bottom panels of Figs.~\ref{fig:cl_y_kappa_agn}, \ref{fig:cl_y_phi_agn}, and \ref{fig:cl_kappa_cmb_gal_notom_agn}).  A joint modelling of these cross-correlations therefore offers an interesting way to constrain the feedback modelling (as well as cosmology).

\subsection{Summary of constraints}
\label{sec:constraints}

We summarize our constraints on the summed mass of neutrinos, $M_\nu$, from the comparisons to tSZ effect, cosmic shear, and CMB lensing data in Fig.~\ref{fig:mnu_constraints}.  Note that for the \planck~lensing-tSZ cross-correlation, we include a second set of constraints accounting for the CIB contamination bias estimated by \citet{Hurier2015} (see Section \ref{sec:shear_lensing_cross}).  In addition, we exclude the constraints obtained from a joint fit to the \planck~2015, ACT and SPT tSZ effect power spectrum data, as these data sets are in strong tension with each other (i.e., the joint fit is poor).

The tests included in Fig.~\ref{fig:mnu_constraints} are as follows. `Planck2015 CMB+BAO' is the 1-sigma constraint on $M_\nu$ that we derive from the the \planck~2015 CMB chains with marginalization over $A_{\rm Lens}$ (see Section \ref{sec:alens}).  `Planck2013+SPT+ACT tSZ PS' refers to a joint fit to the \planck~2013 and SPT and ACT tSZ power spectra (see Section \ref{sec:tsz_power} and Table \ref{tab:mnu_tsz}).  `Planck2013 tSZ PS' and `Planck2015 tSZ PS' refer to fits to the \planck~2013 tSZ power spectrum only and to the \planck~2015 tSZ power spectrum only, respectively (see Section \ref{sec:tsz_power} and Table \ref{tab:mnu_tsz}).  `Planck2015 tSZ PDF' refers to the fit to \planck~2015 tSZ one-point probability distribution function (see Section \ref{sec:ypdf}).  `CFHTLenS 2D' and `CFHTLenS 3D rev' refer to the fits to the 2D cosmic shear data of \citet{Kilbinger2013} and to the 3D tomographic data of \citet{Joudaki2017a}, respectively (see Section \ref{sec:cfht} and Table \ref{tab:mnu_cosmic_shear}).  `KiDS-450' refers to the fit to the cosmic shear tomographic data of \citet{Hildebrandt2017} (see Section \ref{sec:kids} and Table \ref{tab:mnu_cosmic_shear}).  `RCSLenS-Planck2015 tSZ cross' refers to the fit to the RCSLenS galaxy lensing--\planck~2015 tSZ cross-spectrum measurement of \citet{Hojjati2017} (see Section \ref{sec:tsz_shear_cross} and Table \ref{tab:mnu_cross_tests}).  `Planck lensing-tSZ cross' refers to the fit to the \planck~2013 CMB lensing--\planck~2013 tSZ cross-spectrum measurement of \citet{Hill2014a} (see Section \ref{sec:tsz_lensing_cross} and Table \ref{tab:mnu_cross_tests}).  `KiDS 2D-Planck lensing cross' and `KiDS-Planck lensing cross' refer to the fits to the 2D and 3D (respectively) KiDS galaxy lensing--\planck~2015 CMB lensing cross-spectrum measurements of \citet{HarnoisDeraps2017} (see Section \ref{sec:shear_lensing_cross} and Table \ref{tab:mnu_cross_tests}).

When adopting CMB constraints on the other parameters from WMAP9 data (i.e., using the WMAP9-based simulations for the modelling), all of the LSS tests prefer $M_\nu \la 0.3$ eV.  The tSZ effect-only (with the exception of the \planck~2015 constraints) and cosmic shear-only tests show a 2-3 sigma preference for a non-minimal neutrino mass.  The various cross-correlation tests, particularly those involving CMB lensing, are compatible with these constraints but are also compatible with a minimal summed mass.

When adopting CMB constraints on the other parameters from \planck~2015 data (with marginalization over $A_{\rm Lens}$, see discussion in Section \ref{sec:alens}),  the LSS tests are compatible with masses of up to $M_\nu \la 0.5$ eV.  Again, the tSZ effect-only (with the exception of the \planck~2015 tSZ power spectrum constraints) and cosmic shear-only tests show a strong preference for a non-minimal neutrino mass.  The various cross-correlation tests, especially those involving CMB lensing, are not as constraining and are compatible with these tSZ-only and cosmic shear-only results but are also compatible with a minimal summed mass.

We highlight that, with the exception of the \planck~2015 tSZ power spectrum constraints, there is reasonable consistency between the different tests.  Fitting a constant value of $M_\nu$ to the different WMAP9-based (Planck2015/$A_{\rm Lens}$-based) constraints yields a best-fit summed neutrino mass of $0.27\pm0.05$ ($0.40\pm0.05$) eV with a reduced-$\chi^2$ of 0.70 (1.35), if we exclude the discrepant \planck~2015 tSZ power spectrum constraints\footnote{As highlighted previously (footnote 13), a re-analysis of the tSZ power spectrum derived from \planck~2015 data has recently appeared in \citet{Bolliet2017}.  The new measurements are very similar to the \planck~2013 tSZ power spectrum measurements at $\ell \ga 300$ and would therefore yield a constraint on $M_\nu$ that is consistent with the other LSS constraints presented in Fig.~\ref{fig:mnu_constraints}.}.  Formally speaking, our results therefore strongly support a non-minimal neutrino mass.  However, our quoted uncertainties are underestimates given that we have not marginalised over the other relevant cosmological parameters (see Section \ref{sec:discuss}) or observational nuisance parameters (e.g., intrinsic alignments and photometric redshift errors in the case of cosmic shear).  We have, however, considered the theoretical uncertainties in modelling the baryons and concluded that these are sub-dominant at present.  We also note that the uncertainties quoted above are strongly affected by the inclusion of the \planck~2013 tSZ power spectrum constraints, but, as discussed in Section \ref{sec:tsz_power}, there are reasons to believe that the uncertainties in the tSZ measurements are larger than quoted. 

\section{Discussion and Conclusions}
\label{sec:discuss}

We have used self-consistent cosmological hydrodynamical simulations from the \calsim~project to constrain cosmological parameters.  To our knowledge, this is the first time that such simulations, as opposed to dark matter-only simulations, have been used directly to constrain cosmological parameters from LSS data.  Our analysis has avoided the use of many simplifying assumptions that enter into the standard halo model-based approach (e.g., particular forms for the halo mass function and bias, parametric forms for the matter distribution within haloes, the Limber approximation, baryonic effects, etc.).  

An important aspect of our study is that the physical models for stellar and AGN feedback in the simulations have been carefully calibrated to match key observational diagnostics (the galaxy stellar mass function and galaxy group and cluster gas fractions) so that the distribution of baryons and the back reaction of baryons on the total matter distribution are realistic.  The calibration was then a-posteriori checked against multiple observations of the baryon-matter connection (see M17).  We have demonstrated that our calibration approach is insensitive to cosmology (see Section \ref{sec:degen}) i.e., astrophysics and variations in cosmology of interest here are not degenerate when calibrated in this way.  Through the construction of light cones and synthetic observational maps, we have been able to compare the {\it same} model to a range of different data sets (tSZ effect, cosmic shear, CMB lensing, and their various cross-correlations) that probe different aspects of the matter distribution on different scales and at different cosmic epochs.  Our work demonstrates that for current data the effect of baryonic
physics is significant, but that the residual {\it uncertainties} in the baryonic modelling (derived from measurement error in the calibration data) is not. This thus represents a strong proof of concept validating the use of hydrodynamical simulations for LSS cosmology.

Consistent with a number of previous LSS studies, our results generally indicate that there is tension between current LSS data and the primary CMB measured by \planck~{\it when one adopts the minimum possible neutrino mass}, as found by neutrino oscillation experiments.  We have demonstrated, using additional simulations that vary the feedback within the maximum acceptable range (compared to the observational diagnostics), that this conclusion does not change when one accounts for the residual uncertainties (after calibration) in the baryon physics modelling in the simulations.  In contrast with some recent studies, we have found that including a non-minimal summed neutrino mass component {\it can} potentially reconcile this tension and that the constraints from the various tests we have examined are largely consistent with each other (the one exception to this is the \planck~2015 tSZ power spectrum, which has now been revised in \citealt{Bolliet2017}).  See Section \ref{sec:constraints} for a summary of the individual and overall constraints on $M_\nu$.

Our conclusion that massive neutrinos can potentially reconcile the CMB-LSS tension depends strongly on which set of primary CMB constraints are adopted.  Specifically, as discussed in Section \ref{sec:alens}, if one adopts the fiducial analysis, where the amplitude scale factor of the CMB lensing power spectrum, $A_{\rm Lens}$, is fixed to unity when modelling the primary CMB TT power spectrum, \planck+BAO data constrains $M_\nu < 0.21$ eV (95\%).  This is too low to resolve the primary CMB-LSS tension.  However, under these conditions (i.e., with $A_{\rm Lens}$ fixed to unity), it has been demonstrated that the best-fit cosmological parameters derived from the \planck~data are sensitive to the range of multipoles over which one fits the CMB data (e.g., \citealt{Addison2016,Planck2016}).  If allowed to vary, the \planck~data itself favours a higher value of the lensing scale factor (with $A_{\rm Lens} = 1.2\pm0.1$) and this reduces the sensitivity of the derived cosmological constraints to the multipole fitting range.  Under these conditions, the \planck+BAO constraints are not only consistent with a relatively high value of $M_\nu$, they actually marginally prefer it: we derive a best-fit value of $M_\nu = 0.20^{+0.13}_{-0.12}$ eV (68\% C.L.) from the Markov chains.  This has been noted previously by (among others) \citealt{Planck2015_cmb} and \citet{DiValentino2017}.  The inclusion of LSS constraints further strengthens the case for a non-minimal summed neutrino mass, as we have shown here.

While we believe our study has made important progress in examining the current tension and the role that uncertainties in theoretical modelling play, there are also important limitations to consider.  In particular, we have varied only a single cosmological parameter ($M_\nu$) while adopting the best-fit values for the other relevant cosmological parameters from either the WMAP 9-yr or \planck~2015 primary CMB data.  The motivation for this strategy, which was adopted due to the expense of the simulations, is discussed in detail in Section \ref{sec:sims}. Ideally, one would vary all of the cosmological parameters relevant for LSS in the simulations and then compare the constraints with those of the primary CMB in an independent fashion before possibly combining the constraints.  To do this, many more simulations would be required, as would a fast and accurate mechanism to interpolate the predictions for choices of parameter values that were not directly simulated.  Such an approach is beginning to be employed in the context of dark matter-only simulations, such as the Coyote Universe project \citep{Heitmann2010,Heitmann2014}.  Extending this type of approach to full cosmological hydrodynamical simulations will be a challenge, but would have many benefits, including the ability to emulate directly observable quantities (such as the tSZ effect) rather than dark matter-only quantities that the user must convert into observables using simplifying assumptions that we would like to avoid.  

How might our conclusions be altered if we could already perform such an analysis?  While we have shown that the LSS observables (at least the ones we have considered) can generally be fit well by adopting a primary CMB-based cosmology with a freely-varying summed mass of neutrinos, the LSS data would almost certainly be just as well reproduced by adopting a minimal neutrino mass case but with lower values of $\sigma_8$ and/or $\Omega_{\rm m}$; i.e., LSS data alone does not provide a compelling case for massive neutrinos.  This, of course, would result in the well-known tension with the primary CMB.  We expect that including a varying neutrino mass and then {\it jointly} fitting the LSS and primary CMB data (as opposed to fixing all parameters at their primary CMB best-fit values and using the LSS data to determine $M_\nu$, as we have done here) would result in very similar results to the ones we have obtained here, since the primary CMB constraints on the other parameters are more precise than the constraints via current LSS tests.  However, the uncertainties on $M_\nu$ would likely increase somewhat relative to what we have quoted here.  

Going forward, future observatories (e.g., \textsc{Euclid}, LSST, e-ROSITA, Advanced ACTpol, etc.) will be able to place much tighter constraints on a variety of parameters from LSS data alone.  Emulation techniques applied to large-volume cosmological hydrodynamical simulations will surely play a major role in this endeavour.  Furthermore, we note that while uncertainties associated with baryonic physics are currently sub-dominant, they will become critical for future experiments, further increasing the importance of the use of hydrodynamical simulations.

Finally, in the current study we have focused on only a subset of possible LSS tests, involving the tSZ effect, cosmic shear, and CMB lensing.  In future work we plan to compare our simulations with observations of galaxy-galaxy lensing+galaxy clustering, redshift-space distortions, cluster number counts, and lensing peak counts.  In some of these cases, it is likely that larger volumes will need to be simulated, as current observations typically focus on very massive systems (in the case of number counts) or moderately high redshifts (in the case of galaxy-galaxy lensing+galaxy clustering and redshift-space distortions).

%We have made all of the simulated tSZ effect, cosmic shear, and CMB lensing maps (as well as the computed correlation functions, etc.) available online, at the \calsim~website: \url{http://www.astro.ljmu.ac.uk/~igm/BAHAMAS/}.  {\bf IGM: To be put online when paper is accepted.}

\section*{Acknowledgements}

The authors thank the anonymous referee for helpful suggestions that improved the paper.  IGM thanks Nick Battaglia, Sarah Bridle, Barbara Comis, George Efstahthiou, Martin Haehnelt, Colin Hill, Alireza Hojjati, Scott Kay, Amandine Le Brun, and Simon White for helpful discussions.  SB was supported by NASA through Einstein Postdoctoral Fellowship Award Number PF5-160133.  This work was supported by the Netherlands Organisation for Scientific Research (NWO), through VICI grant 639.043.409.  JHD is supported by the European Commission under a Marie-Sk{\l}odowska-Curie European Fellowship (EU project 656869).

This work used the DiRAC Data Centric system at Durham University, operated by the Institute for Computational Cosmology on behalf of the STFC DiRAC HPC Facility (www.dirac.ac.uk). This equipment was funded by BIS National E-infrastructure capital grant ST/K00042X/1, STFC capital grants ST/H008519/1 and ST/K00087X/1, STFC DiRAC Operations grant ST/K003267/1 and Durham University. DiRAC is part of the National E-Infrastructure.

\appendix

\section{Cosmic shear correlation functions}
\label{sec:append_shear}

Here we present the other comparisons to the cosmic shear correlation functions, referred to in Section \ref{sec:results_shear}.  Note that while these figures were not presented in the main text (for brevity), these comparisons are folded into our summed neutrino mass constraints (e.g., in Table \ref{tab:mnu_cosmic_shear}).  

\begin{figure*}
\includegraphics[width=0.75\textwidth]{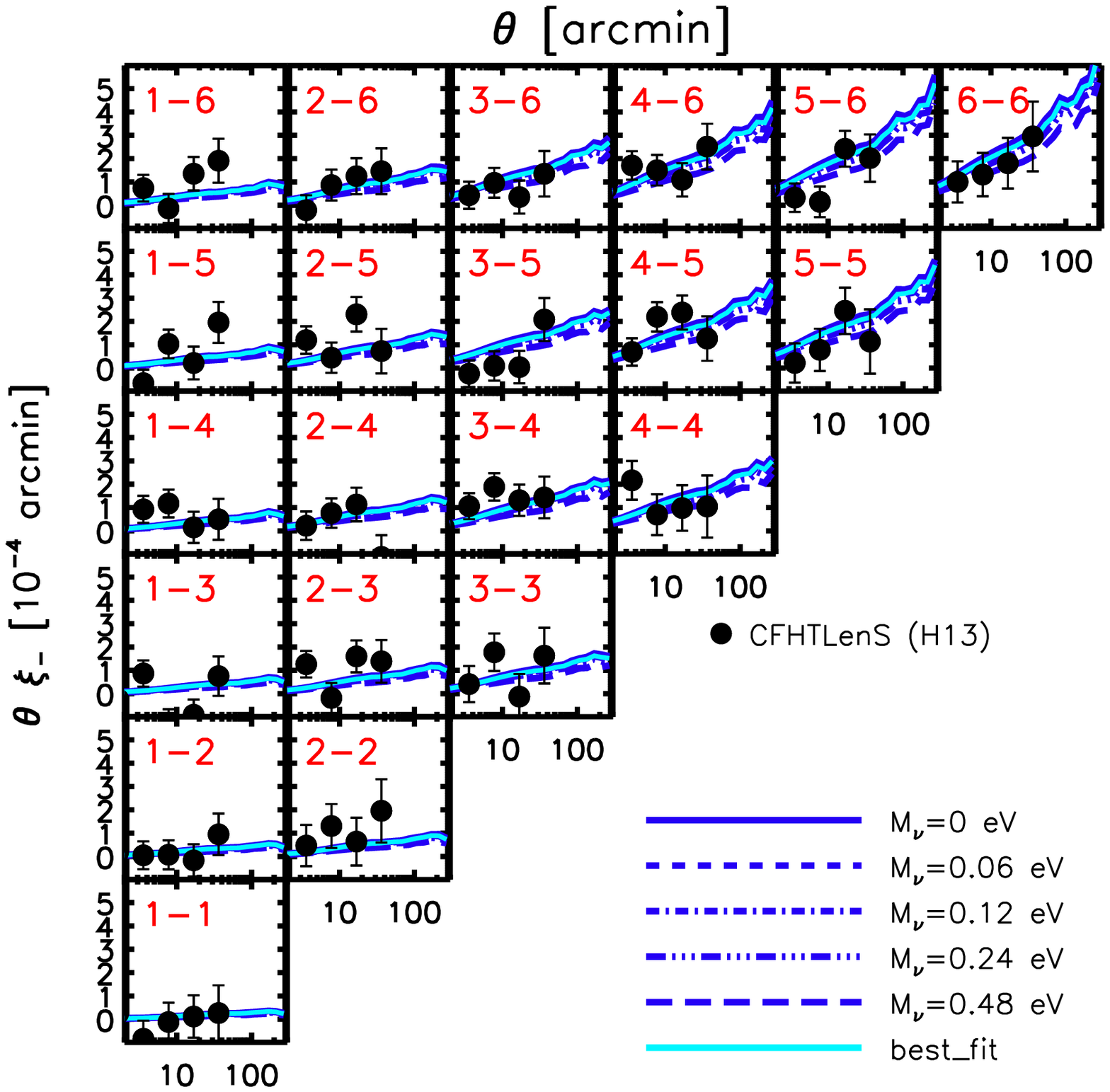}
\caption{\label{fig:xim_cfht}
Comparison of the WMAP9-based predictions to the $\xi_-$ tomographic CFHTLenS shear measurements of \citet{Heymans2013}.
}
\end{figure*}

\begin{figure*}
\includegraphics[width=0.99\columnwidth]{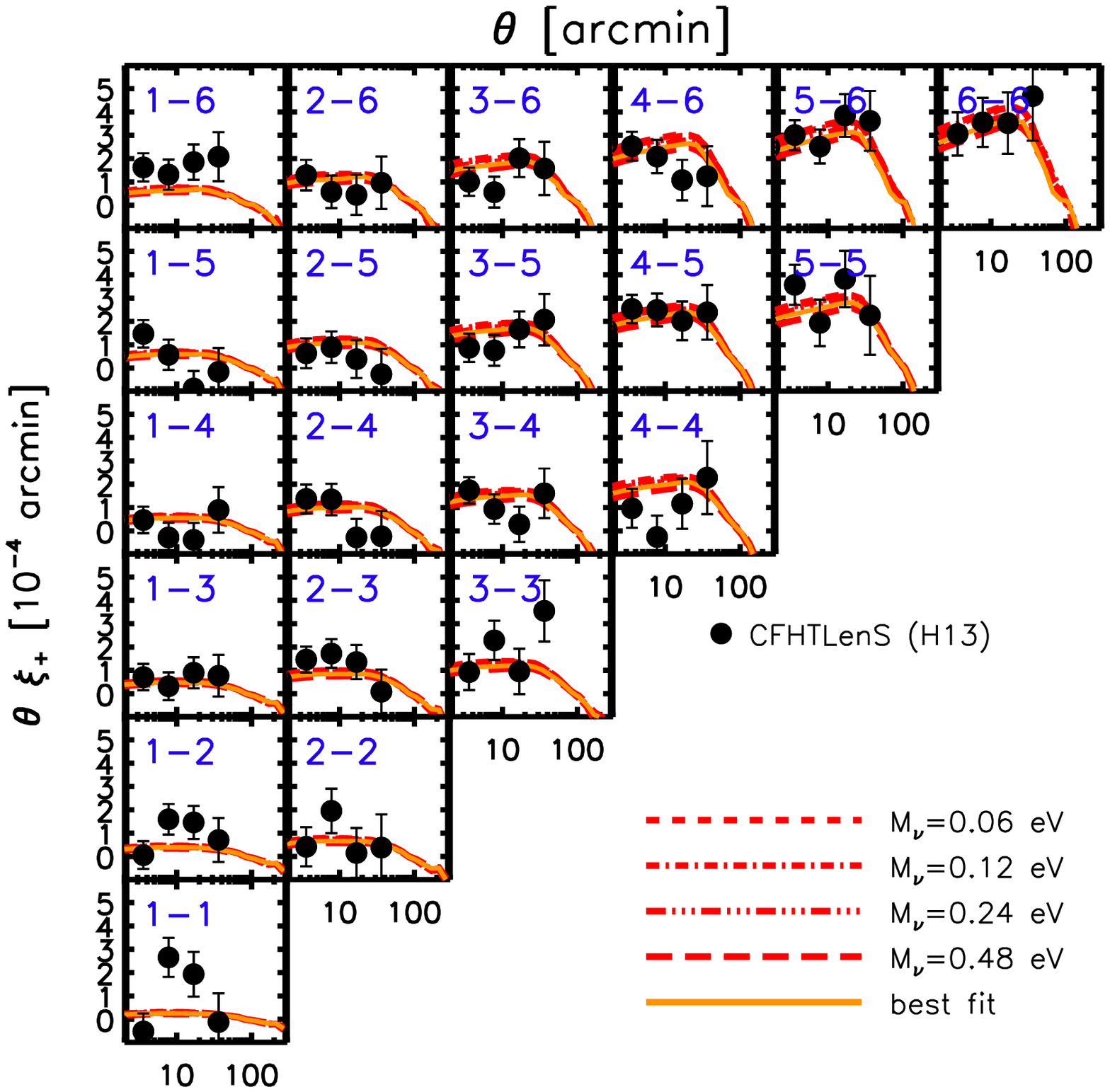}
\includegraphics[width=0.99\columnwidth]{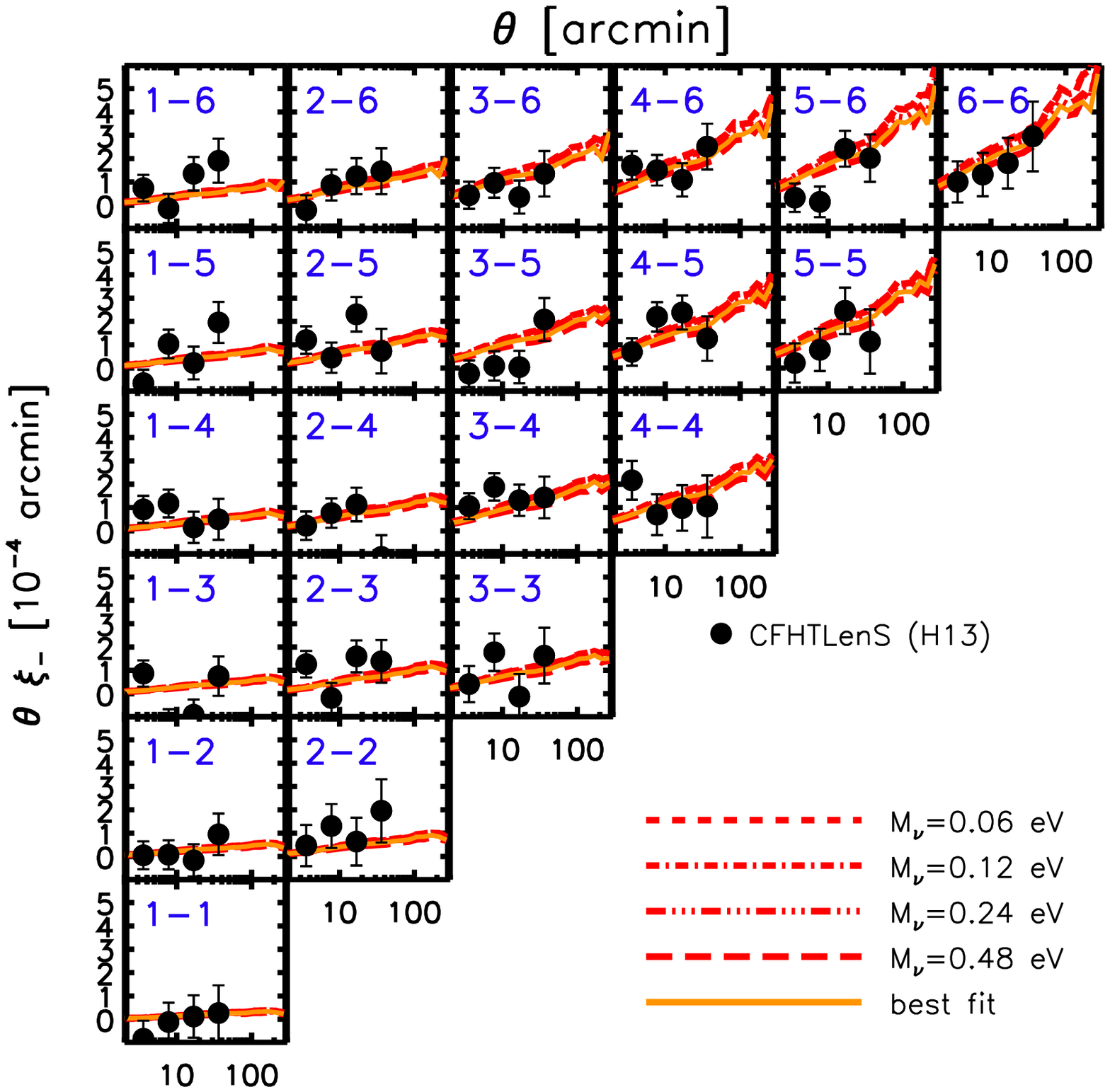}
\caption{\label{fig:xipm_cfht_pla15}
Comparison of the Planck2015/$A_{\rm Lens}$-based predictions to the $\xi_\pm$ tomographic CFHTLenS shear measurements of \citet{Heymans2013}.
}
\end{figure*}

\begin{figure*}
\includegraphics[width=0.95\columnwidth]{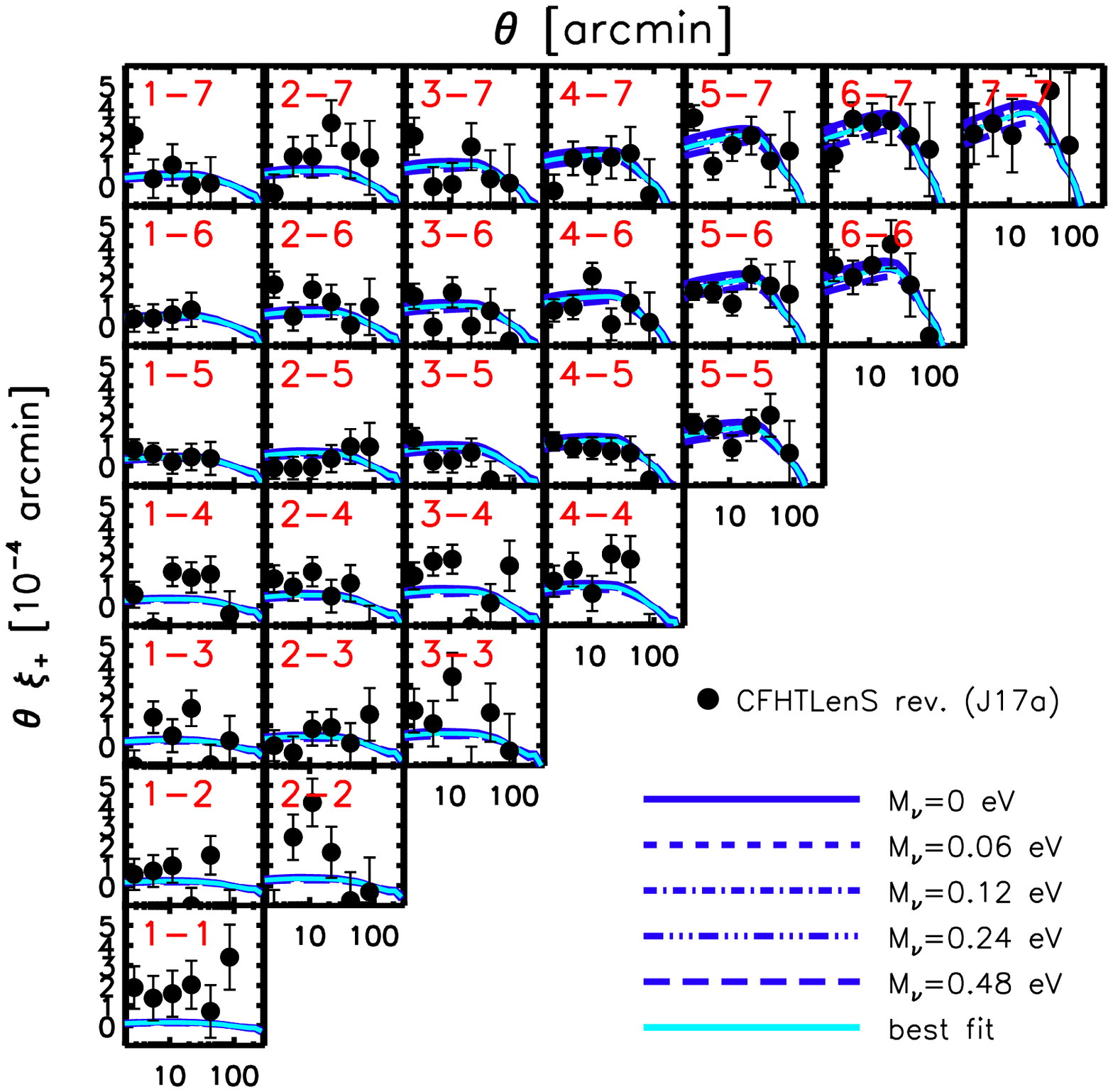}
\includegraphics[width=0.95\columnwidth]{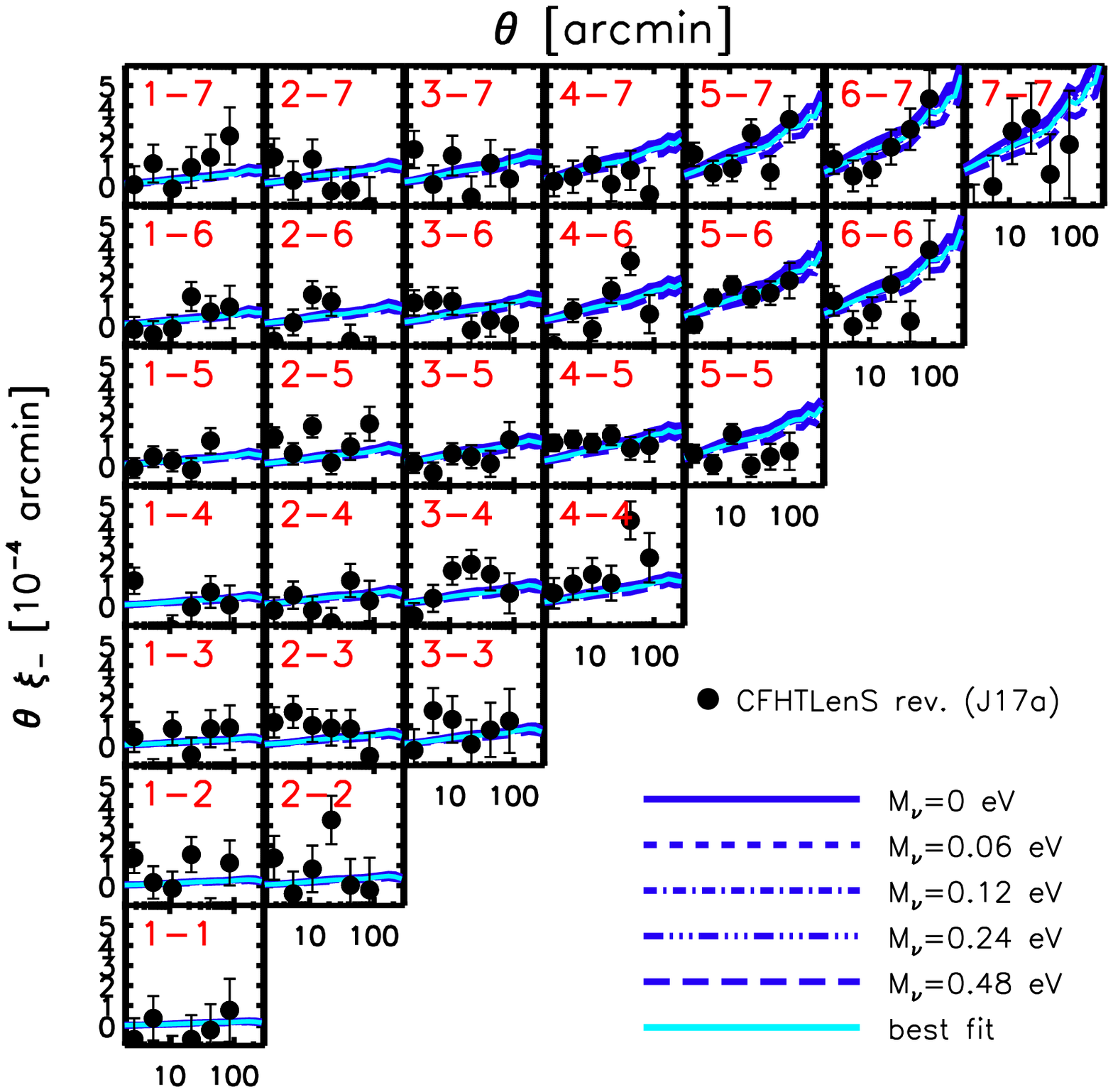}
\caption{\label{fig:xipm_cfht_rev}
Comparison of the WMAP9-based predictions to the $\xi_\pm$ tomographic CFHTLenS shear measurements of \citet{Joudaki2017a}.
}
\end{figure*}

\begin{figure*}
\includegraphics[width=0.95\columnwidth]{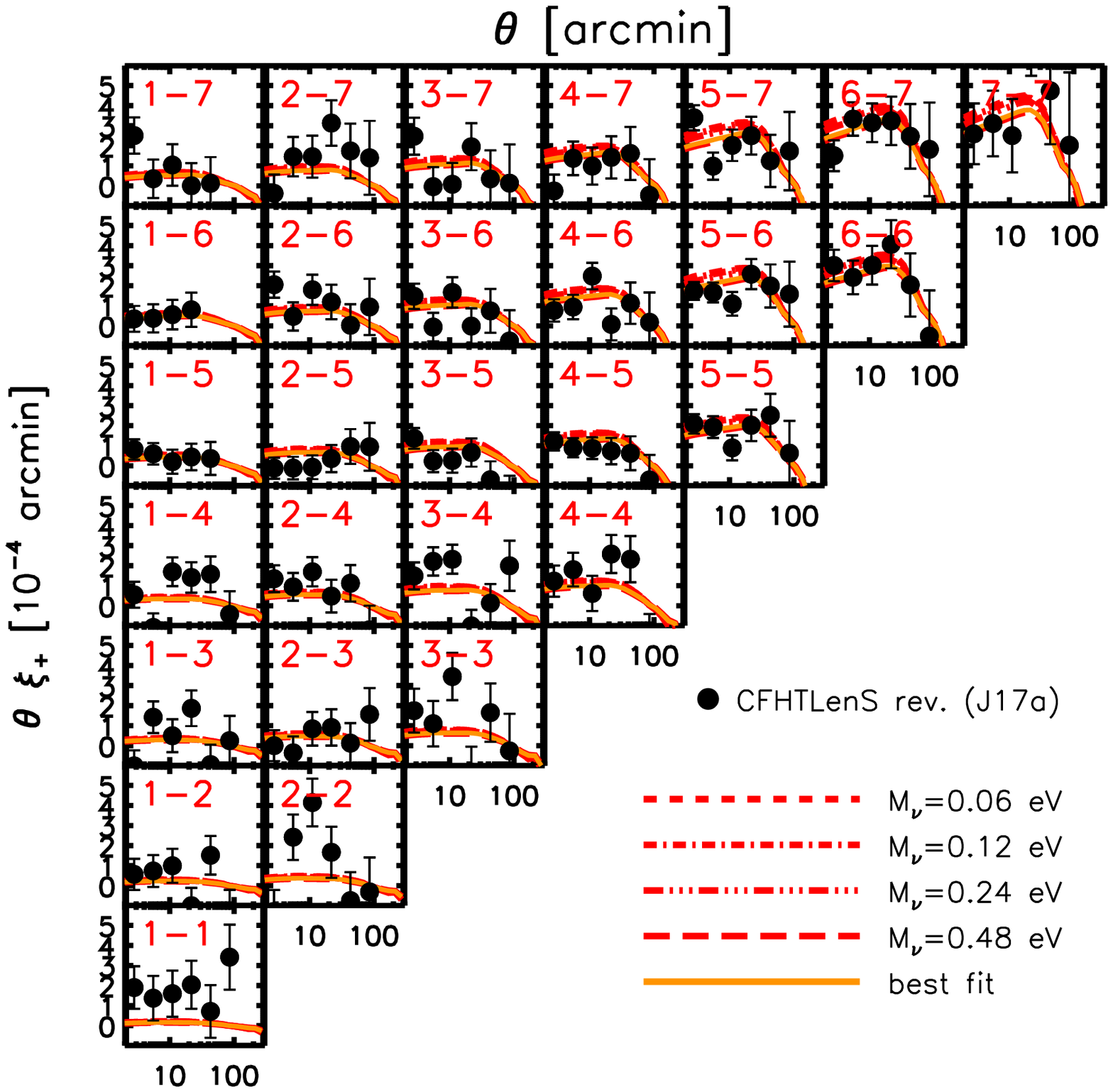}
\includegraphics[width=0.95\columnwidth]{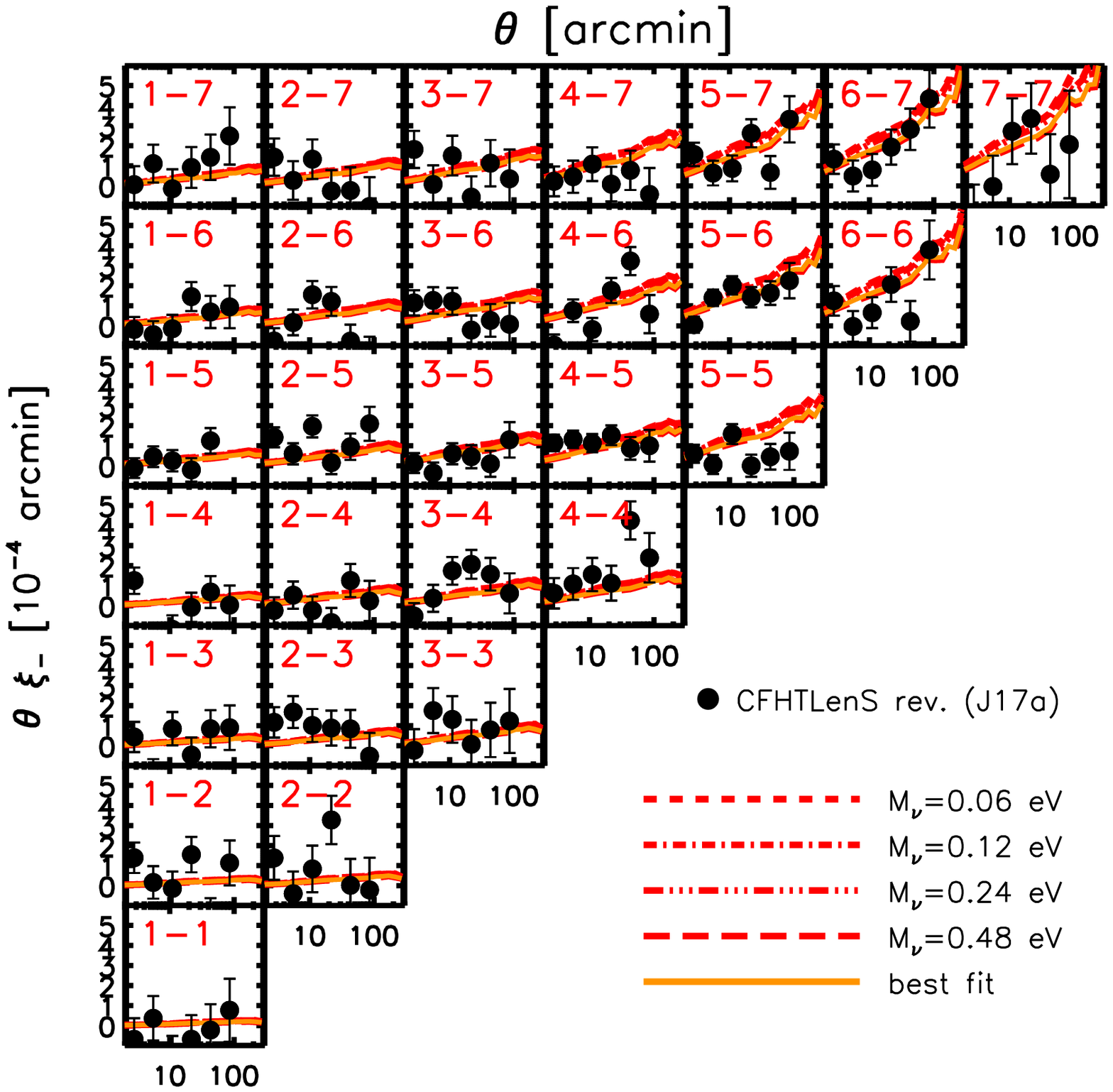}
\caption{\label{fig:xipm_cfht_rev}
Comparison of the Planck2015/$A_{\rm Lens}$-based predictions to the $\xi_\pm$ tomographic CFHTLenS shear measurements of \citet{Joudaki2017a}.
}
\end{figure*}

\begin{figure*}
\includegraphics[width=0.65\textwidth]{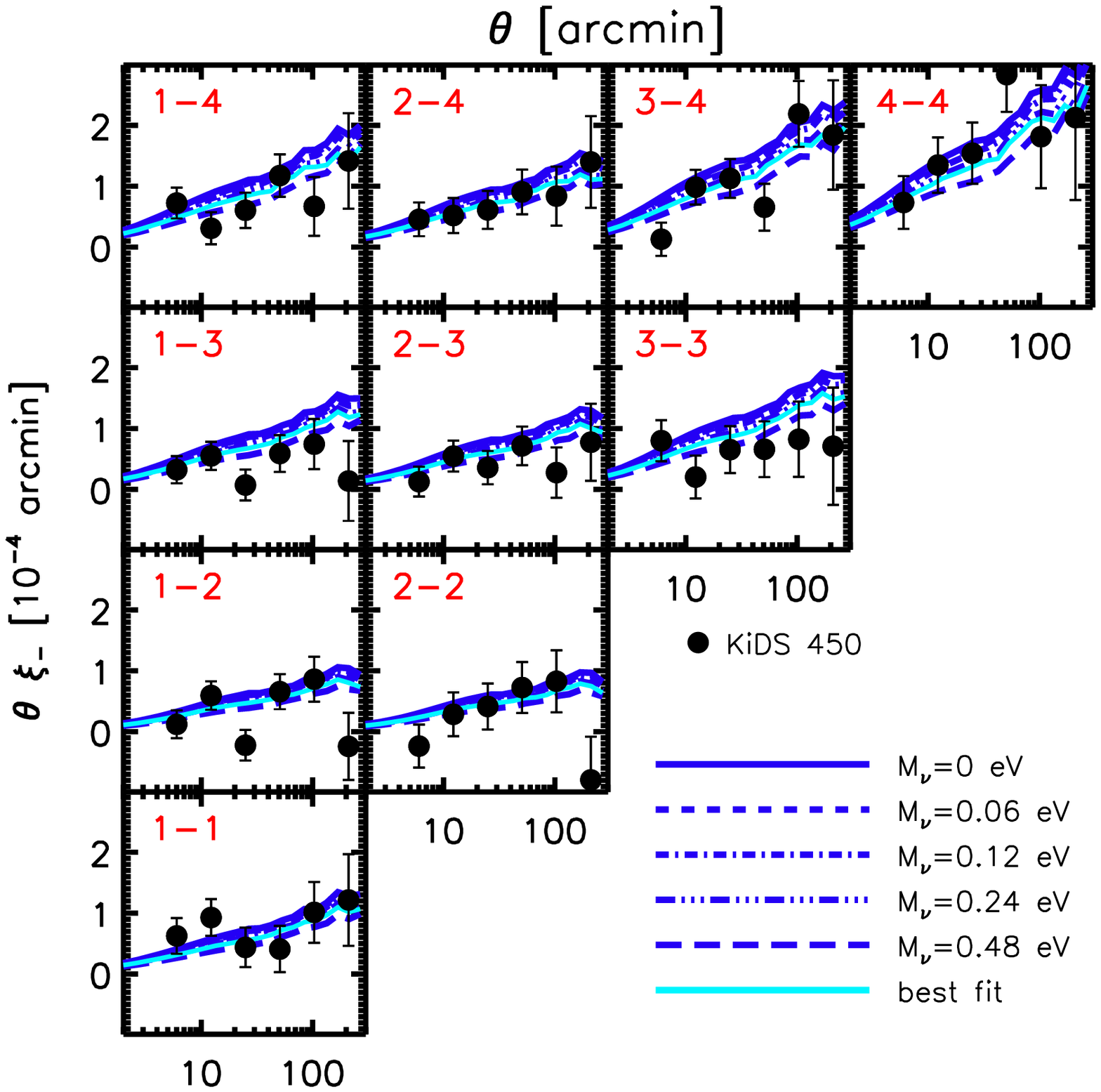}
\caption{\label{fig:xim_kids}
Comparison of the WMAP9-based predictions to the $\xi_-$ tomographic KiDS-450 shear measurements of \citet{Hildebrandt2017}.
}
\end{figure*}

\begin{figure*}
\includegraphics[width=0.99\columnwidth]{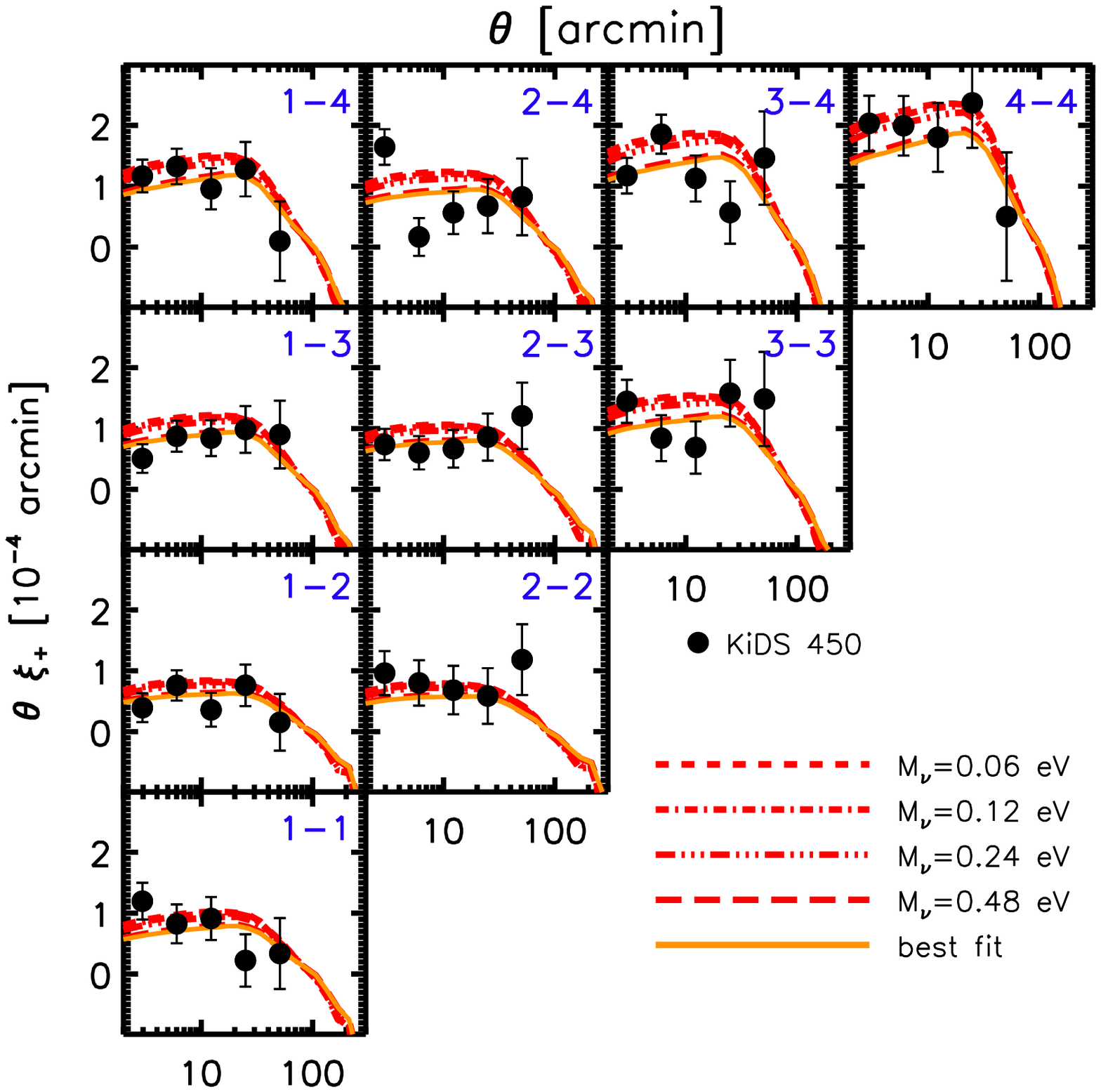}
\includegraphics[width=0.99\columnwidth]{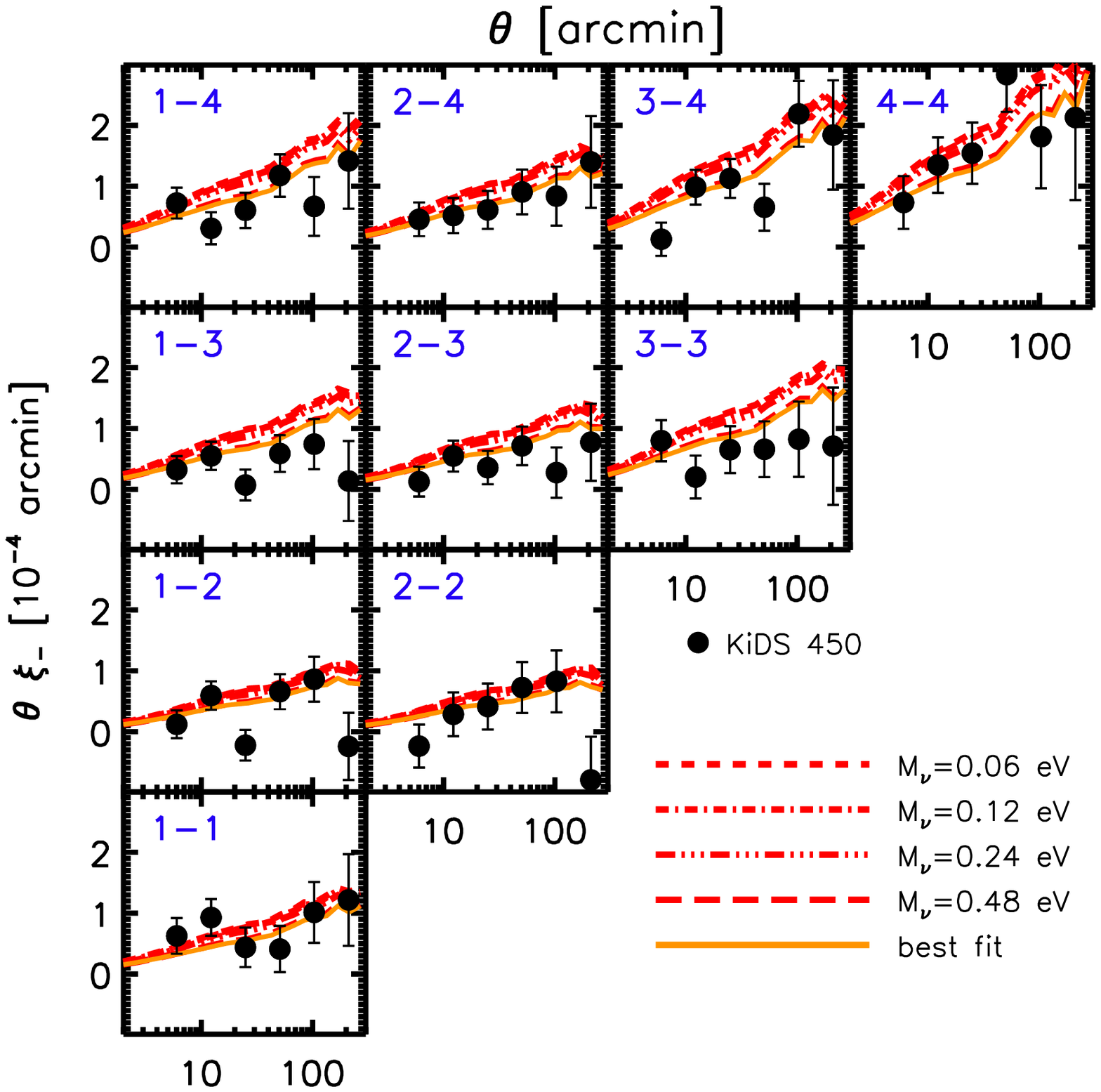}
\caption{\label{fig:xipm_kids_pla15}
Comparison of the Planck2015/$A_{\rm Lens}$-based predictions to the $\xi_\pm$ tomographic KiDS-450 shear measurements of \citet{Hildebrandt2017}.
}
\end{figure*}

\label{lastpage}


\begin{thebibliography}{99}

\bibitem[\protect\citeauthoryear{Addison et al.}{2016}]{Addison2016} Addison G.~E., Huang Y., Watts D.~J., Bennett C.~L., Halpern M., Hinshaw G., Weiland J.~L., 2016, ApJ, 818, 132 

\bibitem[\protect\citeauthoryear{Albrecht et al.}{2006}]{Albrecht2006} Albrecht A., et al., 2006 (arXiv:astro-ph/0609591)

\bibitem[\protect\citeauthoryear{Ali-Ha{\"i}moud \& Bird}{2013}]{Bird2013} Ali-Ha{\"i}moud Y., Bird S., 2013, MNRAS, 428, 3375 

\bibitem[\protect\citeauthoryear{Anderson et al.}{2015}]{Anderson2015} Anderson M.~E., Gaspari M., White S.~D.~M., Wang W., Dai X., 2015, MNRAS, 449, 3806 
 
\bibitem[\protect\citeauthoryear{Bah{\'e}, McCarthy, \& King}{2012}]{Bahe2012} Bah{\'e} Y.~M., McCarthy I.~G., King L.~J., 2012, MNRAS, 421, 1073 

\bibitem[\protect\citeauthoryear{Baldry et al.}{2012}]{Baldry2012} Baldry I.~K., et al., 2012, MNRAS, 421, 621 
  
\bibitem[\protect\citeauthoryear{Barnes et al.}{2017}]{Barnes2017} Barnes D.~J., Kay S.~T., Henson M.~A., McCarthy I.~G., Schaye J., Jenkins A., 2017, MNRAS, 465, 213

\bibitem[\protect\citeauthoryear{Battaglia et al.}{2010}]{Battaglia2010} Battaglia N., Bond J.~R., Pfrommer C., Sievers J.~L., Sijacki D., 2010, ApJ, 725, 91 

\bibitem[\protect\citeauthoryear{Battaglia et al.}{2012}]{Battaglia2012} Battaglia N., Bond J.~R., Pfrommer C., Sievers J.~L., 2012, ApJ, 758, 75 

\bibitem[\protect\citeauthoryear{Battaglia, Hill, \& Murray}{2015}]{Battaglia2015} Battaglia N., Hill J.~C., Murray N., 2015, ApJ, 812, 154 

\bibitem[\protect\citeauthoryear{Battye \& Moss}{2014}]{Battye2014} Battye R.~A., Moss A., 2014, PhRvL, 112, 051303 

\bibitem[\protect\citeauthoryear{Bernardi et al.}{2013}]{Bernardi2013} Bernardi M., Meert A., Sheth R.~K., Vikram V., Huertas-Company M., Mei S., Shankar F., 2013, MNRAS, 436, 697
  
\bibitem[\protect\citeauthoryear{Beutler et al.}{2014}]{Beutler2014} Beutler F., et al., 2014, MNRAS, 444, 3501 

\bibitem[\protect\citeauthoryear{Bird, Viel, \& Haehnelt}{2012}]{Bird2012} Bird S., Viel M., Haehnelt M.~G., 2012, MNRAS, 420, 2551 

\bibitem[\protect\citeauthoryear{Blumenthal et al.}{1984}]{Blumenthal1984} Blumenthal G.~R., Faber S.~M., Primack J.~R., Rees M.~J., 1984, Nature, 311, 517

\bibitem[\protect\citeauthoryear{Bolliet et al.}{2017}]{Bolliet2017} Bolliet B., Comis B., Komatsu E., Mac{\'{\i}}as-P{\'e}rez J.~F., 2017, MNRAS, submitted (arXiv:1712.00788)

\bibitem[\protect\citeauthoryear{Bond, Efstathiou, \& Silk}{1980}]{Bond1980} Bond J.~R., Efstathiou G., Silk J., 1980, PhRvL, 45, 1980

\bibitem[\protect\citeauthoryear{Booth \& Schaye}{2009}]{Booth2009} Booth C.~M., Schaye J., 2009, MNRAS, 398, 53 

\bibitem[\protect\citeauthoryear{Brandbyge \& Hannestad}{2009}]{Brandbyge2009} Brandbyge J., Hannestad S., 2009, JCAP, 5, 002 

\bibitem[\protect\citeauthoryear{Brandbyge et al.}{2008}]{Brandbyge2008} Brandbyge J., Hannestad S., Haugb{\o}lle T., Thomsen B., 2008, JCAP, 8, 020 

\bibitem[\protect\citeauthoryear{Bridle \& King}{2007}]{Bridle2007} Bridle S., King L., 2007, NJPh, 9, 444
  
\bibitem[\protect\citeauthoryear{Bullock et al.}{2001}]{Bullock2001} Bullock J.~S., Kolatt T.~S., Sigad Y., Somerville R.~S., Kravtsov A.~V., Klypin A.~A., Primack J.~R., Dekel A., 2001, MNRAS, 321, 559

\bibitem[\protect\citeauthoryear{Cacciato et al.}{2013}]{Cacciato2013} Cacciato M., van den Bosch F.~C., More S., Mo H., Yang X., 2013, MNRAS, 430, 767 

\bibitem[\protect\citeauthoryear{Calabrese et al.}{2008}]{Calabrese2008} Calabrese E., Slosar A., Melchiorri A., Smoot G.~F., Zahn O., 2008, PhRvD, 77, 123531 

\bibitem[\protect\citeauthoryear{Caldwell et al.}{2016}]{Caldwell2016} Caldwell C.~E., McCarthy I.~G., Baldry I.~K., Collins C.~A., Schaye J., Bird S., 2016, MNRAS, 462, 4117 

\bibitem[\protect\citeauthoryear{Clowe, De Lucia, \& King}{2004}]{Clowe2004} Clowe D., De Lucia G., King L., 2004, MNRAS, 350, 1038 

\bibitem[\protect\citeauthoryear{Correa et al.}{2015}]{Correa2015} Correa C.~A., Wyithe J.~S.~B., Schaye J., Duffy A.~R., 2015, MNRAS, 452, 1217

\bibitem[\protect\citeauthoryear{Crain et al.}{2015}]{Crain2015} Crain R.~A., et al., 2015, MNRAS, 450, 1937

\bibitem[\protect\citeauthoryear{da Silva et al.}{2000}]{daSilva2000} da Silva A.~C., Barbosa D., Liddle A.~R., Thomas P.~A., 2000, MNRAS, 317, 37

\bibitem[\protect\citeauthoryear{Dalla Vecchia \& Schaye}{2008}]{DallaVecchia2008} Dalla Vecchia C., Schaye J., 2008, MNRAS, 387, 1431 

\bibitem[\protect\citeauthoryear{Dark Energy Survey Collaboration et al.}{2016}]{DES2016} Dark Energy Survey Collaboration, et al., 2016, PhRvD, 94, 022001 

\bibitem[\protect\citeauthoryear{Davis et al.}{1985}]{Davis1985} Davis M., Efstathiou G., Frenk C.~S., White S.~D.~M., 1985, ApJ, 292, 371

\bibitem[\protect\citeauthoryear{de Haan et al.}{2016}]{deHaan2016} de Haan T., et al., 2016, ApJ, 832, 95 

\bibitem[\protect\citeauthoryear{Di Valentino et al.}{2017}]{DiValentino2017} Di Valentino E., Melchiorri A., Linder E.~V., Silk J., 2017, PhRvD, 96, 023523

\bibitem[\protect\citeauthoryear{Diemer et al.}{2017}]{Diemer2017} Diemer B., Mansfield P., Kravtsov A.~V., More S., 2017, ApJ, 843, 140 
  
\bibitem[\protect\citeauthoryear{Dolag, Komatsu, \& Sunyaev}{2016}]{Dolag2016} Dolag K., Komatsu E., Sunyaev R., 2016, MNRAS, 463, 1797

\bibitem[\protect\citeauthoryear{Eifler et al.}{2015}]{Eifler2015} Eifler T., Krause E., Dodelson S., Zentner A.~R., Hearin A.~P., Gnedin N.~Y., 2015, MNRAS, 454, 2451 

\bibitem[\protect\citeauthoryear{Eke, Navarro, \& Steinmetz}{2001}]{Eke2001} Eke V.~R., Navarro J.~F., Steinmetz M., 2001, ApJ, 554, 114 
  
\bibitem[\protect\citeauthoryear{Foreman, Becker, \& Wechsler}{2016}]{Foreman2016} Foreman S., Becker M.~R., Wechsler R.~H., 2016, MNRAS, 463, 3326

\bibitem[\protect\citeauthoryear{George et al.}{2015}]{George2015} George E.~M., et al., 2015, ApJ, 799, 177 

\bibitem[\protect\citeauthoryear{Giannantonio et al.}{2016}]{Giannantonio2016} Giannantonio T., et al., 2016, MNRAS, 456, 3213 

\bibitem[\protect\citeauthoryear{Hammami \& Mota}{2015}]{Hammami2015} Hammami A., Mota D.~F., 2015, A\&A, 584, A57 

\bibitem[\protect\citeauthoryear{Hand et al.}{2015}]{Hand2015} Hand N., et al., 2015, PhRvD, 91, 062001

\bibitem[\protect\citeauthoryear{Harnois-D{\'e}raps, Vafaei, \& Van Waerbeke}{2012}]{HarnoisDeraps2012} Harnois-D{\'e}raps J., Vafaei S., Van Waerbeke L., 2012, MNRAS, 426, 1262 

\bibitem[\protect\citeauthoryear{Harnois-D{\'e}raps \& van Waerbeke}{2015}]{HarnoisDeraps2015a} Harnois-D{\'e}raps J., van Waerbeke L., 2015, MNRAS, 450, 2857

\bibitem[\protect\citeauthoryear{Harnois-D{\'e}raps et al.}{2015}]{HarnoisDeraps2015b} Harnois-D{\'e}raps J., van Waerbeke L., Viola M., Heymans C., 2015, MNRAS, 450, 1212 

\bibitem[\protect\citeauthoryear{Harnois-D{\'e}raps et al.}{2016}]{HarnoisDeraps2016} Harnois-D{\'e}raps J., et al., 2016, MNRAS, 460, 434 

\bibitem[\protect\citeauthoryear{Harnois-D{\'e}raps et al.}{2017}]{HarnoisDeraps2017} Harnois-D{\'e}raps J., et al., 2017, MNRAS, 471, 1619

\bibitem[\protect\citeauthoryear{Heitmann et al.}{2010}]{Heitmann2010} Heitmann K., White M., Wagner C., Habib S., Higdon D., 2010, ApJ, 715, 104 
  
\bibitem[\protect\citeauthoryear{Heitmann et al.}{2014}]{Heitmann2014} Heitmann K., Lawrence E., Kwan J., Habib S., Higdon D., 2014, ApJ, 780, 111 

\bibitem[\protect\citeauthoryear{Heymans et al.}{2012}]{Heymans2012} Heymans C., et al., 2012, MNRAS, 427, 146 

\bibitem[\protect\citeauthoryear{Heymans et al.}{2013}]{Heymans2013} Heymans C., et al., 2013, MNRAS, 432, 2433 

\bibitem[\protect\citeauthoryear{Hildebrandt et al.}{2016}]{Hildebrandt2016} Hildebrandt H., et al., 2016, MNRAS, 463, 635

\bibitem[\protect\citeauthoryear{Hildebrandt et al.}{2017}]{Hildebrandt2017} Hildebrandt H., et al., 2017, MNRAS, 465, 1454 

\bibitem[\protect\citeauthoryear{Hill \& Spergel}{2014}]{Hill2014a} Hill J.~C., Spergel D.~N., 2014, JCAP, 2, 030 

\bibitem[\protect\citeauthoryear{Hill et al.}{2014}]{Hill2014b} Hill J.~C., et al., 2014, arXiv, arXiv:1411.8004 

\bibitem[\protect\citeauthoryear{Hinshaw et al.}{2009}]{Hinshaw2009} Hinshaw G., et al., 2009, ApJS, 180, 225

\bibitem[\protect\citeauthoryear{Hojjati et al.}{2017}]{Hojjati2017} Hojjati A., et al., 2017, MNRAS, 471, 1565 

\bibitem[\protect\citeauthoryear{Hojjati et al.}{2015}]{Hojjati2015} Hojjati A., McCarthy I.~G., Harnois-Deraps J., Ma Y.-Z., Van Waerbeke L., Hinshaw G., Le Brun A.~M.~C., 2015, JCAP, 10, 047 

\bibitem[\protect\citeauthoryear{Holder et al.}{2013}]{Holder2013} Holder G.~P., et al., 2013, ApJ, 771, L16 

\bibitem[\protect\citeauthoryear{Hurier}{2015}]{Hurier2015} Hurier G., 2015, A\&A, 575, L11

\bibitem[\protect\citeauthoryear{Jakobs et al.}{2018}]{Jakobs2018} Jakobs A., et al., 2018, MNRAS, submitted (arXiv:1712.05463) 
  
\bibitem[\protect\citeauthoryear{Joudaki et al.}{2017a}]{Joudaki2017a} Joudaki S., et al., 2017a, MNRAS, 465, 2033

\bibitem[\protect\citeauthoryear{Joudaki et al.}{2017b}]{Joudaki2017b} Joudaki S., et al., 2017b, MNRAS, 471, 1259
  
\bibitem[\protect\citeauthoryear{Kaiser}{1987}]{Kaiser1987} Kaiser N., 1987, MNRAS, 227, 1 

\bibitem[\protect\citeauthoryear{Kilbinger et al.}{2013}]{Kilbinger2013} Kilbinger M., et al., 2013, MNRAS, 430, 2200 

\bibitem[\protect\citeauthoryear{Kirk et al.}{2016}]{Kirk2016} Kirk D., et al., 2016, MNRAS, 459, 21 

\bibitem[\protect\citeauthoryear{Kilbinger, Bonnett, \& Coupon}{2014}]{Kilbinger2014} Kilbinger M., Bonnett C., Coupon J., 2014, Astrophysics Source Code Library, record ascl:1402.026

\bibitem[\protect\citeauthoryear{Komatsu \& Kitayama}{1999}]{Komatsu1999} Komatsu E., Kitayama T., 1999, ApJ, 526, L1 

\bibitem[\protect\citeauthoryear{Le Brun et al.}{2014}]{LeBrun2014} Le Brun A.~M.~C., McCarthy I.~G., Schaye J., Ponman T.~J., 2014, MNRAS, 441, 1270

\bibitem[\protect\citeauthoryear{Le Brun, McCarthy, \& Melin}{2015}]{LeBrun2015} Le Brun A.~M.~C., McCarthy I.~G., Melin J.-B., 2015, MNRAS, 451, 3868 

\bibitem[\protect\citeauthoryear{Leauthaud et al.}{2017}]{Leauthaud2017} Leauthaud A., et al., 2017, MNRAS, 467, 3024

\bibitem[\protect\citeauthoryear{Lesgourgues \& Pastor}{2006}]{Lesgourgues2006} Lesgourgues J., Pastor S., 2006, PhR, 429, 307 

\bibitem[\protect\citeauthoryear{Lewis et al.}{2000}]{Lewis2000} Lewis, A., Challinor, A., \& Lasenby, A.\ 2000, \apj, 538, 473 

\bibitem[\protect\citeauthoryear{Li \& White}{2009}]{Li2009} Li C., White S.~D.~M., 2009, MNRAS, 398, 2177 

\bibitem[\protect\citeauthoryear{Lim et al.}{2018}]{Lim2018} Lim S., Mo H., Li R., Liu Y., Ma Y.-Z., Wang H., Yang X., 2018, ApJ, submitted (arXiv:1710.06856)

\bibitem[\protect\citeauthoryear{Lin et al.}{2012}]{Lin2012} Lin Y.-T., Stanford S.~A., Eisenhardt P.~R.~M., Vikhlinin A., Maughan B.~J., Kravtsov A., 2012, ApJ, 745, L3 
  
\bibitem[\protect\citeauthoryear{Liu \& Hill}{2015}]{Liu2015} Liu J., Hill J.~C., 2015, PhRvD, 92, 063517 

\bibitem[\protect\citeauthoryear{Lovisari, Reiprich, \& Schellenberger}{2015}]{Lovisari2015} Lovisari L., Reiprich T.~H., Schellenberger G., 2015, A\&A, 573, A118 
  
\bibitem[\protect\citeauthoryear{Ma \& Bertschinger}{1995}]{Ma1995} Ma C.-P., Bertschinger E., 1995, ApJ, 455, 7 

\bibitem[\protect\citeauthoryear{MacCrann et al.}{2015}]{MacCrann2015} MacCrann N., Zuntz J., Bridle S., Jain B., Becker M.~R., 2015, MNRAS, 451, 2877 

\bibitem[\protect\citeauthoryear{Mandelbaum et al.}{2013}]{Mandelbaum2013} Mandelbaum R., Slosar A., Baldauf T., Seljak U., Hirata C.~M., Nakajima R., Reyes R., Smith R.~E., 2013, MNRAS, 432, 1544

\bibitem[\protect\citeauthoryear{Maughan et al.}{2008}]{Maughan2008} Maughan B.~J., Jones C., Forman W., Van Speybroeck L., 2008, ApJS, 174, 117-135 
  
\bibitem[\protect\citeauthoryear{McCarthy et al.}{2011}]{McCarthy2011} McCarthy I.~G., Schaye J., Bower R.~G., Ponman T.~J., Booth C.~M., Dalla Vecchia C., Springel V., 2011, MNRAS, 412, 1965 

\bibitem[\protect\citeauthoryear{McCarthy et al.}{2014}]{McCarthy2014} McCarthy I.~G., Le Brun A.~M.~C., Schaye J., Holder G.~P., 2014, MNRAS, 440, 3645 

\bibitem[\protect\citeauthoryear{McCarthy et al.}{2017}]{McCarthy2017} McCarthy I.~G., Schaye J., Bird S., Le Brun A.~M.~C., 2017, MNRAS, 465, 2936 

\bibitem[\protect\citeauthoryear{Mead et al.}{2016}]{Mead2016} Mead A.~J., Heymans C., Lombriser L., Peacock J.~A., Steele O.~I., Winther H.~A., 2016, MNRAS, 459, 1468
  
\bibitem[\protect\citeauthoryear{More et al.}{2015}]{More2015} More S., Miyatake H., Mandelbaum R., Takada M., Spergel D.~N., Brownstein J.~R., Schneider D.~P., 2015, ApJ, 806, 2 

\bibitem[\protect\citeauthoryear{Mummery et al.}{2017}]{Mummery2017} Mummery B.~O., McCarthy I.~G., Bird S., Schaye J., 2017, MNRAS, 471, 227

\bibitem[\protect\citeauthoryear{Omori et al.}{2017}]{Omori2017} Omori Y., et al., 2017, ApJ, 849, 124 

\bibitem[\protect\citeauthoryear{Palanque-Delabrouille et al.}{2015}]{Lyman2015} Palanque-Delabrouille N., et al., 2015, JCAP, 11, 011

\bibitem[\protect\citeauthoryear{Peacock \& Dodds}{1994}]{Peacock1994} Peacock J.~A., Dodds S.~J., 1994, MNRAS, 267, 1020

\bibitem[\protect\citeauthoryear{Pearson et al.}{2017}]{Pearson2017} Pearson R.~J., et al., 2017, MNRAS, 469, 3489 
  
\bibitem[\protect\citeauthoryear{Peebles}{1980}]{Peebles1980} Peebles P.~J.~E., 1980, The Large-Scale Structure of the Universe. Princeton Univ.~Press, Princeton, NJ

\bibitem[\protect\citeauthoryear{Pillepich et al.}{2018}]{Pillepich2018} Pillepich A., et al., 2018, MNRAS, 473, 4077

\bibitem[\protect\citeauthoryear{Planck Collaboration XI}{2013}]{Planck2013_stack} Planck Collaboration XI, et al., 2013b, A\&A, 557, A52 
  
\bibitem[\protect\citeauthoryear{Planck Collaboration XVI}{2014}]{Planck2013_cmb} Planck Collaboration XVI, et al., 2014, A\&A, 571, A16 

\bibitem[\protect\citeauthoryear{Planck Collaboration XXI}{2014}]{Planck2013_sz} Planck Collaboration XXI, et al., 2014, A\&A, 571, A21

\bibitem[\protect\citeauthoryear{Planck Collaboration XIII}{2016}]{Planck2015_cmb} Planck Collaboration XIII, et al., 2016, A\&A, 594, A13 

\bibitem[\protect\citeauthoryear{Planck Collaboration XXIV}{2016}]{Planck2015_clusters} Planck Collaboration XXIV, et al., 2016, A\&A, 594, A24 

\bibitem[\protect\citeauthoryear{Planck Collaboration XXII}{2016}]{Planck2015_sz} Planck Collaboration XXII, et al., 2016, A\&A, 594, A22 

\bibitem[\protect\citeauthoryear{Planck Collaboration XV}{2016}]{Planck2015_lensing} Planck Collaboration XV, et al., 2016, A\&A, 594, A15 

\bibitem[\protect\citeauthoryear{Planck Collaboration LI}{2017}]{Planck2016} Planck Collaboration, et al., 2017, A\&A, 607, A95

\bibitem[\protect\citeauthoryear{Planck Collaboration XVI}{2014}]{Planck2013} Planck Collaboration XVI, et al., 2014, A\&A, 571, A16 

\bibitem[\protect\citeauthoryear{Pratt et al.}{2009}]{Pratt2009} Pratt G.~W., Croston J.~H., Arnaud M., B{\"o}hringer H., 2009, A\&A, 498, 361 

\bibitem[\protect\citeauthoryear{Rasmussen \& Ponman}{2009}]{Rasmussen2009} Rasmussen J., Ponman T.~J., 2009, MNRAS, 399, 239 

\bibitem[\protect\citeauthoryear{Riess et al.}{2016}]{Riess2016} Riess A.~G., et al., 2016, ApJ, 826, 56 

\bibitem[\protect\citeauthoryear{Rogers et al.}{2017}]{Rogers2017} Rogers K.~K., Bird S., Peiris H.~V., Pontzen A., Font-Ribera A., Leistedt B., 2017, MNRAS, submitted (arXiv:1706.08532)

\bibitem[\protect\citeauthoryear{Roncarelli et al.}{2006}]{Roncarelli2006} Roncarelli M., Moscardini L., Tozzi P., Borgani S., Cheng L.~M., Diaferio A., Dolag K., Murante G., 2006, MNRAS, 368, 74 

\bibitem[\protect\citeauthoryear{Roncarelli et al.}{2007}]{Roncarelli2007} Roncarelli M., Moscardini L., Borgani S., Dolag K., 2007, MNRAS, 378, 1259 

\bibitem[\protect\citeauthoryear{Schaye \& Dalla Vecchia}{2008}]{Schaye2008} Schaye J., Dalla Vecchia C., 2008, MNRAS, 383, 1210 

\bibitem[\protect\citeauthoryear{Schaye et al.}{2010}]{Schaye2010} Schaye, J., Dalla Vecchia, C., Booth, C.~M., et al.\ 2010, MNRAS, 402, 1536 

\bibitem[\protect\citeauthoryear{Schaye et al.}{2015}]{Schaye2015} Schaye J., et al., 2015, MNRAS, 446, 521

\bibitem[\protect\citeauthoryear{Schneider et al.}{1998}]{Schneider1998} Schneider P., van Waerbeke L., Jain B., Kruse G., 1998, MNRAS, 296, 873 

\bibitem[\protect\citeauthoryear{Schneider \& Teyssier}{2015}]{Schneider2015} Schneider A., Teyssier R., 2015, JCAP, 12, 049 

\bibitem[\protect\citeauthoryear{Sembolini et al.}{2016}]{Sembolini2016} Sembolini F., et al., 2016, MNRAS, 459, 2973

\bibitem[\protect\citeauthoryear{Semboloni et al.}{2011}]{Semboloni2011} Semboloni E., Hoekstra H., Schaye J., van Daalen M.~P., McCarthy I.~G., 2011, MNRAS, 417, 2020 

\bibitem[\protect\citeauthoryear{Semboloni, Hoekstra, \& Schaye}{2013}]{Semboloni2013} Semboloni E., Hoekstra H., Schaye J., 2013, MNRAS, 434, 148 

\bibitem[\protect\citeauthoryear{Sherwin et al.}{2017}]{Sherwin2017} Sherwin B.~D., et al., 2017, PhRvD, 95, 123529

\bibitem[\protect\citeauthoryear{Sievers et al.}{2013}]{Sievers2013} Sievers J.~L., et al., 2013, JCAP, 10, 060 

\bibitem[\protect\citeauthoryear{Singh, Mandelbaum, \& Brownstein}{2017}]{Singh2017} Singh S., Mandelbaum R., Brownstein J.~R., 2017, MNRAS, 464, 2120 

\bibitem[\protect\citeauthoryear{Smith et al.}{2003}]{Smith2003} Smith R.~E., et al., 2003, MNRAS, 341, 1311 

\bibitem[\protect\citeauthoryear{Spergel, Flauger, \& Hlo{\v z}ek}{2015}]{Spergel2015} Spergel D.~N., Flauger R., Hlo{\v z}ek R., 2015, PhRvD, 91, 023518 

\bibitem[\protect\citeauthoryear{Springel, Di Matteo, \& Hernquist}{2005}]{Springel2005a} Springel V., Di Matteo T., Hernquist L., 2005, MNRAS, 361, 776 

\bibitem[\protect\citeauthoryear{Springel}{2005}]{Springel2005b} Springel V., 2005, MNRAS, 364, 1105 

\bibitem[\protect\citeauthoryear{Springel et al.}{2017}]{Springel2017} Springel V., et al., 2017, MNRAS, submitted (arXiv:1707.03397)

\bibitem[\protect\citeauthoryear{Sun et al.}{2009}]{Sun2009} Sun M., Voit G.~M., Donahue M., Jones C., Forman W., Vikhlinin A., 2009, ApJ, 693, 1142 
  
\bibitem[\protect\citeauthoryear{Takahashi et al.}{2012}]{Takahashi2012} Takahashi R., Sato M., Nishimichi T., Taruya A., Oguri M., 2012, ApJ, 761, 152

\bibitem[\protect\citeauthoryear{Troxel et al.}{2017}]{Troxel2017} Troxel M.~A., et al., 2017, PhRvD, submitted (arXiv:1708.01538)

\bibitem[\protect\citeauthoryear{van Daalen et al.}{2011}]{vanDaalen2011} van Daalen M.~P., Schaye J., Booth C.~M., Dalla Vecchia C., 2011, MNRAS, 415, 3649 

\bibitem[\protect\citeauthoryear{van Daalen et al.}{2014}]{vanDaalen2014} van Daalen M.~P., Schaye J., McCarthy I.~G., Booth C.~M., Dalla Vecchia C., 2014, MNRAS, 440, 2997
  
\bibitem[\protect\citeauthoryear{van Engelen et al.}{2015}]{vanEngelen2015} van Engelen A., et al., 2015, ApJ, 808, 7

\bibitem[\protect\citeauthoryear{Van Waerbeke, Hinshaw, \& Murray}{2014}]{vanWaerbeke2014} Van Waerbeke L., Hinshaw G., Murray N., 2014, PhRvD, 89, 023508

\bibitem[\protect\citeauthoryear{Velliscig et al.}{2014}]{Velliscig2014} Velliscig M., van Daalen M.~P., Schaye J., McCarthy I.~G., Cacciato M., Le Brun A.~M.~C., Dalla Vecchia C., 2014, MNRAS, 442, 2641 

\bibitem[\protect\citeauthoryear{Velliscig et al.}{2015}]{Velliscig2015} Velliscig M., et al., 2015, MNRAS, 454, 3328 

\bibitem[\protect\citeauthoryear{Vikhlinin et al.}{2006}]{Vikhlinin2006} Vikhlinin A., Kravtsov A., Forman W., Jones C., Markevitch M., Murray S.~S., Van Speybroeck L., 2006, ApJ, 640, 691 
  
\bibitem[\protect\citeauthoryear{Vogelsberger et al.}{2014}]{Vogelsberger2014} Vogelsberger M., et al., 2014, MNRAS, 444, 1518

\bibitem[\protect\citeauthoryear{Wang et al.}{2016}]{Wang2016} Wang W., White S.~D.~M., Mandelbaum R., Henriques B., Anderson M.~E., Han J., 2016, MNRAS, 456, 2301 

\bibitem[\protect\citeauthoryear{White \& Vale}{2004}]{White2004} White M., Vale C., 2004, APh, 22, 19 

\bibitem[\protect\citeauthoryear{Wiersma, Schaye, \& Smith}{2009}]{Wiersma2009a} Wiersma R.~P.~C., Schaye J., Smith B.~D., 2009, MNRAS, 393, 99 

\bibitem[\protect\citeauthoryear{Wiersma et al.}{2009}]{Wiersma2009b} Wiersma R.~P.~C., Schaye J., Theuns T., Dalla Vecchia C., Tornatore L., 2009, MNRAS, 399, 574 

\bibitem[\protect\citeauthoryear{Wilson et al.}{2012}]{Wilson2012} Wilson M.~J., et al., 2012, PhRvD, 86, 122005

\bibitem[\protect\citeauthoryear{Wyman et al.}{2014}]{Wyman2014} Wyman M., Rudd D.~H., Vanderveld R.~A., Hu W., 2014, PhRvL, 112, 051302 

\bibitem[\protect\citeauthoryear{Y{\`e}che et al.}{2017}]{Lyman2017} Y{\`e}che C., Palanque-Delabrouille N., Baur J., du Mas des Bourboux H., 2017, JCAP, 6, 047

\end{thebibliography}
\end{document}